\documentclass[fleqn,usenatbib]{mnras}

\usepackage{newtxtext,newtxmath}
\usepackage[T1]{fontenc}

\DeclareRobustCommand{\VAN}[3]{#2}
\let\VANthebibliography\thebibliography
\def\thebibliography{\DeclareRobustCommand{\VAN}[3]{##3}\VANthebibliography}

\usepackage{graphicx}	% Including figure files
\usepackage{amsmath}	% Advanced maths commands
\usepackage{pdflscape} %horizontal table page
\usepackage{multirow}
\usepackage{array}
\usepackage{tabularx}
\usepackage{orcidlink}

\newcommand{\aref}[1]{\hyperref[#1]{Appendix~\ref{#1}}}

\title[Kinematic age for PDS~70 system]{Kinematic structure and age of a Scorpius--Centaurus subcomponent: implications for PDS 70 planetary system}

\author[A. Arunima et al.]{
Arunima Arunima\orcidlink{0009-0009-2580-1027}$^{1}$\thanks{E-mail: arunima.arunima@anu.edu.au},
Michael J. Ireland$^{1}$,
Mark R. Krumholz\orcidlink{0000-0003-3893-854X}$^{1}$
\\
$^{1}$Research School of Astronomy and Astrophysics, Australian National University, Canberra ACT 2601, Australia \\
}
\date{Accepted XXX. Received YYY; in original form ZZZ}

\pubyear{\the\year{}}

\begin{document}
\label{firstpage}
\pagerange{\pageref{firstpage}--\pageref{lastpage}}
\maketitle

\begin{abstract}
PDS~70, the only system with two directly imaged protoplanets embedded in their natal disk, is a benchmark for giant-planet formation theory. Interpreting the planetary luminosities requires a precise age for the host star, but the widely adopted isochronal age of $5.4\pm1.0$~Myr is substantially younger than ages inferred for the Upper Centaurus--Lupus (UCL) population with which it is associated.
We address this discrepancy by determining a new kinematic age for PDS~70 using radial velocities for 405 candidate members of the Scorpius--Centaurus T subcomponent from the Wide Field Spectrograph (WiFeS) on the ANU 2.3~m telescope, reducing uncertainties by a factor of $\sim5.5$ compared to \textit{Gaia}.
%We supplement these measurements with H$\alpha$ and lithium equivalent widths as youth diagnostics.
We combine these with \textit{Gaia} DR3 astrometry in the \texttt{Chronostar} kinematic reconstruction framework to derive a posterior-weighted kinematic age of approximately $10$--$15$~Myr for all the T subcomponent stars, with a kinematic age of $\approx 12$--$16$~Myr for PDS 70, ruling out ages as young as the published isochronal estimate.
% are ruled out with high confidence. 
At this kinematic age, PDS~70~\textit{b}'s (and to a lesser extent, PDS~70~\textit{c}'s) luminosity is incompatible with warm-start cooling tracks under current dynamical mass constraints. The most natural interpretation is that PDS~70~\textit{b} formed $\approx10.5$ Myr after its host star, and no less than $\approx7.7$ Myr after it.
%\textcolor{red}{[abstract is still around 10 words too long.]}
\end{abstract}

\begin{keywords}
stars: kinematics and dynamics -- open clusters and associations: individual: Scorpius--Centaurus --  planets and satellites: formation 
% stars: pre-main-sequencestars: pre-main-sequence
\end{keywords}

%%%%%%%%%%%%%%%%%%%%%%%%%%%%%%%%%%%%%%%%%%%%%%%%%%

%%%%%%%%%%%%%%%%% BODY OF PAPER %%%%%%%%%%%%%%%%%%
\section{Introduction}
\label{sec:intro}
% looked at pds 70 age: muller paper cites riaud 2006 (who only say <10 Myr) and pecaut and mamajek (whose table shows pds 70 to be 8 Myr) but the muller paper uses MIST models to find age of $5.4\pm1.0$ Myr

Directly imaged planets are among the most valuable objects in exoplanet science. Unlike radial-velocity or transit discoveries, direct imaging yields resolved photometry and, increasingly, spectroscopy — providing luminosities, atmospheric temperatures, and compositional constraints that are inaccessible for the vast majority of known planets \citep{bowler2016}. Translating these observables into physical properties, however, requires knowing the planet's age.
Evolutionary cooling models predict that a planet's luminosity depends on its mass, time since formation, and entropy at the end of accretion, while the measured emission may also include contributions from accretion and circumplanetary material. Consequently, an error in the adopted age for a planet can lead to incorrect inferences about planet mass and formation pathway.

PDS~70 is a particularly important system in this context. It is a young K-type pre-main-sequence star surrounded by a transition disk containing two directly imaged protoplanets, PDS~70~\textit{b} and PDS~70~\textit{c} \citep{keppler2018,muller2018,haffert2019}. The planets reside within the disc cavity, and circumplanetary material has been detected or inferred around the companions \citep[e.g.][]{haffert2019,benisty2021}. Their luminosities, spectra, and increasingly precise dynamical constraints make PDS~70 a benchmark for testing giant-planet formation and early thermal-evolution models. Yet realising this potential requires an accurate age of the system. 

Unfortunately the age of the host star PDS~70 is currently not well constrained. The most widely adopted estimate comes from the MIST isochrone modelling \citep{choi2016}: \citet{muller2018} derived $5.4\pm1.0$~Myr from the stellar position in the Hertzsprung--Russell diagram relative to pre-main-sequence isochrones. However, PDS~70 is associated with the Upper Centaurus--Lupus (UCL) region of the well-studied Scorpius--Centaurus (Sco--Cen) association, and published age estimates for UCL span approximately $14$--$22$ Myr depending on the adopted membership sample, stellar evolutionary tracks and age diagnostics \citep[e.g.,][]{mamajek2002,pecaut_mamajek2016,squicciarini2021,kerr2021,ratzenboeck2023}. This suggests that PDS~70 might be substantially older than \citeauthor{muller2018}'s estimate. The discrepancy between the star's own isochronal age and the age of the association with which it is affiliated points to a broader problem: pre-main-sequence isochronal ages are notoriously model-dependent, sensitive to convection prescriptions, magnetic activity, photospheric spots, and, for younger ages, assumptions about the accretion history and the entropy content of accreting gas \citep{Hosokawa11a, bell2014,herczeg2015,feiden2016}. An independent age estimate, anchored to kinematics rather than stellar interior models, would place the luminosity-age comparison for PDS~70~\textit{b} and~\textit{c} on firmer ground, especially in the \textit{Gaia} era.

Kinematic ages exploit the fact that a stellar association expands from an initially compact configuration as its members disperse into the Galactic field \citep{blaauw1964,miret-roig2020,chronostar2}.
Rather than tracing individual stars backwards and applying a closest-approach criterion, an approach whose uncertainties grow rapidly with lookback time, modern implementations such as the \texttt{Chronostar} method that we adopt here represent an association as a  mixture of initially compact Gaussian phase-space components, evolve those  components forward through a Galactic potential, and fit their present-day  distributions while accounting for observational uncertainties \citep{chronostar,chronostar2}. By simultaneously fitting multiple kinematic components of a complex association, this approach can resolve spatial and  temporal substructure that single-component traceback methods smear over. These methods are independent of stellar interior physics and thus provide a complementary age scale against which isochronal ages can be calibrated.

The quality of a kinematic age depends critically on the precision of the input radial velocities. The \textit{Gaia} mission has provided precise five-parameter astrometry for millions of stars \citep{gaia_dr3}, but at the spectral types and brightness levels typical of T-Tauri stars, \textit{Gaia} radial velocities are either unavailable or limited to precisions of several km~s$^{-1}$ which is insufficient to resolve the internal kinematics of a structured region like Sco--Cen. Moreover, Sco--Cen is not a single coeval stellar population. \textit{Gaia} astrometry has revealed substantial spatial and kinematic substructure across the association, including multiple clusters and extended stellar chains that show systematic age-position covariance on scales of tens to hundreds of parsecs \citep[e.g.,][]{kerr2021,ratzenboeck2023,chronostar2}. Assigning to PDS~70 the mean age of an entire classical group may therefore obscure the age of its more immediate phase-space environment.

In this paper we therefore revisit the problem of the age of PDS 70, and of the UCL region of Sco-Cen in which it resides. We obtain new spectroscopic observations of candidate members in the region with the Wide Field Spectrograph (WiFeS) on the ANU 2.3~m telescope, from which we derive radial velocities with a median precision of $\sim0.91$~km~s$^{-1}$, a factor of $\sim5.5$ improvement over available \textit{Gaia} radial velocities. We then apply the \texttt{Chronostar} kinematic modelling framework \citep{chronostar,chronostar2} to obtain a new kinematic age estimate for the component of Sco-Cen to which PDS 70 belongs, providing an independent age for PDS~70's immediate phase-space environment. We then examine what this age implies for the luminosities of PDS~70~\textit{b} and \textit{c} relative to planetary cooling models.

The paper is organised as follows. In \autoref{sec:radial_velocities}, we describe our new observations. In \autoref{sec:age_extraction}, we describe the \texttt{Chronostar} framework, construct the six-dimensional input catalogue from our new radial velocities and \textit{Gaia} DR3, and quantify the parameter and stellar-age uncertainties. The implications of this decomposition are examined in \autoref{sec:implications}, focusing first in \autoref{subsec:chronostar_past} on the kinematic decomposition and reconstructed past phase-space of the entire region, and then in \autoref{subsec:pds70} on the implications for PDS~70~\textit{b} and \textit{c}, and the broader implications for the planetary cooling and formation. We conclude in \autoref{sec:conclusions}.
 
\section{Radial velocities}
\label{sec:radial_velocities}

Here we describe our observations and the process by which we extract radial velocities from them.

\subsection{Sample Selection}

Our sample is drawn from Table C1 of the catalogue from \cite{chronostar2}, which identifies probabilistic members of Sco-Cen subgroups using \textit{Gaia} DR2 astrometry and radial velocities. We select all stars assigned to the `T' subcomponent, to which PDS~70 belongs. The `T' subcomponent is a compact UCL group at $\sim 140$ pc with age $\sim 15$ Myr \citep{chronostar2}. The only initial selection cuts are on the \textit{Gaia} radial velocity uncertainty ($\sigma_\text{RV} > 3 \,\text{km s}^{-1}$) and apparent magnitude ($G_0<16.5$ mag), prioritising low mass stars for which the available \textit{Gaia} radial velocities are relatively uncertain or unavailable.

\begin{figure}
    \centering
    \includegraphics[width=\linewidth]{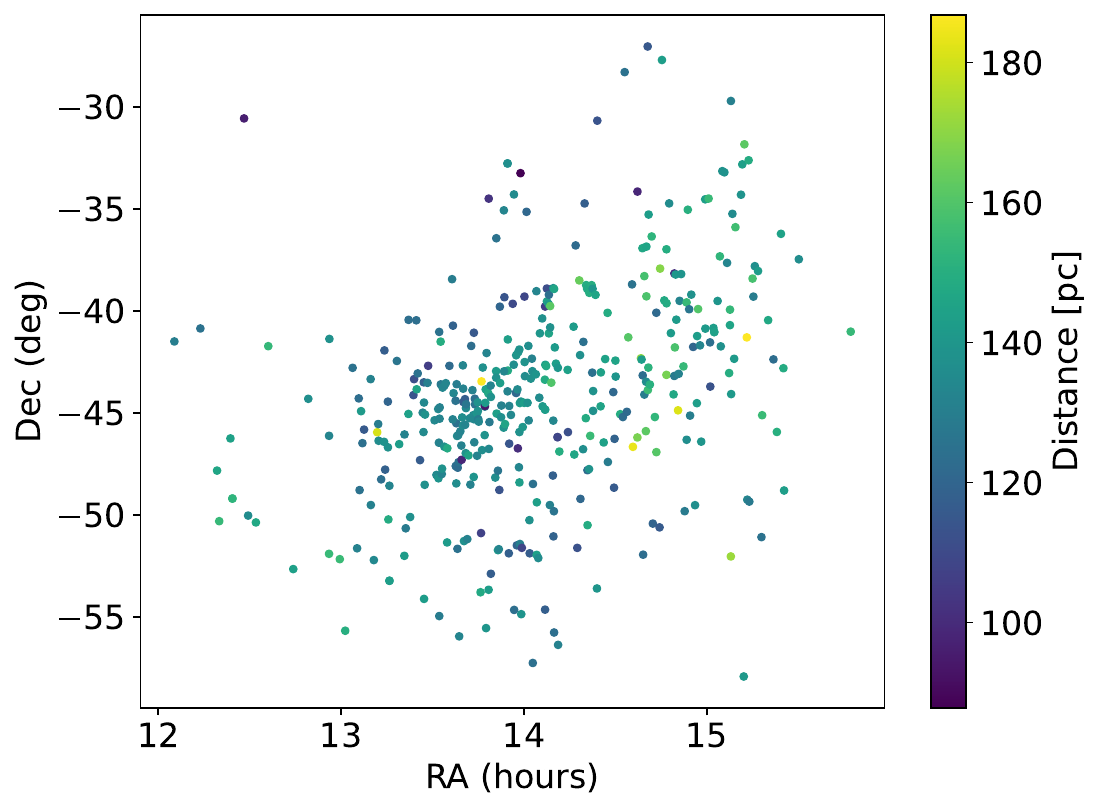}
    \caption{Spatial distribution of the observed sample in the equatorial coordinate system, with the colour bar showing the distance to each star from \textit{Gaia}.}
    \label{fig:spatial_sample}
\end{figure}

To ensure reliable spectroscopic measurements, we remove targets with neighbouring sources within $2.5''$ and with a magnitude difference of $\Delta G < 2.5$ mag. Such systems are not spatially resolved at the angular resolution of the ANU 2.3\,m telescope and if targeted would likely suffer from significant flux contamination in the extracted spectra. After applying this cut we are left with a sample of 735 stars, of which we observe 405. The spatial distribution of the sample, together with the distances to individual stars, is shown in \autoref{fig:spatial_sample}.

% This gives 735 stars that are named T001 to T735. We upload all of these as the observing proposal, however, not all targets were observed due to scheduling and conditions. 

%A sample of 405 targets was observed after applying these selection criteria. The spatial distribution of the sample, together with the distances to individual stars, is shown in \autoref{fig:spatial_sample}.

\begin{figure}
    \centering
    \includegraphics[width=\linewidth]{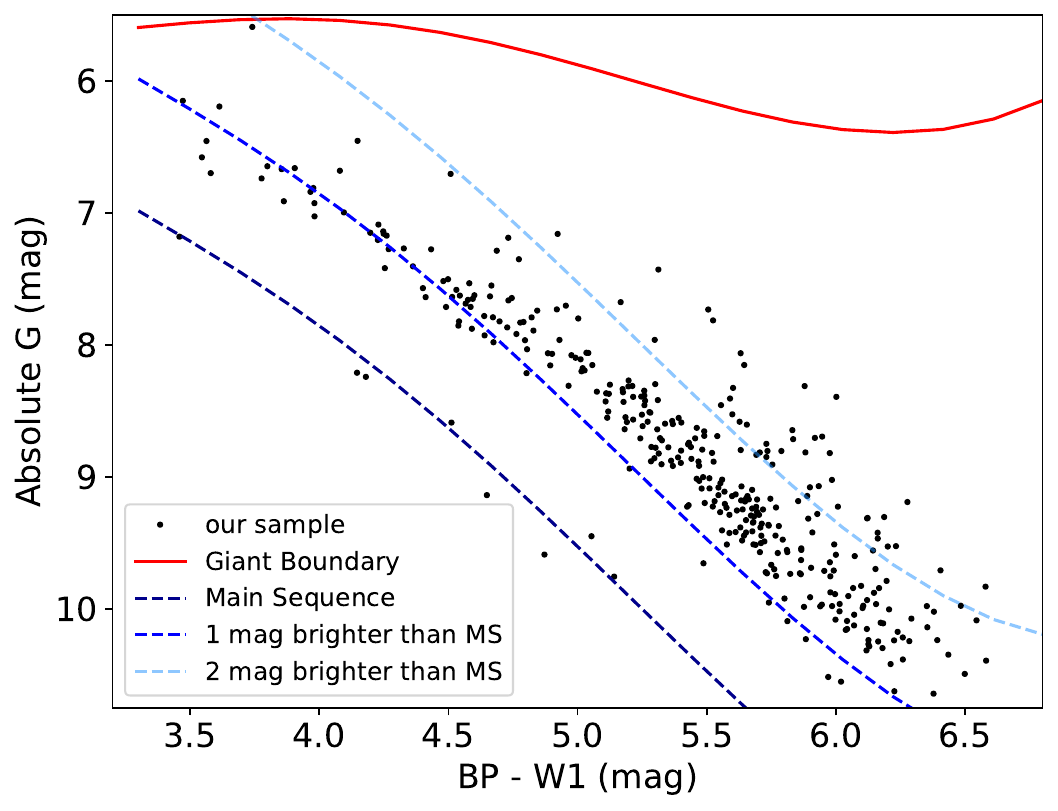}
    \caption{Colour-magnitude diagram of the final sample, zoomed in to show the region containing most of the stars for clarity. The solid red line shows the upper luminosity boundary used to reject giants from the \citet{zerjal21} sample, while the dashed blue lines show the main sequence and 1 and 2 magnitudes brighter than the main sequence in decreasing order of darkness of the colour.}
    \label{fig:sample}
\end{figure}

Following the photometric youth criteria of \citet{zerjal21} (their Section 2.1), we verify that these stars occupy the expected pre-main-sequence locus in the BP-W1 versus $M_G$ diagram (\autoref{fig:sample}). This confirms that most of our sample is consistent with the expected location of young, low-mass Sco-Cen members.

\subsection{Observations}
%\textcolor{red}{\textit{(this is the only part that uses past tense now:)}

Observations were carried out between February and July 2025 over 94.02 hours using the Wide Field Spectrograph (WiFeS) on the ANU 2.3\,m telescope at Siding Spring Observatory with resolving power of 7000 in the red arm covering 5294 -- 7059 \AA. We used the RT560 beam splitter. The exposure times ranged from about 3 to 15 minutes per star based on the magnitude of the star. 
% Neon-Argon lamp frames were taken between two observations of nearby stars together to maintain a stable wavelength calibration.
To minimise drift in the wavelength calibration over the course of the night without incurring the overhead of calibrating every individual exposure, we grouped stars that were close together on the sky (and observed consecutively), and took a single neon-argon arc lamp frame between each such pair, which is then applied to both adjacent observations thus maintaining a stable wavelength calibration.
%}

The raw data are reduced using the \texttt{PyWiFeS} pipeline \citep{pywifes2014}\footnote{\url{https://pywifes-pipeline.readthedocs.io/en/latest/}},
%(the newer version's paper isn't out yet?), 
which performs bias subtraction, flat-fielding, wavelength calibration, and flux calibration to produce fully reduced three-dimensional data cubes.

\subsection{Spectral Extraction}

For the majority of targets, we also use \texttt{PyWiFes} to extract spectra using the automated detection and aperture extraction routines, which we briefly summarise here for the reader's convenience. We identify sources in the datacube using a Gaussian-convolved detection image formed from the spectral median of the spaxels, and extract spectra within a circular aperture centred on the detected source position. We carry out sky subtraction using an annular region surrounding the source aperture. For each wavelength slice, we compute the source flux by summing within the aperture, estimate the sky background using the median flux within the annulus and subtract, and propagate the variance by summing the variance within the source aperture and including the contribution from the sky estimate uncertainty.

% For a small number of faint sources, 
% We are doing this for all sources with automatic extraction 
We adopt modified source detection and extraction parameters relative to the default pipeline configuration in order to improve recovery and mitigate contamination from nearby objects. The modified parameters are:

\begin{itemize}
    \item Detection threshold: $\sigma = 2$ (reduced from the default value to improve sensitivity to faint targets),
    \item Extraction radius: $r = 2.5''$ (reduced from $5''$ to minimise contamination from nearby sources),
    \item Sky estimation: median statistic (instead of weighted mean) to reduce the impact of outliers and nearby contaminating sources.
\end{itemize}

We validate these modifications through visual inspection of detection plots and extracted spectra. Using this automated pipeline, we obtain the source flux, variance, and sky-subtracted spectra for the majority of targets.

For a subset of targets ($23$ objects), the automated extraction either fails to detect the source or produces contaminated spectra due to nearby objects. For these cases, we implement a custom extraction procedure that closely reproduces the \texttt{PyWiFeS} extraction algorithm but with manually supplied target coordinates instead of automatic source detection. For these sources we manually supply target coordinates (Right Ascension, Declination), convert to pixel coordinates using the datacube world coordinate system (WCS), and define the source aperture and sky annulus at these positions.
%The resulting spectra include the source flux, variance, and sky spectrum, written to FITS files with headers matching the format produced by the automatic \texttt{PyWiFeS} extraction and expected by the downstream radial-velocity pipeline.

\subsection{Radial Velocity Determination}
\label{subsec:rad_vel}

We measure radial velocities using a template-matching approach, consisting of three main steps: spectral cleaning and wavelength interpolation, template matching and velocity fitting, and sky-line-based zero-point correction.

\subsubsection{Spectral cleaning and wavelength interpolation}
\label{subsubsec:spec_cleaning_wave_interpolation}
We construct the wavelength solution of each reduced spectrum from the linear dispersion keywords in the FITS header, namely the reference wavelength of the first pixel (\texttt{CRVAL1}) and the wavelength step per pixel (\texttt{CDELT1}). We take the per-pixel flux uncertainty as the square root of the variance array produced by the reduction pipeline. We then discard pixels with unreliable flux or uncertainty: those with non-finite or non-positive flux, those with non-finite or non-positive uncertainty, and those with an uncertainty greater than twice the median uncertainty of the spectrum. 
We also remove all pixels with wavelengths blueward of 5900\,\AA, using only the red-arm spectra, where the signal-to-noise ratio of the red-arm spectra drops steeply for our cool targets. These criteria provide a robust rejection of low-quality pixels and residual reduction artefacts.
% of each spliced spectrum (5900 -- 7059\,\AA). This avoids contamination from the blue arm and the overlap region between the two arms\footnote{For our cool, low-mass targets the blue arm has substantially lower signal-to-noise ratio and contributes little velocity information relative to the molecular and atomic features in the red. Excluding it also avoids the splice region between the two arms, where differences in flux calibration, wavelength solution, and instrumental resolution between the arms can introduce systematic errors, and ensures that all fitted pixels share a single wavelength calibration on an air wavelength scale.}.

We then resample the retained pixels onto a logarithmically spaced wavelength grid using cubic-spline interpolation of both flux and uncertainty, so that a Doppler shift corresponds to a constant shift in pixels. The grid spans the minimum to maximum retained wavelength with $N = \lfloor c\ln(\lambda_\text{max}/\lambda_\text{min})/\Delta v_0 \rfloor + 1$ pixels, where $\Delta v_0 = 1.5$\,km\,s$^{-1}$, giving a velocity step of $\Delta v \approx 1.5$\,km\,s$^{-1}$ per pixel.
On this grid we define a mask of pixels excluded from all subsequent fitting. It removes 50 pixels at each end of the spectrum and the following regions affected by telluric absorption, airglow emission, and strong stellar emission:
% Na\,D sky emission (5875 -- 5905\,\AA), this is removed in the edge pixel cuts anyway
[O\,I] $\lambda6300$ (6285 -- 6325\,\AA), [O\,I] $\lambda6363$ (6340 -- 6385\,\AA), H$\alpha$ (6550 -- 6575\,\AA, masked in all spectra regardless of whether it is in emission), and the O$_2$ B band (6860 -- 6940\,\AA).

% We first wavelength-calibrate each observed spectrum using and then interpolate onto a logarithmically spaced wavelength grid, so that Doppler shifts correspond to constant pixel shifts in velocity space. We next remove pixels with unreliable uncertainties, including those with unusually large values, non-finite flux or variance, and all wavelengths below 5900 \AA. We also mask regions affected by strong telluric absorption, airglow emission, and other contamination, including the O$_2$ A band (6860 -- 6940 \AA), prominent [O\,I] lines (6285 -- 6325 \AA, 6340 -- 6385 \AA) and H$\alpha$ emission (6550 -- 6575 \AA).
% start at 5600 ang
%remove 6850 - end all telluric
% something funny going on at 5910 -- 5990, 6470 -- 6490 \AA
%removing these didn't make any difference in the two T501 rad velocity measurements' diff so keeping the description same as the ones applied to the rest. duplicate velocities are fixed by doing a sky correction.
To remove the continuum, we first divide each resampled spectrum by its median flux. We then fit an unweighted fourth-order polynomial in wavelength to the spectrum and identify outlying pixels, which are dominated by absorption lines, as those whose residuals from the fit deviate from the median residual by more than $2.5\sigma$, where $\sigma$ is the standard deviation of the residuals. We refit the polynomial to the remaining pixels and repeat this procedure for a fixed total of five iterations. The set of rejected pixels is unchanged after five iterations for 423 of the 436 spectra in our sample (and after at most seven iterations for all spectra; for the remaining 13, the five-iteration continuum differs from the fully converged continuum by less than 0.2 per cent). We then divide the spectrum by the final polynomial.
A fourth-order polynomial cannot follow continuum structure on scales of tens of \AA, such as residual flux-calibration features. We therefore compute a running mean of the normalised spectrum using a 501-pixel boxcar ($\approx 750$\,km\,s$^{-1}$, or $\approx 16$\,\AA\ at 6500\,\AA) and subtract it from the spectrum. The result is a zero-mean spectrum in which structure on scales broader than the boxcar is suppressed, so that the cross-correlation is driven by spectral lines rather than broad continuum variations.
We propagate the flux uncertainties through the division steps by dividing them by the same median flux and polynomial continuum, i.e.\ $\sigma_{f/C} = \sigma_f / C$, treating the continuum as noise-free since it is fit to many pixels. We do not modify the uncertainties for the running-mean subtraction: the mean over 501 pixels (approximately 18 independent resolution elements) has an uncertainty of $\approx 1/\sqrt{18}\sigma_f=0.24\sigma_f$, which increases the per-pixel uncertainty by only $\approx 3$ per cent when added in quadrature. Any residual misestimate of the uncertainties is absorbed by the $\chi^2$ rescaling described in \autoref{subsubsec:template_rad_vel}.

% To suppress continuum structure, we then normalise each spectrum iteratively by fitting a low-order polynomial continuum to the logarithmically re-binned spectrum and rejecting outlying pixels using sigma clipping. We repeat this procedure until the continuum fit converges. We then divide the spectrum by the converged continuum model, and a smoothed residual continuum component was removed to minimise sensitivity to broad spectral slopes. The associated uncertainty array was normalised in the same way.

\subsubsection{Template matching and radial velocity fit}
\label{subsubsec:template_rad_vel}

We next extract radial velocities by cross-correlating the observed spectra with a spectral template library. For our library we employ the NewEra grid of synthetic stellar spectra \citep{hauschildt2025}, spanning a range of effective temperatures $T_\text{eff}=(2800-5000\,K)$ in intervals of $100\,K$ and surface gravities $\log g =4, 4.5$  at solar metallicity ($Z=0$).
We choose these templates to match the expected cool, low-mass nature of the observed Sco-Cen targets.
% https://www.aanda.org/articles/aa/full_html/2025/06/aa54171-25/aa54171-25.html
We convert the template wavelengths from vacuum to air using the relation of \citet{birch1994} and convolve each template spectrum to the approximate WiFeS instrumental resolution. We then interpolate it onto the logarithmic wavelength grid and normalise it with the same median, polynomial, and running-mean procedure that we apply to the observed spectra. 

The convolution ensures that template line profiles reflect the instrumental broadening of the data, allowing the radial velocity fit to depend on Doppler shift rather than differences in spectral resolution. We implement the convolution step by using a Gaussian of fixed width in wavelength, with full width at half maximum, $\text{FWHM} = 0.9$\,\AA\ and $\sigma = \text{FWHM}/2.355 = 0.38$\,\AA, which is the instrumental FWHM of the WiFeS red arm at 6300\,\AA\ and corresponds to a resolving power $R = \lambda/\text{FWHM} = 7000$ at that wavelength. Because the kernel width is constant in wavelength, the effective resolving power varies across our fitted range, from $R \approx 6600$ at 5900\,\AA\ to $R \approx 7600$ at 6850\,\AA.

% applying a Gaussian line spread function with resolving power $R \approx 7000$ at the reference wavelength, corresponding to a full width at half maximum of $\text{FWHM} = \lambda/R$ and a Gaussian sigma of $\sigma = \text{FWHM}/2.355$. We choose this $R$ value because the instrumenal FWHM is approximately 0.9 \AA at $\lambda = 6300 $ \AA, which corresponds to $R=7000$.

For each target, we cross-correlate each of the convolved, interpolated templates with the observed spectrum via fast Fourier transform, and select the template that produces the highest maximum cross-correlation as the best match. The location of the maximum, within $\pm 200$ pixels ($\pm 300$\,km\,s$^{-1}$) of zero lag, then provides a coarse radial velocity estimate and a first-pass identification of the most suitable stellar template. We then refine the velocity using a maximum-likelihood fit for the shift $s$ (in pixels) of the best-matching template relative to the observed spectrum. 
In this step, we shift the template using the Fourier shift theorem, evaluating the likelihood only over unmasked pixels. We minimise the negative log-likelihood
\begin{equation}
    \mathcal{L}(s) \equiv -\ln L(s) = -\sum_i \ln\left[\exp\left(-\tfrac{1}{2}r_i^2\right) + \epsilon\right], \qquad r_i = \frac{f_i - m_i(s)}{\sigma_i},
    \label{eq:lnL}
\end{equation}
where $f_i$ and $\sigma_i$ are the normalised observed flux and its uncertainty in pixel $i$, $m_i(s)$ is the shifted template, and the sum runs over unmasked pixels. The constant $\epsilon = 10^{-4}$ makes the likelihood robust to outliers by capping the contribution of any single pixel at $\ln(1/\epsilon) \approx 9.2$. We minimise $\mathcal{L}$ using bounded Brent minimisation via \texttt{scipy.optimize.fminbound} over shifts within $\pm 5$\,km\,s$^{-1}$ of the coarse velocity.

Next we iteratively remove outlying pixels. Using the best-fit shift, we compute the normalised residuals $r_i$ of \autoref{eq:lnL} for all pixels in the original mask, exclude those with $|r_i| \geq 3$, and re-minimise $\mathcal{L}$ over the remaining pixels within $\pm 1$ pixel of the previous best fit. We repeat this, recomputing the rejected pixels from the original mask at each iteration, and stop when the number of retained pixels is unchanged between two successive iterations, when fewer than 50 pixels remain, or after a maximum of 10 iterations. The best-fit velocity is then $v = s\,\Delta v$.

% . We also apply iterative $3\sigma$ clipping to remove outlier pixels, and repeat the fit until the residuals stabilise.

After finding the best fit, we estimate radial velocity uncertainties. This is done by getting uncertainty on $s$ by evaluating the curvature of $\mathcal{L}$ at the best-fit shift $\hat{s}$, using the central finite-difference estimate of the second derivative,
\begin{equation}
    \sigma_s^2 = \left(\frac{\mathrm{d}^2\mathcal{L}}{\mathrm{d}s^2}\right)^{-1} \approx \frac{h^2}{\mathcal{L}(\hat{s}+h) + \mathcal{L}(\hat{s}-h) - 2\mathcal{L}(\hat{s})}.
\end{equation}
Because $\mathcal{L}$ is evaluated numerically, the second difference can be zero or negative for very small $h$, when the change in $\mathcal{L}$ is below numerical precision or dominated by small-scale structure in the likelihood surface. We therefore try step sizes of $h = 0.25$, 0.5, 1, 2, and 4 pixels (0.375 -- 6\,km\,s$^{-1}$) in increasing order, and adopt the estimate from the smallest step for which the denominator is positive and greater than $10^{-12}$; we do not combine estimates from different step sizes. The velocity uncertainty is $\sigma_v = \sigma_s\,\Delta v$.
% using multiple finite-difference step sizes and convert the inverse Hessian estimate into an uncertainty in velocity. 

% If this curvature-based estimate was unstable or non-finite, we fall back to a profile-likelihood method, in which we scan the likelihood around the best fit and defined the $1\sigma$ interval as that over which $\Delta(-\ln L) = 0.5$. 

% Finally, the formal uncertainty was empirically re-scaled using the residual scatter of the best-fitting model. Specifically, we computed the mean squared normalised residuals over the final masked pixels and applied an additional factor involving the instrument resolution ($R=7000$) and the logarithmic velocity sampling ($\Delta v$), yielding a final uncertainty estimate that accounts for both residual template mismatch and the finite number of independent resolution elements.  
Finally, we rescale the formal uncertainty to account for two effects that the likelihood in \autoref{eq:lnL} does not capture. First, residual template mismatch and any misestimate of the pipeline uncertainties cause the residuals to scatter by more than the per-pixel flux uncertainties $\sigma_i$ imply: we correct for this using the mean squared normalised residual over the $N_\text{pix}$ pixels retained after clipping, $\overline{\chi^2} = N_\text{pix}^{-1}\sum_i r_i^2$. Second, \autoref{eq:lnL} treats every pixel as independent, whereas our logarithmic grid oversamples the spectrum: one resolution element spans $c/(R\,\Delta v) \approx 28.5$ grid pixels for $R = 7000$ and $\Delta v = 1.5$\,km\,s$^{-1}$, so neighbouring pixels are strongly correlated and the likelihood overstates the information content of the spectrum by approximately this factor. Our final uncertainty is therefore
\begin{equation}
    \sigma_{v,\text{final}} = \sigma_v \left(\overline{\chi^2}\,\frac{c}{R\,\Delta v}\right)^{1/2}.
\end{equation}

\subsubsection{Sky-line correction}
\label{subsubsec:sky_corr}
We assign a final radial velocity for each spectrum by applying two corrections to the best-fit Doppler shift computed as described above. We first correct for the barycentric velocity computed by the reduction pipeline and stored in the \texttt{RADVEL} header keyword, and second add an additional sky-line-based correction to account for small zero-point shifts in the wavelength solution. 
These zero-point shifts can vary on timescales shorter than the cadence at which arc lamps were taken, hence the need for a sky-line-based correction.

We measure the positions of several strong night-sky emission lines in each spectrum: the [O\,I] lines at 5577.35, 6300.31, and 6363.78 \AA~and the OH Meinel line at 6863.97 \AA. 
For each line, we estimate a local continuum, and discard lines whose peak height is less than three times the standard deviation of the continuum. We then fit each line with a Gaussian profile, discarding fits with unphysical widths (FWHM outside the range 2 -- 140\,km\,s$^{-1}$), low signal-to-noise ratios (SNR$_j <5$), or excessively large shifts ($|\Delta v_j| > 50$\,km\,s$^{-1}$), where $\Delta v_j$ is the velocity offset defined as $\Delta v_j = c\,(\lambda_{\text{fit},j} - \lambda_{\text{rest},j})/\lambda_{\text{rest},j}$.

To combine the individual line velocity offset measurements into a single correction,
%The per-line velocity offsets were then combined into a single sky correction using an SNR-based weighted mean. Before averaging, 
we first perform robust outlier rejection by computing the median of the offsets and the median absolute deviation (MAD), estimating the standard deviation as $\sigma \approx 1.4826 \times \text{MAD}$ (with a floor of 0.5 km/s), and rejecting any line whose offset differs from the median by more than 2.5$\sigma$; the exception is if this clipping removes all but one line, in which case we keep all lines to avoid being overly aggressive. After this rejection procedure, we compute our final correction as a SNR-squared-weighted mean of the individual line velocity offsets, take the uncertainty on the correction to be the weighted standard deviation of the individual line offsets, reduced by a factor of $\sqrt{n}$, where $n$ is the number of retained lines.
%The final uncertainty was estimated from the line-to-line scatter of the retained lines, scaled by $\sqrt{n}$.
We then apply the resulting correction to the barycentre-corrected best-fit Doppler solution, combining the uncertainties from that solution and our sky-line correction in quadrature.

%We apply the sky correction was uniformly to every spectrum, whether or not the target had multiple observations. The corrected stellar radial velocity was computed by subtracting the sky-line correction from the previously obtained radial velocity. The sky-correction uncertainty was combined in quadrature with the formal radial-velocity uncertainty. 

\subsubsection{Multi-epoch targets}
\label{subsubsec:multi_epoch}

We observed 27 of our targets on more than one night (24 on two nights, two on three, and one on four; 58 spectra in total). These repeated targets provide an opportunity to verify the accuracy of the sky-line correction. For each target, we quantify the consistency of its epochs using the most discrepant pair of epochs $(i,j)$, i.e.\ the pair that maximises
\begin{equation}
    \frac{|v_i - v_j|}{\sqrt{\sigma_i^2 + \sigma_j^2}},
    \label{eq:pair_sigma}
\end{equation}
where $v_i$ and $\sigma_i$ are the radial velocity and its uncertainty at epoch $i$. Before the sky-line correction we use the velocities and their uncertainties of \autoref{subsubsec:template_rad_vel}; after it we use the corrected velocities and the uncertainties that include $\sigma_\text{sky}$. If our uncertainties are correct and the stars' velocities are constant, the quantity in \autoref{eq:pair_sigma} should follow a half-normal distribution with unit scale, so that ${\approx}95$ per cent of targets lie below 2 and ${\approx}1$ per cent above 2.5.

\begin{figure}
    \centering
    \includegraphics[width=\linewidth]{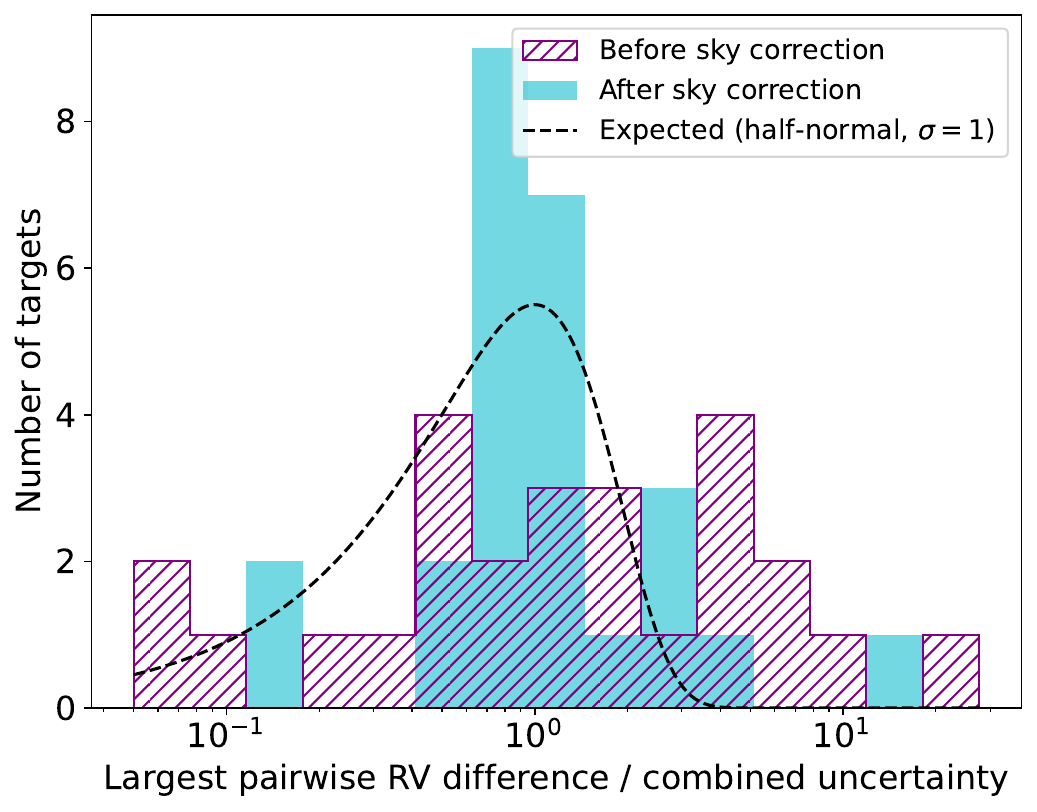}
    \caption{Distribution, over the 27 targets with repeat observations, of the largest pairwise radial-velocity difference between epochs in units of the combined uncertainty of that pair (\autoref{eq:pair_sigma}), before (purple, hatched) and after (cyan, filled) the sky-line correction. The dashed curve shows the half-normal distribution with unit scale expected if the uncertainties are correct and the velocities of each target constant between epochs, as for a single star.}
    \label{fig:multiple}
\end{figure}

Before the sky-line correction, the repeat observations are inconsistent at a level well beyond this expectation (\autoref{fig:multiple}). Only 16 of the 27 targets agree to within $2\sigma$, and the largest discrepancy is $23\sigma$. The median velocity difference between epochs is 1.5\,km\,s$^{-1}$, compared with a median combined uncertainty of 1.2\,km\,s$^{-1}$. After the correction, 22 of the 27 targets agree to within $2\sigma$, and the median difference between epochs falls to 1.1\,km\,s$^{-1}$. This improvement indicates that most of the original discrepancies were caused by instrumental shifts in the wavelength solution rather than by changes in the stars' velocities.

Three targets remain discrepant at more than $2.5\sigma$ after the correction: T029 ($13\sigma$; a velocity difference of 6.8\,km\,s$^{-1}$), T027 ($3.7\sigma$; 2.9\,km\,s$^{-1}$), and T619 ($2.7\sigma$; 5.3\,km\,s$^{-1}$ between the most discrepant two of its three epochs), compared with ${\approx}0.3$ expected by chance. These may be spectroscopic binaries. All 25 repeat targets with a \textit{Gaia} DR3 RUWE measurement, including these three, have RUWE $< 1.4$ (maximum 1.36), the threshold commonly used to flag likely unresolved binaries \citep{lindegren2018}; the remaining two lack a RUWE measurement. However, RUWE is insensitive to short-period binaries whose orbits produce no detectable astrometric signal, so a low RUWE does not rule out binarity for these targets.

For targets with repeat observations, we combine the epochs into a single radial velocity $\bar{v}$ using an inverse-variance weighted mean. We adopt as its uncertainty the larger of the formal uncertainty of this mean and the weighted standard deviation of the individual epochs, $[\sum_k w_k (v_k - \bar{v})^2 / \sum_k w_k]^{1/2}$ with $w_k = \sigma_k^{-2}$, ensuring that for the small fraction of targets where the radial velocities remain significantly inconsistent over multiple epochs, such as those above, we are not assigning artificially small uncertainties. 

\subsubsection{Comparison to \textit{Gaia} radial velocities}

\begin{figure}
    \centering
    \includegraphics[width=\linewidth]{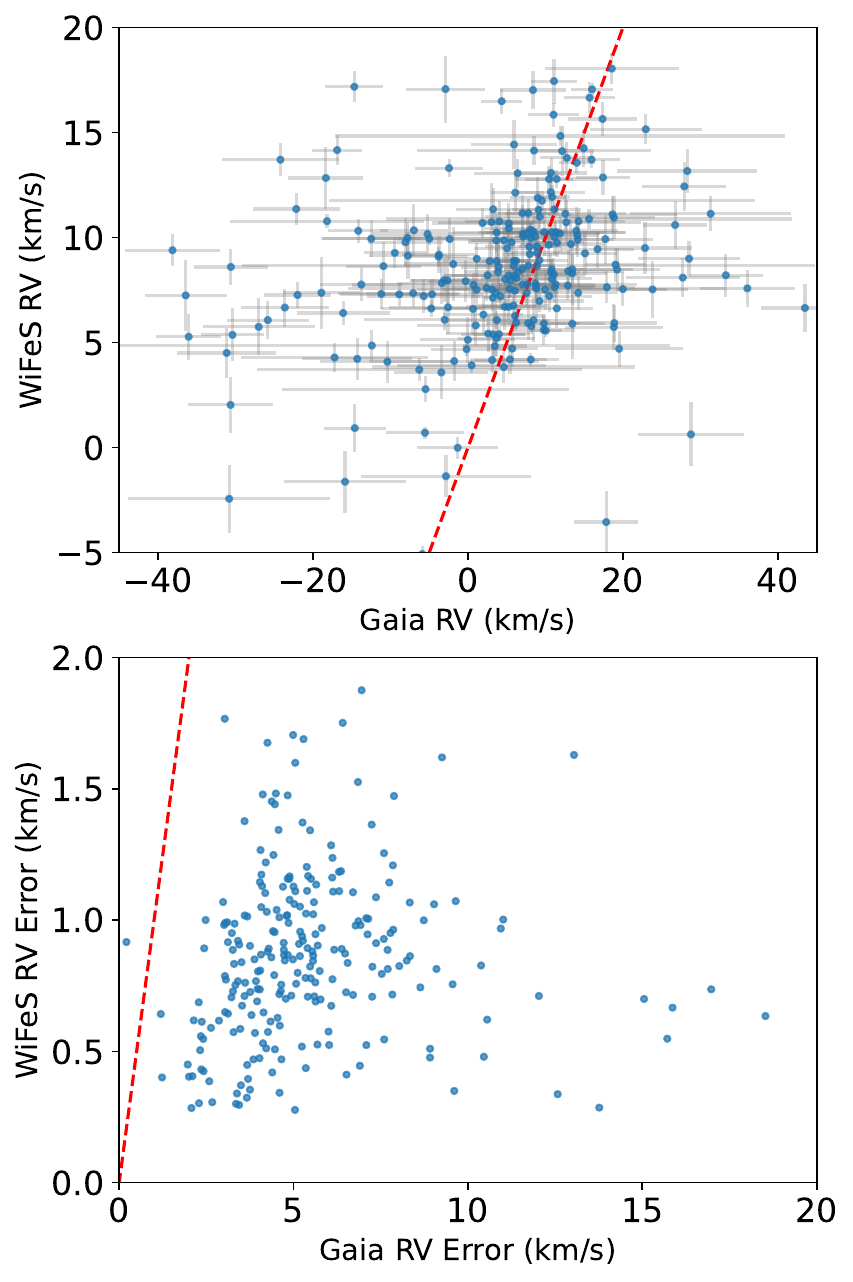}
    \caption{Top: \textit{Gaia} vs WiFeS radial velocities. Bottom: The uncertainties in the WiFeS and \textit{Gaia} radial velocities. The red dashed lines show the one-to-one relation.}
    \label{fig:rad_vel}
\end{figure}

\autoref{fig:rad_vel} compares our final WiFeS radial velocities and uncertainties with the \textit{Gaia} radial velocities and their uncertainties available for 254 of our 405 targets. 
The WiFeS radial velocities mostly lie on the one-to-one relation (red, dashed line) agreeing with the \textit{Gaia} radial velocities, but with noticeably smaller scatter. This is also reflected in the uncertainties: for these 254 targets, the median \textit{Gaia} radial velocity uncertainty is 4.98 km s$^{-1}$, compared with a median WiFeS uncertainty of 0.85 km s$^{-1}$ (0.91 km s$^{-1}$ for all targets), an improvement of a factor of about 6 (5.5 for all targets in the sample). The \textit{Gaia} radial velocity uncertainties range from 0.21 to 38.45 km s$^{-1}$ for our sample, whereas the WiFeS uncertainties are much more tightly concentrated. 
%Because the sample was selected to be kinematically coherent, the new measurements provide substantially improved precision for stars with uncertain or unavailable \textit{Gaia} radial velocities.

\subsection{Youth indicators}
\label{subsec:youth}
Spectroscopic youth indicators provide a useful check on the age and membership of our observed sample. Because the T component is expected to be a young population, its members should exhibit youth signatures such as chromospheric activity and lithium absorption. 

% \textcolor{red}{should i put some spectra plots- for fitting, for h alpha and li lines? No, can put them if the referee asks}
% - check the binary things with the stars which dont agree with gaia dr2 velocities.
\subsubsection{H$\alpha$}
Strong emission across the Balmer series, particularly the H$\alpha$ line at 6564.6 \AA, is associated with chromospheric activity in young T Tauri stars \citep{lyra_portodemello_2005,kurosawa2006,soderblom2014}. We calculate the equivalent width of the H$\alpha$ line directly from the extracted spectra of the stars by fitting a local continuum and integrating the line profile. We estimate the continuum by fitting a linear model to two line-free windows on either side of H$\alpha$, 6530--6550 \AA\,and 6580--6600 \AA, then compute the equivalent width by integrating over 6550--6580 \AA\, and subtracting this value from the linear model integrated over the same interval. With this convention, negative equivalent widths correspond to emission and positive values to absorption. We estimate uncertainties from the scatter of the continuum-normalised flux in the continuum window, and propagate across the integration interval. 

\begin{figure}
    \centering
    \includegraphics[width=\linewidth]{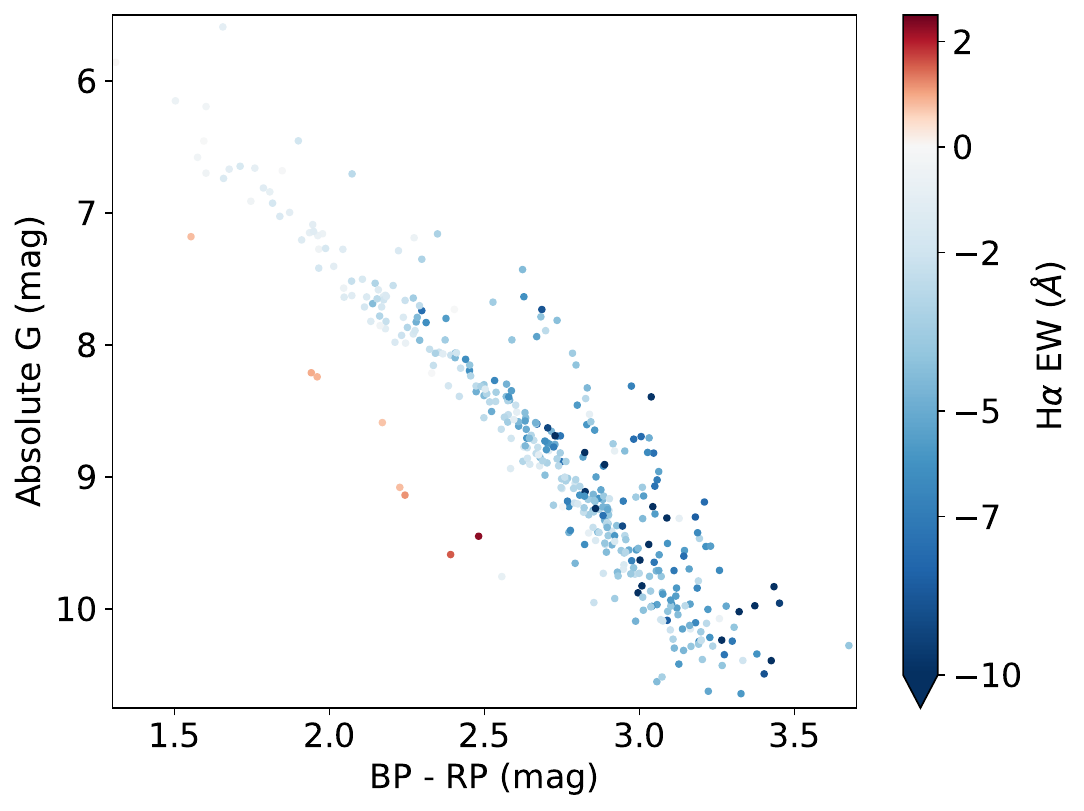}
    \caption{Colour-magnitude diagram of the observed sample, with the colour scale showing H$\alpha$ equivalent width in \AA. Most of the sample shows H$\alpha$ emission, consistent with youth.}
    \label{fig:halpha}
\end{figure}

We show the resulting H$\alpha$ equivalent widths in \autoref{fig:halpha}. We classified stars with H$\alpha$ equivalent widths below \(-3\) \AA\,as ``young stars''. For stars with multiple observations, we combine the equivalent widths and their uncertainties using inverse variance weighting, just as for the radial velocities. However, if any one of individual observations satisfy the H$\alpha$ equivalent width youth criteria, we label the star as young. This extra proviso only affects one star, whose combined equivalent width of $-2.92 \pm 0.91$ \AA\, sits almost exactly at our adopted threshold. In total we find that 237 out of 405 stars in our sample are classified as ``young'' based on their H$\alpha$ equivalent width, corresponding to 58.5\% of the sample; we refer to these as H$\alpha$-young stars. The median H$\alpha$ equivalent width over the full sample (not just the H$\alpha$-young stars) is $-3.50 \pm 0.27$ \AA. \autoref{fig:halpha} shows that the young stars are predominantly located above the main sequence, as expected for a young low-mass Sco-Cen population. By contrast, the few stars with positive or near-zero equivalent widths sit nearly on the main sequence.

\subsubsection{Lithium}
\label{subsubsec:li}

The Li feature at 6707.856 \AA\ is a well-established youth indicator in late-type stars because lithium is progressively depleted during pre-main-sequence evolution and on the main sequence \citep{zerjal21,eagles2023,GutierrezAlbarran2024}. For a population with an age of $\lesssim 20$ Myr, many low-mass members are therefore expected to retain strong lithium absorption, making the line a useful discriminator between young Sco-Cen members and older field contaminants.

\begin{figure}
    \centering
    \includegraphics[width=\linewidth]{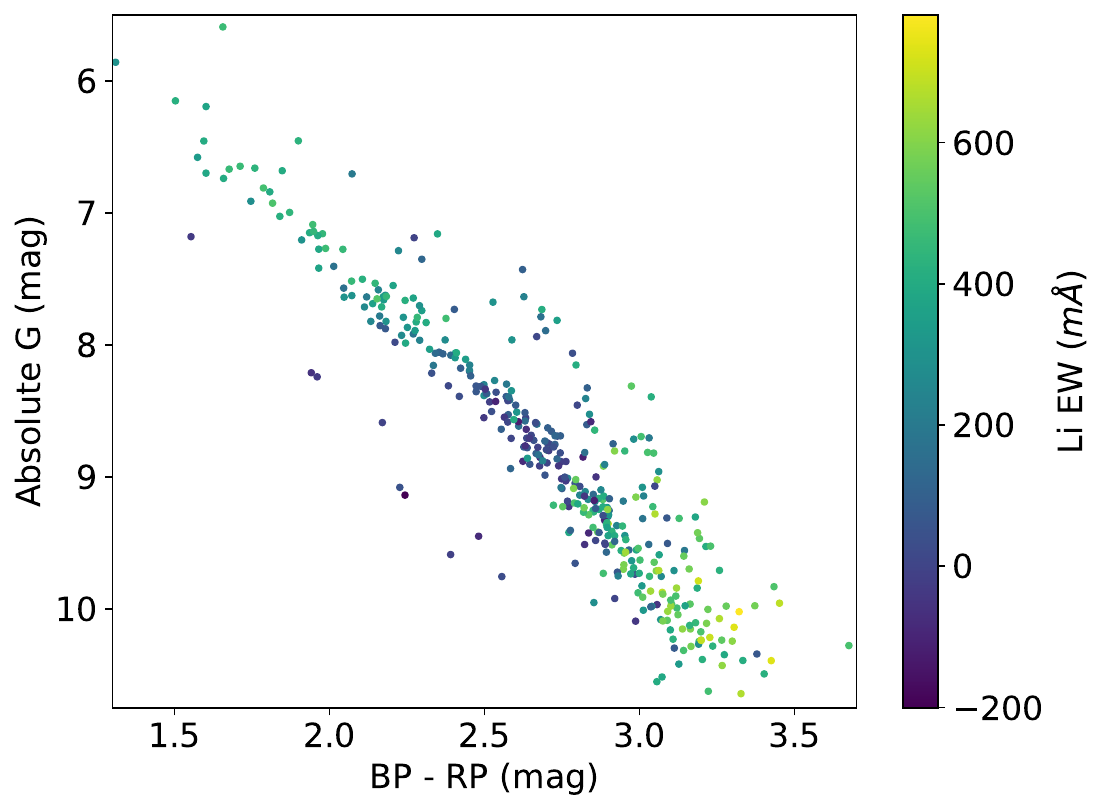}
    \caption{Colour-magnitude diagram of the observed sample, with the colour scale showing the equivalent width of Li 6708 \AA\  feature in m\AA. Most of the sample shows Li absorption, consistent with youth.}
    \label{fig:li_ew}
\end{figure}

We measured the Li equivalent width directly from the extracted spectra using much the same method as for the H$\alpha$ equivalent width: we estimate a continuum with a second-order polynomial fitted to the sideband regions 6700--6706 \AA\ and 6709.5--6715 \AA, then compute the equivalent width by integrating over a narrow window centred on 6707.856 \AA\ with a half-width of 1.4 \AA, following the same wavelength range adopted by \citet{zerjal21} for WiFeS spectra. At the WiFeS resolution of $R \sim 7000$, the Li doublet is unresolved and can therefore be treated as a single blended feature. We estimate the uncertainty from the scatter of the continuum-normalised flux in the continuum windows, propagated across the integration interval. The median Li equivalent width is 300.4 m\AA, with a median uncertainty of 70.6 m\AA. The resulting measurements are shown in \autoref{fig:li_ew}.

\begin{figure}
    \centering
    \includegraphics[width=\linewidth]{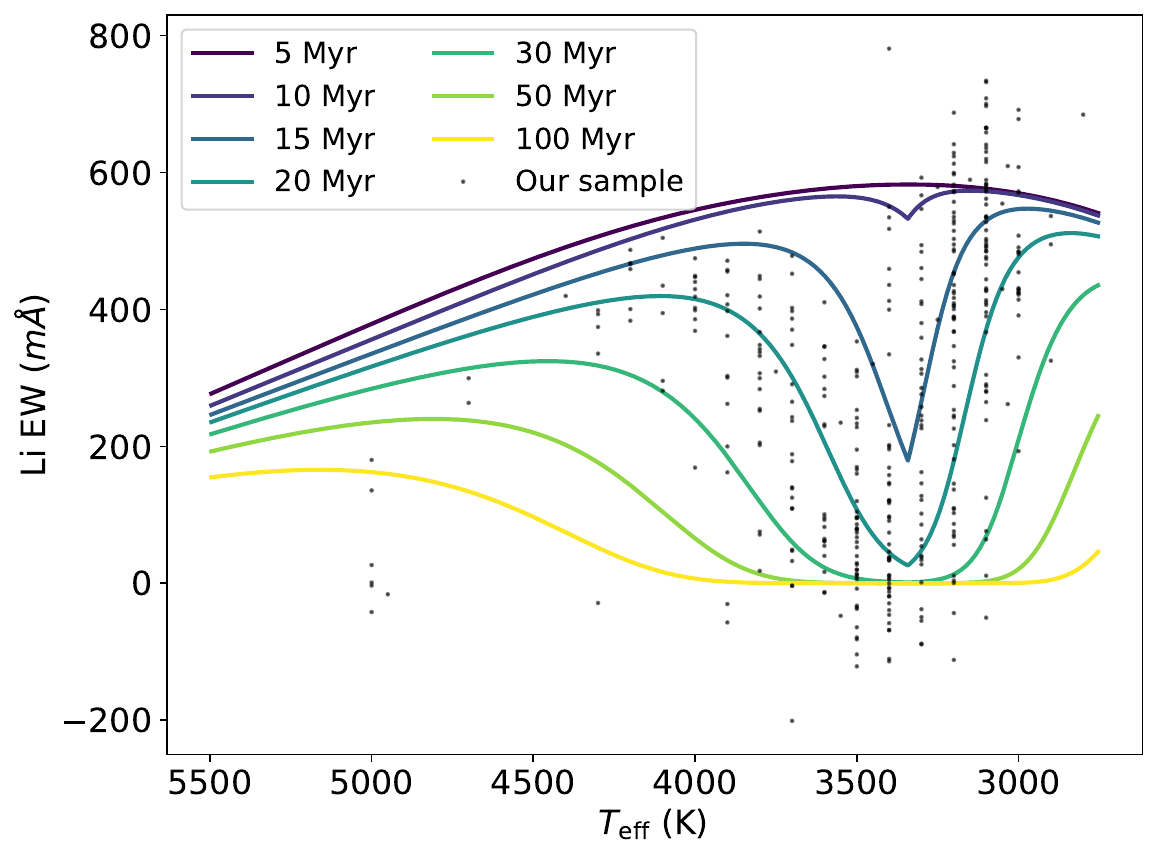}
    \caption{Lithium equivalent width and effective temperature isochrones from the EAGLES model \citep{eagles2023} with darker colours showing younger ages as mentioned in the legend. The sample stars are overplotted using their measured equivalent widths and effective temperature. (can put shaded area around isochrones for uncertainty)}
    \label{fig:li_isochrone}
\end{figure}

To classify stars as young using lithium, we first required a detection significance of at least $2\sigma$. Measurements below this threshold were treated as non-detections. For significant detections, we compare the measured Li equivalent with a temperature-dependent youth threshold based on \cite{eagles2023}, which empirically models lithium depletion using stars from the Gaia-ESO survey. The model lithium equivalent width as a function of $T_\text{eff}$ and age is shown in \autoref{fig:li_isochrone}, along with the stars from our sample. Stars with Li equivalent widths above the threshold at their effective temperatures (for age of 50 Myr) were classified as Li-young. For stars with repeated observations, the Li equivalent widths and their uncertainties were combined using inverse-variance weighting, in the same way as the radial velocities and H$\alpha$ measurements.

% oyr sample has ew width error whose 95th percentile is 144 mAng [which you can show Mike on the plot saved with our sample points on it]

\begin{figure}
    \centering
    \includegraphics[width=\linewidth]{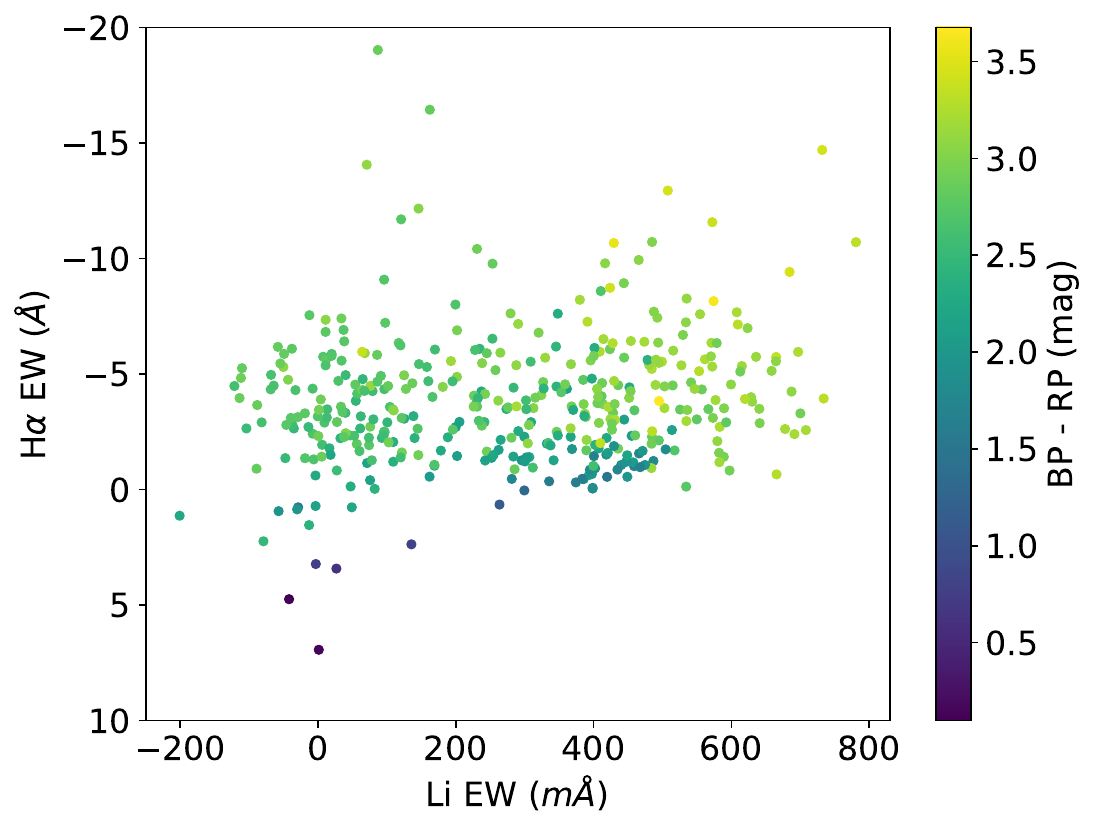}
    \caption{Youth indicators studied in this work, H$\alpha$ emission and Li absorption line are positively correlated across our sample. The colour bar shows BP-RP.}
    \label{fig:ha_li_corr}
\end{figure}

\begin{figure}
    \centering
    \includegraphics[width=\linewidth]{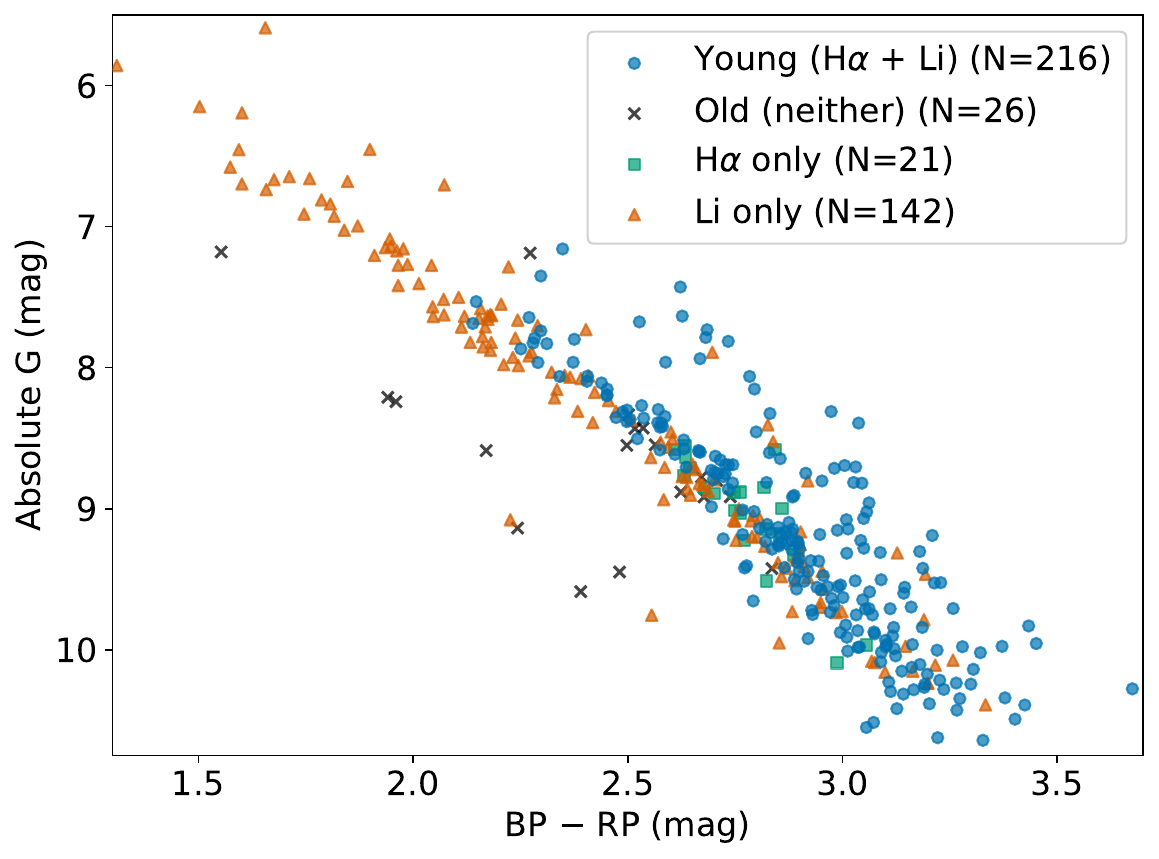}
    \caption{Colour-magnitude diagram showing the observed sample classified by the combined H$\alpha$ and lithium youth criteria. Stars classified as old by both diagnostics are preferentially located below the pre-main-sequence locus.}
    \label{fig:all_youth}
\end{figure}

In total, 358 out of 405 stars satisfy the lithium youth criterion, corresponding to 88.4\% of the sample. Most stars classified as Li-young are consistent with the expected young Sco-Cen population, while stars with weak or undetected lithium are more likely to be contaminants or uncertain members. This is also reflected in \autoref{fig:ha_li_corr}, where the H$\alpha$ and Li youth indicators are positively correlated. \autoref{fig:all_youth} shows the combined youth classifications on a colour--magnitude diagram, where stars classified as old, meaning not young by either criterion, are preferentially located below the pre-main-sequence locus.

The full table of measured and derived quantities for the observed sample is provided electronically. A portion of the table is shown here (\autoref{tab:full_sample}) for guidance, with the complete version including all 405 targets available in machine-readable form. The table lists the target identifier, \textit{Gaia} source identifier, equatorial coordinates, \textit{Gaia} photometry and astrometry, \textit{Gaia} radial velocities and uncertainties, proper motions, \textit{Gaia}-based distances \citep{bailerjones2021} and absolute magnitudes, the derived effective temperature, the WiFeS radial velocity and uncertainty, the H$\alpha$ and Li equivalent widths and their uncertainties, and the resulting youth classifications from each diagnostic.

\begin{table*}
    \centering
    \begin{tabular}{llllllllllllll}
    \hline\hline
    target\_id & gaia\_source\_id & ra & dec & $T_{\rm eff}$ & RV$_{\rm WiFeS}$ & $\sigma_{\rm RV}$ & EW$_{\rm H\alpha}$ & $\sigma_{\rm H\alpha}$ & EW$_{\rm Li}$ & $\sigma_{\rm Li}$ & youth$_{\rm H\alpha}$ & youth$_{\rm Li}$ \\
    &&[h:m:s]&[deg]&[K]&[km/s]&[km/s]&[\AA]&[\AA]&[m\AA]&[m\AA]&&\\
    \hline
    T001 & 6148981399239232000 & 12:05:16.3 & -41.50 & 3300 & 14.12 & 1.17 & -4.55 & 0.34 & 54.07 & 139.81 & True & True \\
    T002 & 6149247652852119168 & 12:13:48.1 & -40.87 & 3800 & 14.84 & 0.48 & -1.15 & 0.14 & 71.22 & 27.54 & False & True \\
    T003 & 6130406279862056960 & 12:19:16.6 & -47.83 & 5000 & 14.19 & 0.68 & 6.93 & 0.11 & 1.09 & 1.44 & False & False \\
    T004 & 6126648703176939392 & 12:19:59.3 & -50.31 & 4300 & 15.85 & 0.57 & -0.63 & 0.18 & 393.07 & 19.40 & False & True \\
    T007 & 6131343961421868928 & 12:23:40.3 & -46.26 & 3900 & 17.19 & 0.73 & -1.72 & 0.21 & 262.16 & 33.93 & False & True \\
    T008 & 6126856545236553984 & 12:24:17.8 & -49.20 & 3900 & 16.67 & 0.73 & -4.36 & 0.20 & 361.27 & 37.76 & True & True \\
    T009 & 6126856476518517888 & 12:24:20.5 & -49.20 & 5000 & 17.06 & 0.31 & 2.19 & 0.12 & 180.17 & 17.82 & False & False \\
    T012 & 3471256722554862208 & 12:28:08.0 & -30.57 & 3600 & -14.52 & 0.89 & 1.53 & 0.23 & -12.91 & 67.24 & False & False \\
    T013 & 6127444509077164928 & 12:29:28.2 & -50.04 & 3500 & 15.65 & 0.79 & -2.50 & 0.25 & 96.76 & 67.09 & False & True \\
    T014 & 6127207633041772416 & 12:32:01.7 & -50.37 & 3600 & 17.03 & 0.89 & -4.79 & 0.22 & 64.92 & 73.70 & True & True \\
    \hline
    \end{tabular}
    \caption{Excerpt from the full electronic table of measured and derived properties for the observed sample. Some of the columns removed here for clarity are available in the complete machine-readable table, which contains information for all 405 sample stars.}
    \label{tab:full_sample}
\end{table*}

\section{Kinematic analysis and age extraction}
\label{sec:age_extraction}
In this section, we describe how we use the phase-space modelling framework \texttt{Chronostar} to estimate the kinematic age of the T component and to investigate its kinematic substructure.

\subsection{\texttt{Chronostar}}
\label{subsec:chronostar}
Here, we briefly summarise the aspects of \texttt{Chronostar} relevant to our analysis. A complete description of the original method, its tests and application to $\beta$ Pictoris moving group are provided by \citep{chronostar}, while the subsequent developments and application to Sco-Cen are described in \citep{chronostar2}.

\texttt{Chronostar} is a Bayesian framework designed for the purpose of using present phase-space distribution of stars to identify kinematically coherent stellar populations, estimate their membership probabilities, and infer their kinematic ages.
Unlike simple traceback methods that integrate individual stellar orbits backwards in time, \texttt{Chronostar} models the initial phase-space distribution of a population and evolves it forward in a Galactic potential to the present day to find the best match to the observed data. This approach accounts for the uncertainties associated with the observed phase-space coordinates and avoids treating the rapidly increasing uncertainty of independently back-traced stellar orbits as the primary age estimator.

\texttt{Chronostar} models an association as one or more independent components in six-dimensional Cartesian phase-space, represented by the coordinates $(X,Y,Z,U,V,W)$. The spatial coordinate system is centred on the Sun's position projected onto the Galactic plane. The $X$, $Y$, and $Z$ axes point towards the Galactic centre, in the direction of Galactic rotation, and towards the Galactic North, respectively, while the velocity coordinates $U$, $V$, and $W$ have the same orientation and are defined relative to the local standard of rest. Each component is characterised by an initial centroid position in this phase space, an isotropic initial spatial dispersion $\sigma_{\mathrm{XYZ}}$, an isotropic initial velocity dispersion $\sigma_{\mathrm{UVW}}$, and a birth time (or equivalently an age); determining the values of these parameters for each component is the primary aim of the fitting procedure.

The initial time assigned to a component corresponds to the point in time at which the stars that are members of that component become gravitationally unbound and begin moving through the Galaxy on ballistic orbits. To find the expected distribution in phase space for such ballistically-moving stars, a proposed component is evolved forward to the present day; following \citet{chronostar2}, we calculate this evolution using an analytic epicyclic approximation for components with proposed ages $<30$ Myr, and using numerical orbit integration with \texttt{GALPY} \citep{bovy2015} and the \texttt{MWPotential2014} Galactic potential for older components. The choice of Galactic potential has only a small effect on the inferred properties over the timescales relevant to this analysis \citep{chronostar2}.

Once \texttt{Chronostar} has computed the expected present-day phase-space distribution for each proposed component, it evaluates the likelihood that each observed star belongs to each component. This likelihood is calculated from the overlap between the observed six-dimensional phase-space distribution of the star and the modelled present-day distribution of the component. Measurement uncertainties, including the covariance between the astrometric quantities, are incorporated into this overlap. The fit therefore accounts for the finite precision of the observed positions and velocities rather than treating the measured phase-space coordinates as exact.

Membership probabilities and component parameters are inferred simultaneously using the Expectation-Maximisation (EM) algorithm. In this algorithm, one starts with a proposed initial set of component parameters and then calculates the probability that each star belongs to each component, as we have just described; this is the expectation part of the algorithm. In the maximisation step, we hold the membership probabilities fixed and then update the parameters describing each component (centroid phase space position, dispersion, and age) by maximising the likelihood using the derivative-free Nelder--Mead algorithm \citep{neldermead1965}.
The expectation and maximisation steps are then repeated until the fit converges. This procedure resolves the circular dependence between identifying members and determining the properties of the components, and yields a final, converged set of parameters for each component and membership probabilities for each star.

%For a single component, the initial state model is specified by nine scalar parameters: the six phase-space coordinates of the initial centroid ($X_0,Y_0,Z_0,U_0,V_0,W_0$), the initial spatial and velocity dispersions $\sigma_{\mathrm{XYZ}}$ and $\sigma_{\mathrm{UVW}}$, and the component age $t$.
%The initial centroid and dispersions are estimated from the data used to initialise the fit. After this, the likelihood maximisation is performed using the derivative-free Nelder--Mead algorithm \citep{neldermead1965}, which replaces the Markov-chain Monte Carlo maximisation used in the original implementation and provides a substantially faster optimisation \citep{chronostar2}.

The number of components required to describe an association best is not known a priori. We therefore use this algorithm to fit models with successively larger number of components and compare them using the Bayesian Information Criteria (BIC),
\begin{equation}
    \mathrm{BIC} = k \ln(n) - 2 \ln\hat{\mathcal{L}}
\end{equation}
where $k$ is the number of parameters (nine per component -- the six phase-space coordinates of the centroid, the spatial and velocity dispersions, and the age), $n$ is the number of stars, and $\hat{\mathcal{L}}$ is the maximum likelihood (that is the likelihood of the data evaluated at the best-fitting parameter values). The BIC penalises the increased complexity of models while checking for an improvement in likelihood. A model with a lower BIC is preferred and hence, the process of incrementally adding more components is stopped when adding a further component no longer improves the BIC.

Thus the final output of the \texttt{Chronostar} analysis is the preferred number of components, the probability that each star belongs to each component, and for each component the initial phase-space centroid coordinates $(X_0,Y_0,Z_0,U_0,V_0,W_0)$, the initial spatial and velocity dispersions ($\sigma_\mathrm{XYZ}, \sigma_\mathrm{UVW}$), and the kinematic age $t$. These quantities allow us to examine both the age of the T component and the degree to which the input sample contains additional kinematic substructure.

\subsection{Data}
\label{subsec:chronostar_data}

The input catalogue for the \texttt{Chronostar} analysis is based on the probabilistic membership list for the T component from \citet{chronostar2}. This contains candidate members identified using the phase-space information available at the time of its construction. We supplement the radial velocities in that dataset with the new WiFeS measurements presented in this work and with \textit{Gaia} DR3 radial velocities that became available subsequently.

The final input sample contains 605 probable T component members with measured radial velocities. Since the purpose of this analysis is to estimate a kinematic age rather than to perform a blind search for new members, we restrict the analysis to stars with available radial velocities. The input data comprise the stellar positions, parallaxes, proper motions, radial velocities, and their associated uncertainties. Positions, parallaxes, and proper motions are taken from \textit{Gaia} DR3, while radial velocities are obtained from \textit{Gaia} DR3, the WiFeS spectra analysed in this work, or the catalogue of \citet{chronostar2}. We select the radial velocity with the smallest quoted uncertainty, using WiFeS spectral fit measurement in preference to \textit{Gaia} and \citet{chronostar2} in the event of a tie. For \textit{Gaia} measurements, we retain the catalogue designation (DR2 or DR3) and prioritise an available \textit{Gaia} DR3 measurement over the DR2-based value, irrespective of the quoted uncertainties. Because \texttt{Chronostar} marginalises over uncertainties automatically, there is no difficulty in combining our new data set with improved radial velocity uncertainties with the remaining stars that we did not observe and their larger \textit{Gaia} uncertainties. These measurements are transformed into the Cartesian phase-space coordinates $(X,Y,Z,U,V,W)$ required by \texttt{Chronostar}.

Some contamination by older field stars is expected in a probabilistic membership catalogue, particularly because the original selection included stars without measured radial velocities. Hence, we also construct a second input sample from which we exclude stars classified as old by both the H$\alpha$ and lithium criteria (\autoref{subsec:youth}). This selection leaves 580 stars and provides a test of the sensitivity of inferred kinematic structure and age to possible contamination. We run all our analyses on both these samples. 

\subsection{Model fitting and component classification}
\label{subsec:model_component}

We carry out \texttt{Chronostar} fits with to both samples, and find that for the full 605-star sample with no exclusions we obtain the lowest BIC for fourteen components, while for the 580-star sample that excludes stars lacking either H$\alpha$ or Li youth indicators the lowest BIC has fifteen components. However, several of these components contain very few stars, so to reduce clutter in what follows we will ignore components with fewer than ten stars assigned to them, where we consider a star to be assigned to the component for which its membership probability is highest. This criterion excludes three of the fourteen components for the fit to the full catalogue, and six of the fifteen components for the fit to the catalogue where stars not classified as young are excluded.

To gain additional insight and as a check of how our results depend on the number of components, we also carry out fits of the full 605-star sample where we fix the number of components to two and three. The two-component model provides the simplest decomposition, comprising one broad background-like component that absorbs contaminants or stars that otherwise do not fit (see below), and one component associated with the T population. The three-component fit adds a single level of complexity by decomposing the T population into two parts. When comparing results between fits with different numbers of components, we identify corresponding structures using their membership distributions and their locations in position--velocity space. However, such identifications are not always possible, since there is not guaranteed to be a one-to-one mapping between the components identified in an $N$-component fit and those identified in an $M$-component fit, even if the underlying input stellar catalogues are identical.

\begin{figure}
    \centering
    \includegraphics[width=\linewidth]{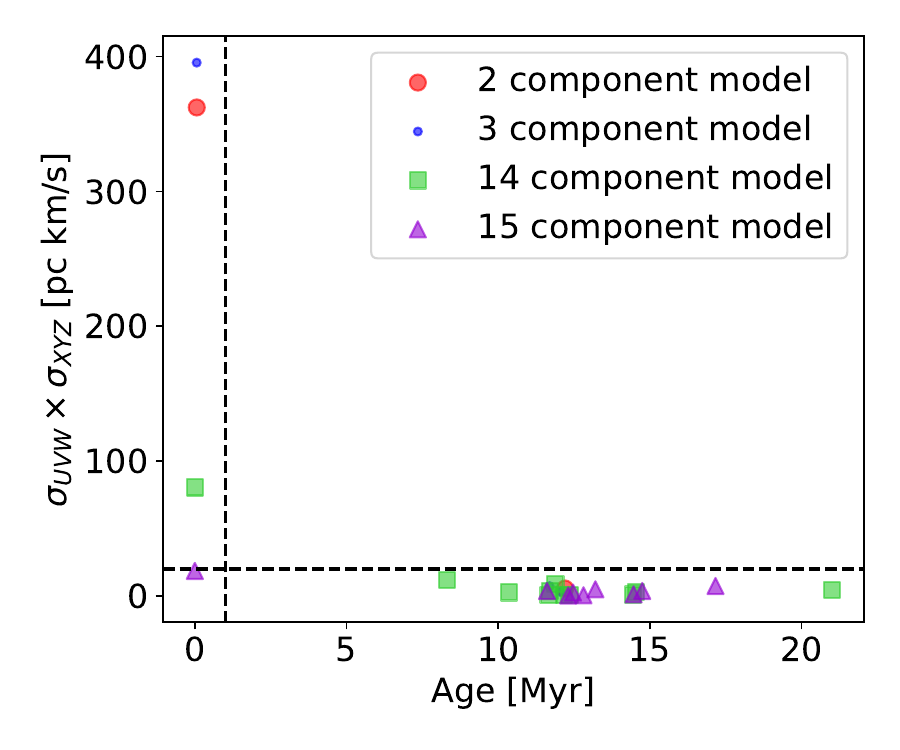}
    \caption{Kinematic ages and initial phase-space extent of fitted \texttt{Chronostar} components. The vertical dashed line marks $t=1$ Myr, and the horizontal dashed line marks $\sigma_{\mathrm{XYZ}}\sigma_{\mathrm{UVW}} =20~\mathrm{pc\,km\,s^{-1}}$; we classify components above or to the left of these lines as backgrounds. Colours and symbols (as labelled) distinguish the component-number models: the 15-component model fit to the 580-star cleaned sample, the 14-component fit to the 605-star full sample, and the additional 2- and 3-component fits to the full sample for comparison. Note that only components with more than 10 members have been plotted.}
    \label{fig:bg_limits}
\end{figure}

\autoref{fig:bg_limits} shows the kinematic ages and initial phase-space extents $\sigma_{\rm XYZ}\sigma_{\rm UVW}$ of the components we recover from all fits (omitting those with fewer than 10 stars). The figure shows that fit components can be broadly divided into two groups: components with older ages and narrow phase-space distributions, and components with near-zero ages and very broad phase space distributions. For each fit, there is only one component in the latter category.
 
This dichotomy is expected: the input catalogue is dominated by genuine T-component members, but its purity is not perfect, and thus we also expect some level of contamination from field stars and potentially stars associated with neighbouring kinematic structures in the wider UCL region. These stars are assigned a very young age because they lack the fundamental kinematic signature that \texttt{Chronostar} detects and uses to estimate kinematic ages: as a group of stars born in a compact phase space volume orbit through the Galactic disc, their positions and velocities become correlated in a specific way determined by the shape of the Galactic potential. A group of unrelated contaminant stars will have near-zero position-age correlation, and for this reason \texttt{Chronostar} assigns them to a component with near-zero age; the distribution then needs to be very broad in phase space to capture the large volume over which these uncorrelated stars are spread.

We refer to the components in this category as background components, since they are unlikely to be true T-component stars at all. Formally, we identify a component as likely background when either 
\begin{equation}
    t < 1~\mathrm{Myr},
\end{equation}
or
\begin{equation}
    \sigma_{\mathrm{XYZ}}\sigma_{\mathrm{UVW}} >
    20~\mathrm{pc\,km\,s^{-1}}.
\end{equation}
We indicate these cuts by the dashed lines in \autoref{fig:bg_limits}. We emphasise that the presence of these background components 
%These criteria identify components with either a very young fitted kinematic age or broad initial phase-space distribution. A young fitted age for a broad field-like component 
should not be interpreted as evidence for recent star formation. Instead, the near-zero age assigned to these components simply reflects that the stars that comprise them show essentially no correlation between their positions and velocities.
%it can arise because an unrelated field-star population is not expected to share a compact common birth configuration. Within the \texttt{Chronostar} model, a young, already extended Gaussian component can provide a higher-likelihood phenomenological representation of such a population. Note that they are applied only after fitting, as a descriptive classification scheme for interpretation. 

\subsection{Parameter uncertainties}
\label{subsec:uncertainties}

For each model, we compute formal uncertainties on the fitted component parameters from the local curvature of the negative weighted log-likelihood (returned by \texttt{Chronostar} as the component loss function) around the optimum with the final membership probabilities held fixed. Specifically, for each component we have nine best-fit parameters
\begin{equation}
    \boldsymbol{\theta}
    =
    (X_0,Y_0,Z_0,U_0,V_0,W_0,
    \sigma_{\mathrm{XYZ}},\sigma_{\mathrm{UVW}},t).
\end{equation}
To compute the uncertainty on these, we evaluate the Hessian matrix 
\begin{equation}
    H_{ab}
    =
    \left.
    \frac{\partial^2 \mathcal{J}}
    {\partial \theta_a \partial \theta_b}
    \right|_{\hat{\boldsymbol{\theta}}},
\end{equation}
where $\mathcal{J}$ is the negative log-likelihood (weighted by the fixed membership probabilities) and
$\hat{\boldsymbol{\theta}}$ is the best-fitting parameter vector. We evaluate these derivatives numerically using centred finite differences with respect to the nine parameters. Assuming that $\mathcal{J}$ is locally quadratic near the optimum, the parameter covariance matrix is then given by
\begin{equation}
    \boldsymbol{\Sigma}_{\theta}
    \simeq
    \boldsymbol{H}^{-1}.
\end{equation}
The formal $1\sigma$ uncertainty on parameter $\theta_a$ is then
\begin{equation}
    \sigma_{\theta_a}
    =
    \sqrt{
        \left(
            \boldsymbol{\Sigma}_{\theta}
        \right)_{aa}
    }.
\end{equation}
In particular, the formal uncertainty on the component age is
\begin{equation}
    \sigma_t
    =
    \sqrt{
        \left(
            \boldsymbol{\Sigma}_{\theta}
        \right)_{tt}
    }.
\end{equation}

These uncertainties are conditional on the final membership probabilities and on the selected number of components. They quantify the local statistical uncertainty of the fitted component parameters and do not include uncertainty associated with membership reassignment, model selection, sample contamination, or departures from the assumed isotropic Gaussian phase-space model. For this reason, we also make a second, independent estimate of the uncertainty based on the observation that our treatment of components as having a single age is certainly an oversimplification, since in reality it would be unphysical for the star formation within a structure to be coordinated on timescale smaller than the signal crossing time across the structure. To quantify this point, we also compute the crossing time at birth for each component, 
\begin{equation}
    t_{\mathrm{cross}} = \frac{\sigma_{\mathrm{XYZ}}}
    {\sigma_{\mathrm{UVW}}},
\end{equation}
We will treat this a second source of uncertainty in component ages in what follows, and whenever we require an age uncertainty for a component we will take it to be the quadrature sum of $\sigma_t$ and $t_\mathrm{cross}$; in practice, the latter term almost always dominates.

\subsection{Stellar kinematic ages and uncertainties}
\label{ssec:stellar_ages}

%We quantify the sensitivity of the inferred component ages to modelling choices by comparing results across the two-, three- and BIC-preferred multi-component, and between the full and youth selected input samples. We quantify model dependence on a star-by-star basis, rather than by attempting to match numerical component labels across independent fits.
For each star and each fitted model, we define a posterior-weighted
kinematic age,
\begin{equation}
    t_{\mathrm{kin},i}
    =
    \sum_{j=1}^{N_{\mathrm{comp}}}
    p_{ij} t_j,
\end{equation}
where $p_{ij}$ is the posterior membership probability of star $i$ in component $j$, and $t_j$ is the fitted kinematic age of that component. This quantity represents the age implied for an individual star by a given mixture-model decomposition; it is not an independently measured stellar age.

We assign uncertainties to these age estimates via the usual error propagation rules for Gaussian uncertainties, i.e., 
\begin{equation}
    \sigma_{t_{\mathrm{kin},i}} = \sqrt{\sum_{j=1}^{N_\mathrm{comp}}p_{ij}^2 \sigma_j^2},
    \label{eq:tkin_uncertainty_crossing}
\end{equation}
where $\sigma_j$ is the uncertainty on the age of component $j$, which, as explained in \autoref{subsec:uncertainties}, we take to be the quadrature sum of the formal statistical uncertainty and the crossing time\footnote{For a small number of components with limited statistics, the Hessian uncertainty is not defined, and hence no formal statistical uncertainty is available. For those components, we set the statistical uncertainty to zero and use the crossing time alone as the age uncertainty. This is a reasonable approximation because the crossing time dominates the total uncertainty in all cases. As shown in \autoref{tab:component_properties}, this occurs for only three components across all of the models considered.}.

It is also interesting to compare this to a second estimate of uncertainty, which we derive by comparing the ages assigned by our fits using different numbers of components. We compute the inferred kinematic age $t_{\mathrm{kin},i}$ from each of our model fits -- the two-, three-, and fourteen-component fits for all stars, and also for the 580 stars retained in the youth-selected sample, we additionally compare the age inferred from the 15-component fit. We then calculate the standard deviation of $t_{\mathrm{kin},i}$ across the available model fits for each star as one estimate of our systematic uncertainty. Using this approach we find a median uncertainty of $0.56$ Myr, with a 16th--84th percentile range of $0.16-1.33$ Myr. These numbers are generally small compared to the uncertainties derived primarily from crossing times (\autoref{eq:tkin_uncertainty_crossing}), so we conclude that the dominant source of uncertainty in stellar kinematic ages does not arise from the systematics of how many components we use to describe the T stellar population, but instead from the intrinsic limitation that we cannot expect to pin down stellar kinematic ages to better than a crossing time of the structures in which those stars formed.

\subsection{Results: present-day kinematic decomposition}
\label{subsec:results-chronostar}

\begin{figure*}
    \centering
    \includegraphics[width=\linewidth]{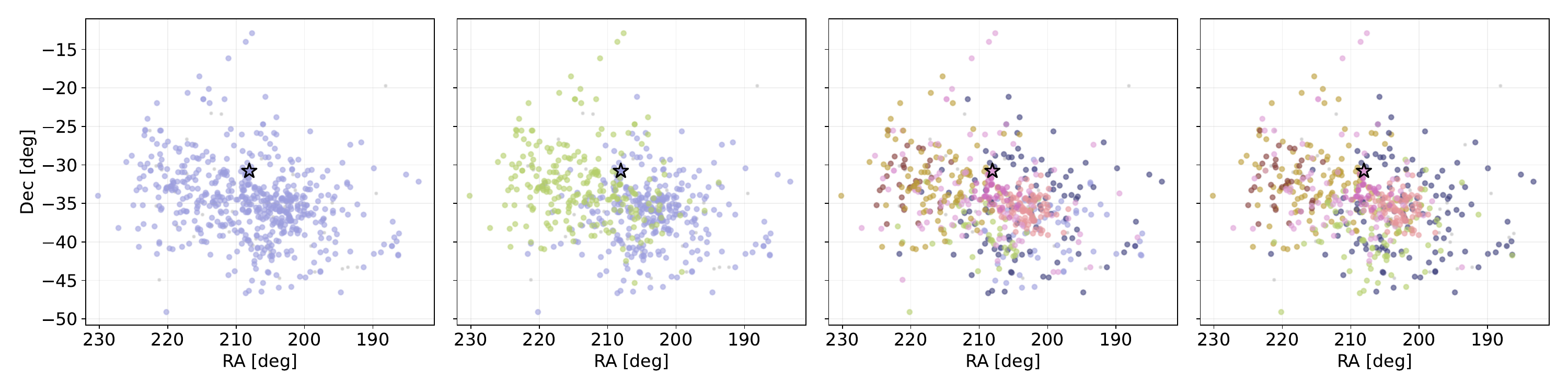}
    \caption{Present-day on-sky distribution (right ascension [deg] and declination [deg]) of the input stars for the two-, three-, 14-, and 15-component \texttt{Chronostar} models, shown
    from left to right. Each star is coloured by the component to which it is assigned, while grey points denote stars that are either unassigned or are assigned to components classified as likely background -- see main text for details. Only components with more than ten assigned stars are shown. Similar colours are used for approximately corresponding structures in the 14- and 15-component models where possible. The large star symbol marks PDS~70, with its colour showing its component.}
    \label{fig:ra_dec}
\end{figure*}

\begin{figure*}
    \centering
    \includegraphics[width=\linewidth]{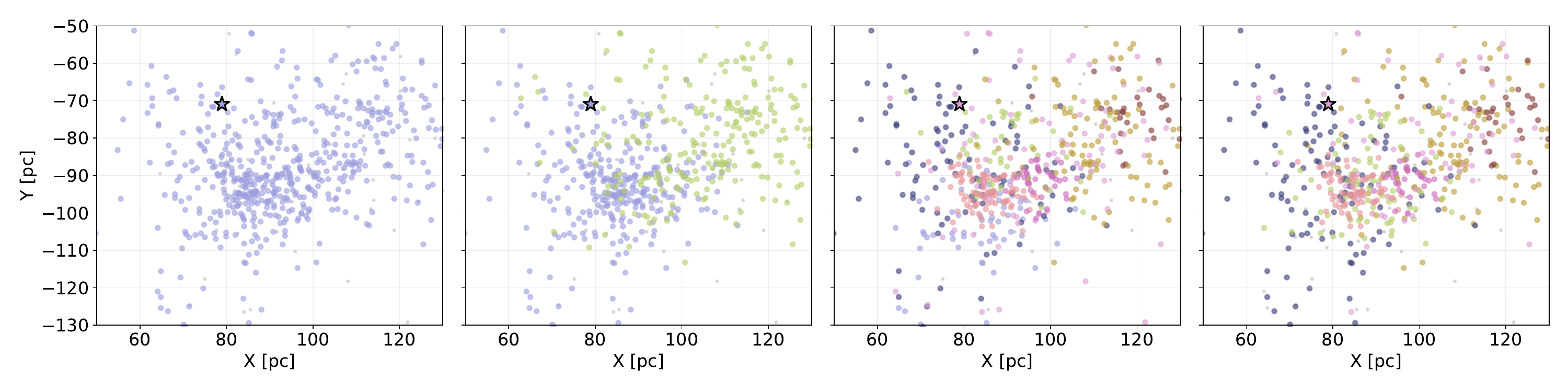}
    \caption{Same as \autoref{fig:ra_dec}, but showing the present-day
    heliocentric Cartesian spatial distribution in $X$ [pc] and $Y$ [pc].}
    \label{fig:XY}
\end{figure*}

\begin{figure*}
    \centering
    \includegraphics[width=\linewidth]{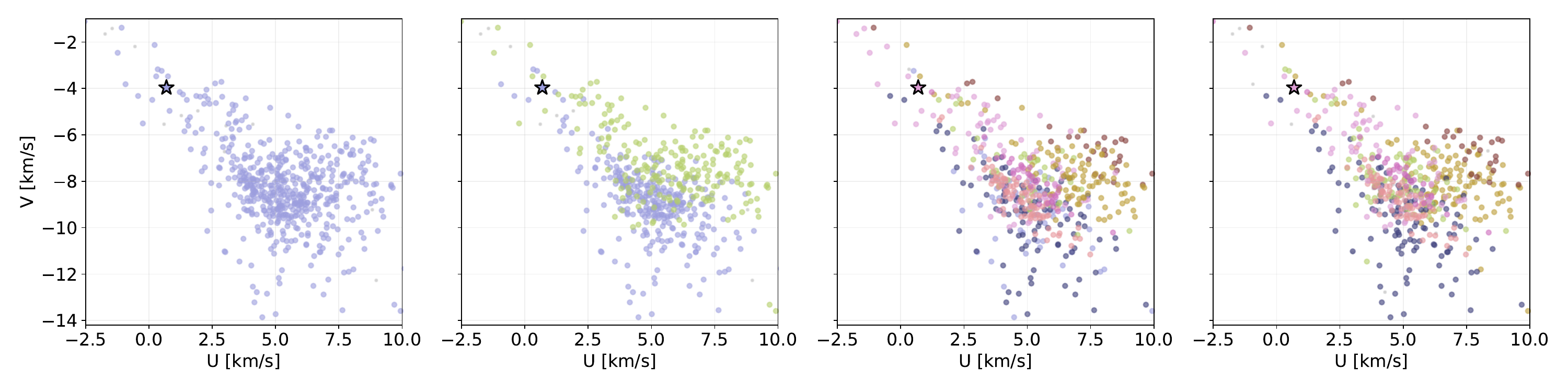}
    \caption{Same as \autoref{fig:ra_dec}, but showing the present-day
    heliocentric Cartesian velocity distribution in $U$ [km/s] and $V$ [km/s].}
    \label{fig:UV}
\end{figure*}

\autoref{fig:ra_dec}, \autoref{fig:XY}, and \autoref{fig:UV} show the decomposition of the input stars at their present-day positions projected into sky coordinates, $XY$ coordinates, and $UV$ coordinates, respectively. The four columns, from left to right, show the two-component, three-component, 14-component, and 15-component models, respectively. For visual clarity, we use approximately matched colours for apparently corresponding structures in the 14-component full-sample fit and the 15-component youth-selected fit. These colour correspondences are intended only as a visual aid and do not imply a one-to-one formal component match. In these visualisations we assign each star to the component for which it has the highest membership probability if that probability is $\geq 0.3$, or mark the star as unassigned otherwise. We show unassigned stars, and stars assigned to components flagged as likely background, in grey. The large star symbol marks PDS~70, coloured according to its component assignment in each model.

In the two-component model, most stars are assigned to a single broad kinematic component. This component forms an elongated structure on the sky and occupies the dominant concentration in both the $X$--$Y$ and $U$--$V$ projections. The second component is classified as likely background according to the post-fit criteria defined in \autoref{subsec:model_component}. The three-component model separates the main population into two kinematically distinct components in addition to a broad background component. 

The BIC-preferred 14-component model resolves further structure within the input catalogue. Several components overlap in projected position but are offset from one another in velocity, while other components occupy distinct regions of the spatial distribution. The 15-component fit to the youth-selected sample produces a qualitatively similar decomposition, although the relative populations and component boundaries change somewhat after removing stars classified as old. 

\autoref{tab:component_properties} lists the fitted properties (with corresponding uncertainties) of every component recovered in the four representative fits: the fitted kinematic age, the birth phase-space centroid, and the initial spatial and velocity dispersions. We also report the number of assigned members, the derived crossing time, and whether the component is classified as background. The uncertainties listed  are the formal $1\sigma$ statistical uncertainties derived from the inverse-Hessian, holding the final membership probabilities fixed (see \autoref{subsec:uncertainties}); absence of uncertainties for a component indicates that the Hessian was near-singular and could not be inverted. Components with membership counts $<10$, marked with a dagger, are included for completeness but should be treated with extreme caution because they are dominated by small-number statistics. These components are excluded from every visualisation throughout this paper and from the interpretation in
\autoref{sec:implications}.

\begin{landscape}
\begin{table}
    \centering
    \footnotesize
    \caption{Fitted \texttt{Chronostar} component properties for the four
    representative fits. Counts is the number of stars assigned to
    the component using the maximum-posterior assignment with
    $p_{\rm max}\geq0.30$; a dagger ($^{\dagger}$) marks components with
    Counts $<10$, whose parameters are dominated by small-number
    statistics and are given for completeness only (see
    \autoref{subsec:model_component}). Uncertainties are formal
    inverse-Hessian $1\sigma$ uncertainties conditional on the fitted
    model; absence of an uncertainty denotes non-convergence of the Hessian. BG denotes the
    post-fit likely-background classification described in
    \autoref{subsec:model_component}. $t_{\rm cross}$ is the birth
    crossing time.}
    \label{tab:component_properties}
    \begin{tabular}{llrcrrrrrrrrrrr}
        \hline
        Sample & Model & Comp. & BG & Counts &
        $t$ & $X_0$ & $Y_0$ & $Z_0$  &
        $U_0$  & $V_0$  &
        $W_0$  & $\sigma_{\rm XYZ}$ &
        $\sigma_{\rm UVW}$ & $t_{\rm cross}$ \\
        && &  &  &
         [Myr] &  [pc] & [pc] &  [pc] &
        [$\mathrm{km\,s^{-1}}$] &  [$\mathrm{km\,s^{-1}}$] &
         [$\mathrm{km\,s^{-1}}$] &  [pc] &
         [$\mathrm{km\,s^{-1}}$] &  [Myr]\\
        \hline
Full & 2 & 0 & Y & 48 & $0.06\pm0.07$ & $97.8\pm2.7$ & $-91.1\pm2.6$ & $64.2\pm2.6$ & $7.49\pm2.32$ & $-8.62\pm2.30$ & $1.79\pm2.91$ & $18.03\pm1.06$ & $20.10\pm1.27$ & 0.90 \\
Full & 2 & 1 & N & 557 & $12.20\pm0.13$ & $34.4\pm0.3$ & $10.9\pm1.2$ & $17.5\pm0.7$ & $1.64\pm0.07$ & $-10.05\pm0.06$ & $4.92\pm0.05$ & $4.89\pm0.12$ & $1.13\pm0.02$ & 4.34 \\
Full & 3 & 0 & Y & 44 & $0.06\pm0.07$ & $96.8\pm2.9$ & $-92.1\pm2.9$ & $64.1\pm2.8$ & $7.94\pm2.66$ & $-8.76\pm2.41$ & $1.86\pm3.31$ & $18.40\pm1.16$ & $21.50\pm1.44$ & 0.86 \\
Full & 3 & 1 & N & 323 & $12.04\pm0.14$ & $31.9\pm0.3$ & $14.0\pm1.4$ & $17.0\pm0.7$ & $1.15\pm0.08$ & $-10.57\pm0.07$ & $4.72\pm0.06$ & $3.22\pm0.10$ & $0.96\pm0.02$ & 3.36 \\
Full & 3 & 2 & N & 238 & $11.69\pm0.29$ & $40.0\pm0.9$ & $0.7\pm2.5$ & $22.0\pm1.6$ & $2.40\pm0.12$ & $-9.38\pm0.08$ & $5.09\pm0.08$ & $6.77\pm0.23$ & $1.07\pm0.03$ & 6.31 \\
Full & 14 & 0 & N & 103 & $11.70\pm0.22$ & $29.7\pm0.5$ & $21.5\pm2.2$ & $15.6\pm1.1$ & $1.20\pm0.14$ & $-10.94\pm0.13$ & $4.71\pm0.11$ & $3.38\pm0.18$ & $1.13\pm0.05$ & 2.99 \\
Full & 14 & 1 & N & 64 & $10.35\pm0.26$ & $31.6\pm0.6$ & $-8.4\pm2.6$ & $25.0\pm1.2$ & $1.48\pm0.16$ & $-10.94\pm0.14$ & $4.41\pm0.13$ & $2.77\pm0.19$ & $0.96\pm0.05$ & 2.89 \\
Full & 14 & 2 & N & 46 & $11.66\pm0.22$ & $37.3\pm0.5$ & $1.1\pm1.7$ & $18.5\pm0.8$ & $1.47\pm0.15$ & $-9.16\pm0.14$ & $3.77\pm0.12$ & $1.23\pm0.11$ & $0.75\pm0.05$ & 1.63 \\
Full & 14 & 3 & N & 101 & $14.54\pm0.35$ & $26.7\pm0.8$ & $36.7\pm3.3$ & $3.4\pm2.0$ & $2.49\pm0.16$ & $-10.36\pm0.11$ & $5.68\pm0.08$ & $3.83\pm0.21$ & $0.84\pm0.04$ & 4.58 \\
Full & 14 & 4 & N & 31 & $14.44\pm0.53$ & $25.0\pm1.7$ & $21.4\pm4.4$ & $4.9\pm2.8$ & $3.45\pm0.24$ & $-9.33\pm0.16$ & $5.42\pm0.10$ & $1.86\pm0.23$ & $0.55\pm0.04$ & 3.39 \\
Full & 14 & 5 & N & 92 & $12.35\pm0.34$ & $35.5\pm0.4$ & $9.7\pm3.1$ & $18.9\pm1.6$ & $0.71\pm0.13$ & $-10.17\pm0.08$ & $4.88\pm0.05$ & $1.26\pm0.09$ & $0.36\pm0.02$ & 3.52 \\
Full & 14 & 6 & N & 36 & $12.15\pm1.50$ & $40.3\pm2.3$ & $-1.7\pm12.9$ & $24.8\pm7.9$ & $1.40\pm0.48$ & $-9.53\pm0.13$ & $5.18\pm0.22$ & $2.25\pm0.21$ & $0.23\pm0.02$ & 9.59 \\
Full & 14 & 7 & N & 68 & $11.88\pm0.68$ & $45.0\pm1.5$ & $1.6\pm5.6$ & $21.2\pm3.5$ & $1.42\pm0.25$ & $-9.24\pm0.14$ & $4.91\pm0.14$ & $8.56\pm0.47$ & $1.01\pm0.05$ & 8.48 \\
Full & 14 & 8 & N & 14 & $21.00\pm1.09$ & $69.7\pm3.6$ & $-5.0\pm4.2$ & $-9.1\pm5.3$ & $-2.48\pm0.34$ & $-5.55\pm0.30$ & $4.89\pm0.22$ & $5.08\pm0.66$ & $0.92\pm0.09$ & 5.53 \\
Full & 14 & 9 & N & 14 & $8.30\pm1.76$ & $99.6\pm5.7$ & $-123.9\pm7.5$ & $66.1\pm6.2$ & $-3.78\pm0.23$ & $0.14\pm0.27$ & $1.44\pm0.71$ & $21.74\pm2.36$ & $0.55\pm0.08$ & 39.73 \\
Full & 14 & 10 & Y & 15 & $0.00\pm0.73$ & $94.4\pm10.3$ & $-91.7\pm11.2$ & $66.8\pm6.9$ & $-14.90\pm1.29$ & $11.28\pm1.26$ & $-6.95\pm1.24$ & $17.56\pm2.37$ & $4.58\pm0.55$ & 3.83 \\
Full & 14 & 11 & N & 6$^{\dagger}$ & $8.62\pm0.52$ & $9.7\pm3.2$ & $29.3\pm7.1$ & $16.2\pm3.3$ & $6.76\pm0.72$ & $-14.45\pm0.67$ & $6.16\pm0.62$ & $2.76\pm0.54$ & $1.68\pm0.26$ & 1.64 \\
Full & 14 & 12 & N & 7$^{\dagger}$ & $3.79\pm0.12$ & $7.2\pm1.9$ & $7.8\pm2.4$ & $27.8\pm1.0$ & $20.87\pm1.46$ & $-26.12\pm1.45$ & $9.88\pm1.55$ & $1.36\pm0.26$ & $4.09\pm0.64$ & 0.33 \\
Full & 14 & 13 & N & 6$^{\dagger}$ & $1.70\pm0.04$ & $2.1\pm1.4$ & $4.9\pm1.4$ & $25.8\pm0.6$ & $59.17\pm2.20$ & $-58.28\pm2.17$ & $21.57\pm3.74$ & $0.72\pm0.16$ & $9.05\pm1.52$ & 0.08 \\
Youth-sel. & 15 & 0 & N & 116 & $11.61\pm0.20$ & $29.5\pm0.4$ & $19.2\pm2.0$ & $16.4\pm1.0$ & $1.08\pm0.14$ & $-11.07\pm0.12$ & $4.61\pm0.11$ & $3.23\pm0.16$ & $1.14\pm0.05$ & 2.85 \\
Youth-sel. & 15 & 1 & N & 79 & $12.46\pm0.27$ & $34.8\pm0.5$ & $7.0\pm2.3$ & $16.2\pm1.2$ & $1.00\pm0.13$ & $-9.75\pm0.11$ & $4.30\pm0.09$ & $2.94\pm0.17$ & $0.85\pm0.04$ & 3.45 \\
Youth-sel. & 15 & 2 & N & 126 & $14.75\pm0.34$ & $26.5\pm0.7$ & $39.4\pm3.2$ & $2.7\pm1.9$ & $2.16\pm0.15$ & $-10.39\pm0.10$ & $5.58\pm0.08$ & $4.28\pm0.21$ & $0.87\pm0.03$ & 4.94 \\
Youth-sel. & 15 & 3 & N & 32 & $14.46\pm0.59$ & $24.7\pm1.8$ & $21.8\pm4.9$ & $4.7\pm3.2$ & $3.44\pm0.25$ & $-9.36\pm0.16$ & $5.47\pm0.10$ & $2.16\pm0.28$ & $0.55\pm0.04$ & 3.93 \\
Youth-sel. & 15 & 4 & N & 82 & $12.31\pm0.36$ & $35.5\pm0.5$ & $9.4\pm3.3$ & $19.2\pm1.8$ & $0.73\pm0.14$ & $-10.16\pm0.09$ & $4.88\pm0.06$ & $1.15\pm0.09$ & $0.34\pm0.02$ & 3.41 \\
Youth-sel. & 15 & 5 & N & 32 & $12.81\pm1.69$ & $39.9\pm2.0$ & $3.4\pm14.5$ & $21.8\pm9.1$ & $1.17\pm0.54$ & $-9.50\pm0.12$ & $5.27\pm0.22$ & $2.16\pm0.22$ & $0.22\pm0.02$ & 9.88 \\
Youth-sel. & 15 & 6 & N & 38 & $17.16\pm0.89$ & $48.3\pm1.8$ & $19.6\pm5.8$ & $-2.0\pm4.8$ & $-0.38\pm0.29$ & $-7.63\pm0.19$ & $5.18\pm0.15$ & $7.52\pm0.68$ & $0.98\pm0.07$ & 7.64 \\
Youth-sel. & 15 & 7 & N & 31 & $13.20\pm0.71$ & $46.3\pm1.3$ & $23.4\pm6.6$ & $13.0\pm3.7$ & $0.85\pm0.30$ & $-10.27\pm0.19$ & $4.96\pm0.16$ & $5.45\pm0.47$ & $0.90\pm0.06$ & 6.03 \\
Youth-sel. & 15 & 8 & Y & 15 & $0.00\pm4.85$ & $89.6\pm9.5$ & $-98.1\pm18.3$ & $66.5\pm13.6$ & $-4.30\pm0.39$ & $1.16\pm0.42$ & $-2.57\pm1.94$ & $18.04\pm2.49$ & $1.03\pm0.18$ & 17.53 \\
Youth-sel. & 15 & 9 & N & 1$^{\dagger}$ & $20.00$ & $173.7$ & $-671.9$ & $93.8$ & $-9.97$ & $23.13$ & $4.08$ & $0.00$ & $0.00$ & 0.01 \\
Youth-sel. & 15 & 10 & Y & 9$^{\dagger}$ & $0.00$ & $95.9$ & $-84.9$ & $62.5$ & $-15.35$ & $10.14$ & $-6.29$ & $10.77$ & $1.52$ & 7.07 \\
Youth-sel. & 15 & 11 & N & 7$^{\dagger}$ & $8.66\pm0.53$ & $9.8\pm3.2$ & $29.3\pm7.2$ & $16.3\pm3.3$ & $6.64\pm0.70$ & $-14.35\pm0.65$ & $6.17\pm0.60$ & $2.81\pm0.54$ & $1.65\pm0.25$ & 1.70 \\
Youth-sel. & 15 & 12 & N & 7$^{\dagger}$ & $3.79\pm0.12$ & $7.2\pm1.9$ & $7.8\pm2.4$ & $27.8\pm1.0$ & $20.89\pm1.47$ & $-26.11\pm1.48$ & $9.89\pm1.55$ & $1.36\pm0.26$ & $4.09\pm0.64$ & 0.33 \\
Youth-sel. & 15 & 13 & N & 1$^{\dagger}$ & $1.10$ & $-0.6$ & $5.0$ & $24.4$ & $78.69$ & $-70.20$ & $22.09$ & $0.00$ & $0.00$ & 0.97 \\
Youth-sel. & 15 & 14 & N & 4$^{\dagger}$ & $1.65\pm0.05$ & $4.4\pm2.3$ & $2.9\pm2.0$ & $27.2\pm0.9$ & $66.64\pm2.39$ & $-57.18\pm2.39$ & $24.86\pm2.54$ & $0.46\pm0.14$ & $4.94\pm1.02$ & 0.09 \\
        \hline
    \end{tabular}
\end{table}
\end{landscape}

\begin{table*}
    \centering
    \footnotesize
    \caption{Excerpt of the machine-readable star-by-star membership
    catalogue. For each representative fit, Comp ($p$) column gives the two most probable components and their
    posterior membership probabilities; $t$ is the posterior-weighted
    kinematic age inferred from that fit while $\sigma_t$ is the uncertainty on the kinematic stellar age
    (\autoref{ssec:stellar_ages}). `--' indicates that the star
    was excluded from the youth-selected sample and therefore has no
    15-component entry. The full catalogue of 605 stars, including
    heliocentric Cartesian phase-space coordinates and the complete
    posterior probability vector for every component in every fit, is
    available in machine-readable form online.}
    \label{tab:membership_catalogue}
    \begin{tabular}{lccccccccc}
        \hline
        \multirow{3}{*}{Gaia DR3 source ID} &
        \multicolumn{2}{c}{2-component} &
        \multicolumn{2}{c}{3-component} &
        \multicolumn{2}{c}{14-component} &
        \multicolumn{2}{c}{15-component} \\
        & Comp ($p$)  & $t\pm\sigma_t$ &
        Comp ($p$)  & $t\pm\sigma_t$  &
        Comp ($p$) & $t\pm\sigma_t$ &
        Comp ($p$)  & $t\pm\sigma_t$  \\
        &&[Myr]&&[Myr]&&[Myr] &&[Myr]  \\
        \hline
6113000946373209216 & 1 (1.0) / 0 (0.0) & $12.2 \pm 4.3 $ & 1 (0.6) / 2 (0.4) & $11.9 \pm 3.1$  & 1 (0.6) / 3 (0.3) & $11.7 \pm 2.4$ & 2 (0.6) / 1 (0.4) & $13.8 \pm 3.0$ \\
6102652720892605696 & 0 (1.0) / 1 (0.0) & $0.1 \pm 0.9 $ & 0 (1.0) / 2 (0.0) & $0.1 \pm 0.8$  & 13 (1.0) / 12 (0.0) & $1.7 \pm 0.1$ & 14 (1.0) / 12 (0.0) & $1.6 \pm 0.1$ \\
6086302089635583104 & 0 (1.0) / 1 (0.0) & $0.1 \pm 0.9 $ & 0 (1.0) / 2 (0.0) & $0.1 \pm 0.8$  & 8 (1.0) / 10 (0.0) & $21.0 \pm 5.6$ & --  / --  & --  \\
\multicolumn{9}{c}{\ldots} \\
        \hline
    \end{tabular}
\end{table*}

An excerpt from the machine-readable \texttt{Chronostar} membership catalogue is given in \autoref{tab:membership_catalogue}. For each star, it lists the Gaia DR3 source ID, the two most probable components and their posterior probabilities for each representative fit along with the posterior-weighted kinematic age inferred from the fit (\autoref{subsec:uncertainties}). Stars excluded from the youth-selected sample have no corresponding 15-component entry, denoted by `--'. The full table, available online, additionally lists the six-dimensional Cartesian phase-space coordinates, complete posterior membership probability vector for every component in every fit (34 columns in total: 2 for the two-component fit, 3 for the three-component fit, 14 for the 14-component fit, and 15 for the 15-component fit), together with the star-by-star standard deviation of the posterior-weighted age across models

\autoref{tab:membership_catalogue} illustrates the range of membership behaviour in the catalogue. The first star has probability mass split between two components in the three-, 14-, and 15-component fits (e.g.\ $p_1=0.568$ and $p_2=0.262$ in the 14-component fit), demonstrating why we use the full posterior-weighted age rather than a single maximum-probability assignment when characterising individual stars (\autoref{subsec:uncertainties}). The second star is assigned to a likely background component with $p_1\geq0.98$ in every fit. The third was excluded from the youth-selected sample; its 14-component age ($t=21.00$~Myr) differs sharply from its two- and three-component ages ($t\approx0.06$~Myr), illustrating the largest source of star-by-star model-dependent scatter discussed in \autoref{subsec:uncertainties}.

The present-day distributions are not cleanly separable in any individual two-dimensional projection. This is expected because the \texttt{Chronostar} components are fitted in the full six-dimensional phase space, rather than in right ascension, Cartesian position, or velocity alone. Consequently, two components that overlap substantially in one projection may remain well separated when all phase-space dimensions and observational uncertainties are considered simultaneously.

More importantly, each fitted component is constrained to begin as an isotropic Gaussian in position and velocity space. A physically extended, filamentary, or dynamically complex population may, therefore, be approximated by several fitted Gaussian components of similar kinematic ages, rather than by a single component. We accordingly do not interpret individual fitted components as distinct, instantaneous `star-forming event'. They are better understood as mathematically convenient building blocks of the mixture model that need not correspond one-to-one with real, discrete episodes of star formation, and whose boundaries in phase space are not quite physically meaningful in themselves.

The direct result of the \texttt{Chronostar} analysis is therefore a probabilistic decomposition of the input catalogue into kinematically coherent Gaussian components,  together with their inferred birth parameters, kinematic ages, and posterior membership probabilities. Whether this decomposition maps onto a coherent star-formation history is a separate question, which we address in the following section.
%together with its implications for the kinematic age of PDS~70 and the formation of its planets.

\section{Implications of the kinematic analysis}
\label{sec:implications}

\subsection{Past phase-space structure and star formation history}
\label{subsec:chronostar_past}

The present-day component maps show that the input catalogue is kinematically structured, but the physical interpretation of the fitted Gaussian components requires consideration of their inferred past configurations. We therefore use the fitted component birth parameters to reconstruct the model-implied phase-space distribution at a common lookback time.

We define a positive lookback time $\tau$, with $\tau=0$ corresponding to the present day. A component $j$ with a fitted age $t_j$ is evolved forward from its fitted birth configuration for a time
\begin{equation}
    \Delta t_j = t_j - \tau,
\end{equation}
and is shown only when $\tau \leq t_j$. Thus, no component is propagated backward beyond its inferred birth time.

\begin{figure}
    \centering
    \includegraphics[width=0.95\linewidth]{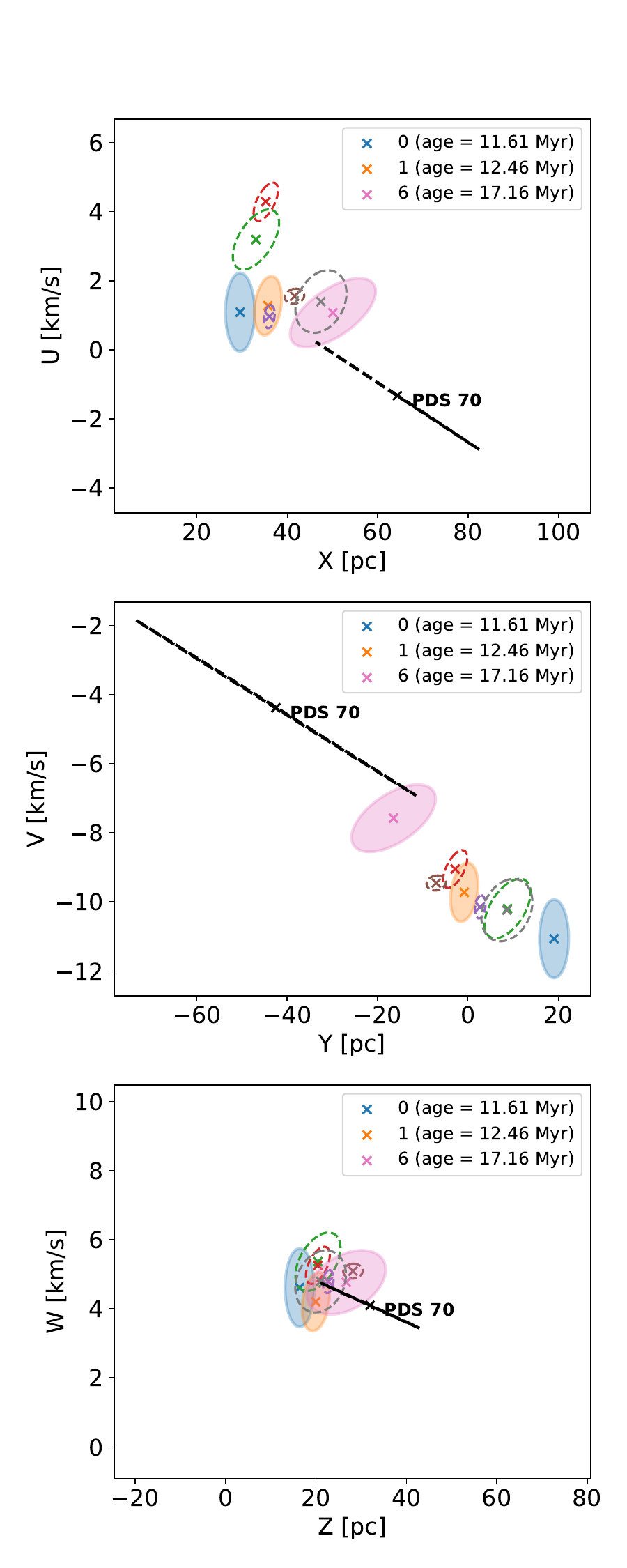}
    \caption{Position--velocity projections of the non-background components in the 15-component fit at a lookback time of 11.6 Myr.
    Ellipses show projected intrinsic $1\sigma$ component covariances and crosses show component means. The black cross and dashed ellipse show the propagated mean and $1\sigma$ phase-space uncertainty of PDS~70 after tracing its present-day state backward by 11.6 Myr. Labels denote the components with the highest posterior probabilities of containing PDS~70 along with their fitted kinematic ages. The horizontal and vertical axis scales are the same in all panels.}
    \label{fig:pds70_phase_space}
\end{figure}

\begin{figure}
    \centering
    \includegraphics[width=0.95\linewidth]{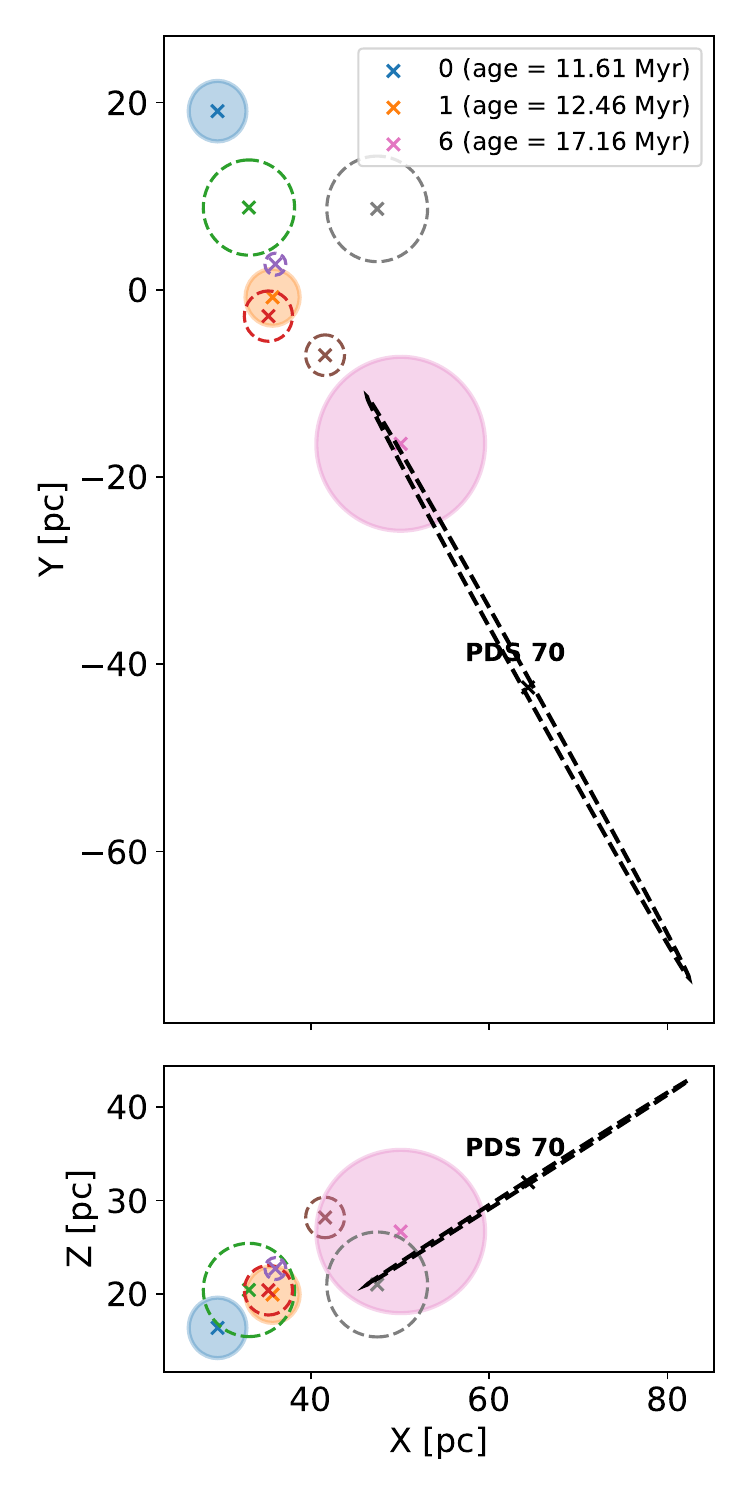}
    \caption{Same as \autoref{fig:pds70_phase_space}, but now showing the configuration at a lookback time of 11.6 Myr in position-position space rather than position-velocity space. The axis scales are isotropic, so the figure shows the true shape of the distribution.}
    \label{fig:pds70_position}
\end{figure}

\autoref{fig:pds70_phase_space} and \autoref{fig:pds70_position} show the reconstructed configuration of the 15-component youth-selected fit at $\tau=11.6$ Myr. We select this time because it is the minimum fitted kinematic age among the non-background components with at least ten
maximum-probability assigned members. Thus our choice of time is one where every component has
already formed, and we show the components that formed earlier than 11.6 Myr ago with their properties evolved forward from the fitted birth configurations to their configurations at $\tau=11.6$ Myr.
For comparison, we also reconstruct the 14-component model at the same lookback time, including only components with fitted ages $t_j\geq11.6$~Myr so that no component is propagated backward beyond its fitted birth time. The two reconstructions therefore provide a like-for-like comparison of the phase-space structure implied by the full and youth-selected samples. We show the position-velocity and position-position reconstructions for the fourteen-component case in \aref{app:14comp}.

%All position projections use a common linear scale in pc, and the velocity projections use a common linear scale in km/s, allowing the relative spatial and kinematic extents to be compared directly between projections and models. 
% should i mention Großschedl, J. E., et al.: A&A, 709, A181 (2026) paper - they do a linear fit in position velocity diagram and find sort of `onset time of expansion' and ours look linear-ish too.

\subsubsection{Morphology}
\label{subsubsec:morphology}

To quantify the morphology of the reconstructed distribution, we calculate the membership-weighted three-dimensional spatial covariance of the evolved component mixture. 
For component $j$, let $\boldsymbol{\mu}_{x,j}$ and
$\mathbf{C}_{x,j}$ denote its evolved spatial mean and $3\times3$ spatial
covariance matrix at the selected lookback time. 
The relative contribution of component $j$ is defined using its posterior
effective membership,
\begin{equation}
    N_{{\rm eff},j}
    =
    \sum_i p_{ij},
\end{equation}
where $p_{ij}$ is the posterior membership probability of star $i$ in
component $j$. The normalised component weight is
\begin{equation}
    w_j
    =
    \frac{N_{{\rm eff},j}}
    {\sum_kN_{{\rm eff},k}}.
\end{equation}

The mixture mean is
\begin{equation}
    \bar{\boldsymbol{x}}
    =
    \frac{\sum_j w_j\boldsymbol{\mu}_{x,j}}
    {\sum_j w_j},
\end{equation}
and the corresponding total spatial covariance is
\begin{equation}
    \mathbf{C}_{x,\mathrm{mix}}
    =
    \frac{1}{\sum_j w_j}
    \sum_j w_j
    \left[
        \mathbf{C}_{x,j}
        +
        \left(
            \boldsymbol{\mu}_{x,j}
            -
            \bar{\boldsymbol{x}}
        \right)
        \left(
            \boldsymbol{\mu}_{x,j}
            -
            \bar{\boldsymbol{x}}
        \right)^{\mathsf T}
    \right],
    \label{eq:mixture_covariance}
\end{equation}
with $\sum_j w_j = 1$.
The total covariance follows the law of total covariance for a finite
Gaussian mixture. The first term in \autoref{eq:mixture_covariance}
is the membership-weighted mean internal covariance of the evolved
components, whereas the second is the covariance of their evolved
centroids about the mixture mean.

We diagonalise $\mathbf{C}_{x,\mathrm{mix}}$ and order its eigenvalues
as $\lambda_1\geq\lambda_2\geq\lambda_3$, with corresponding eigenvectors $\boldsymbol{e}_1, \boldsymbol{e}_2, \boldsymbol{e}_3$ which indicate the orientation of the largest, middle, and shortest axis. The corresponding principal-axis
dispersions are $\sigma_k=\sqrt{\lambda_k}$. We define the
major-to-transverse aspect ratio as
\begin{equation}
    A_{\mathrm{transverse}}
    =
    \frac{\sigma_1}
    {\sqrt{(\sigma_2^2+\sigma_3^2)/2}},
\end{equation}
and define Gaussian-equivalent characteristic extents along and
perpendicular to the major axis as
\begin{equation}
    L_{2\sigma}=4\sigma_1,
    \qquad
    W_{2\sigma}
    =
    4\sqrt{
        \frac{\sigma_2^2+\sigma_3^2}{2}
    }.
\end{equation}
These are characteristic extents of the fitted mixture distribution and
should not be interpreted as literal end-to-end dimensions of a physical
filament.

For the 15-component youth-selected fit, the ordered spatial covariance
eigenvalues are
\begin{equation}
    (\lambda_1,\lambda_2,\lambda_3)
    =
    (153.62,32.91,20.66)~\mathrm{pc^2},
\end{equation}
corresponding to principal-axis dispersions of
\begin{equation}
    (\sigma_1,\sigma_2,\sigma_3)
    =
    (12.39,5.74,4.54)~\mathrm{pc}.
\end{equation}
The reconstructed distribution has a Gaussian-equivalent major-axis extent
of $L_{2\sigma}=49.58$~pc and transverse width of
$W_{2\sigma}=20.7$~pc, giving a major-to-transverse aspect ratio of
$A_{\mathrm{transverse}}=2.39$.

At the same lookback time, the 14-component full-sample fit has
\begin{equation}
    (\lambda_1,\lambda_2,\lambda_3)
    =
    (245.57,30.59,24.94)~\mathrm{pc^2},
\end{equation}
or principal-axis dispersions of 
\begin{equation}
    (\sigma_1,\sigma_2,\sigma_3)
    =
    (15.67,5.53,4.99)~\mathrm{pc}.
\end{equation}
It has $L_{2\sigma}=62.7$~pc,
$W_{2\sigma}=21.1$~pc, and
$A_{\mathrm{transverse}}=2.97$.

%The results for the two reconstructions are therefore quite similar, and paint a consistent picture: 11.6 Myr ago the stars that now comprise the UCL region were arranged in a loose filamentary structure $\sim 12-16$ pc long, and with an aspect ratio of roughly $A_{\rm transverse} \approx 2.4 - 3$. The two minor principle axis sizes 
%transverse widths are nearly identical between the two reconstructions,
%whereas the full-sample fit has a larger major-axis extent. The resulting
%major-to-transverse aspect ratios of $2.39$ and $2.97$ demonstrate a
%substantial elongation that persists after removing stars classified as
%old. In both fits, the two minor principal-axis dispersions 
%are similar,
%indicating an approximately cylindrical rather than sheet-like
%configuration. 
The major-axis directions are also nearly identical in the two fits:
\begin{equation}
    \boldsymbol{e}_{1,14}
    =
    (0.446,-0.853,0.269),
\end{equation}
and
\begin{equation}
    \boldsymbol{e}_{1,15}
    =
    (0.453,-0.855,0.250),
\end{equation}
differing by approximately $1.15^\circ$. Their inclinations relative to the
Galactic plane are $15.6^\circ$ and $14.5^\circ$ for the 14-component and
15-component fits, respectively. The in-plane-to-vertical extent ratios,
$R_{\rm plane/Z}=1.78$ and $1.75$, further show that the reconstructed
structure is preferentially extended within the Galactic plane.

The two fits therefore paint a consistent picture of the reconstructed configuration: 11.6 Myr ago the stars that now comprise UCL were organised into an elongated, approximately
cylindrical structure predominantly aligned with the Galactic plane. Its
morphology is consistent with a thick filament-like or chain-like stellar
configuration in the plane. We note that this configuration is qualitatively quite similar to the characteristic molecular filament configurations identified by \citet{Hu26a} in their study of giant molecular filaments. 

\subsubsection{Velocity structure}
\label{subsubsec:velocity_structure}

To assess whether the reconstructed elongated configuration has a coherent longitudinal velocity field, we project each component centroid onto the major spatial principal axis. For component $j$, we define the coordinate along the major axis as
\begin{equation}
    s_j
    =
    \left(
        \boldsymbol{\mu}_{x,j}
        -
        \bar{\boldsymbol{x}}
    \right)
    \cdot
    \boldsymbol{e}_1,
\end{equation}
where $\boldsymbol{\mu}_{x,j}$ is the evolved spatial centroid at $\tau=11.6$~Myr, $\bar{\boldsymbol{x}}$ is the membership-weighted mixture mean, and $\boldsymbol{e}_1$ is the first spatial principal axis. We similarly define the component velocity parallel to this axis as
\begin{equation}
    v_{\parallel,j}
    =
    \boldsymbol{\mu}_{v,j}
    \cdot
    \boldsymbol{e}_1,
\end{equation}
where $\boldsymbol{\mu}_{v,j}=(U_j,V_j,W_j)$ is the evolved component
velocity centroid.

\begin{figure}
    \centering
    \includegraphics[width=\linewidth]{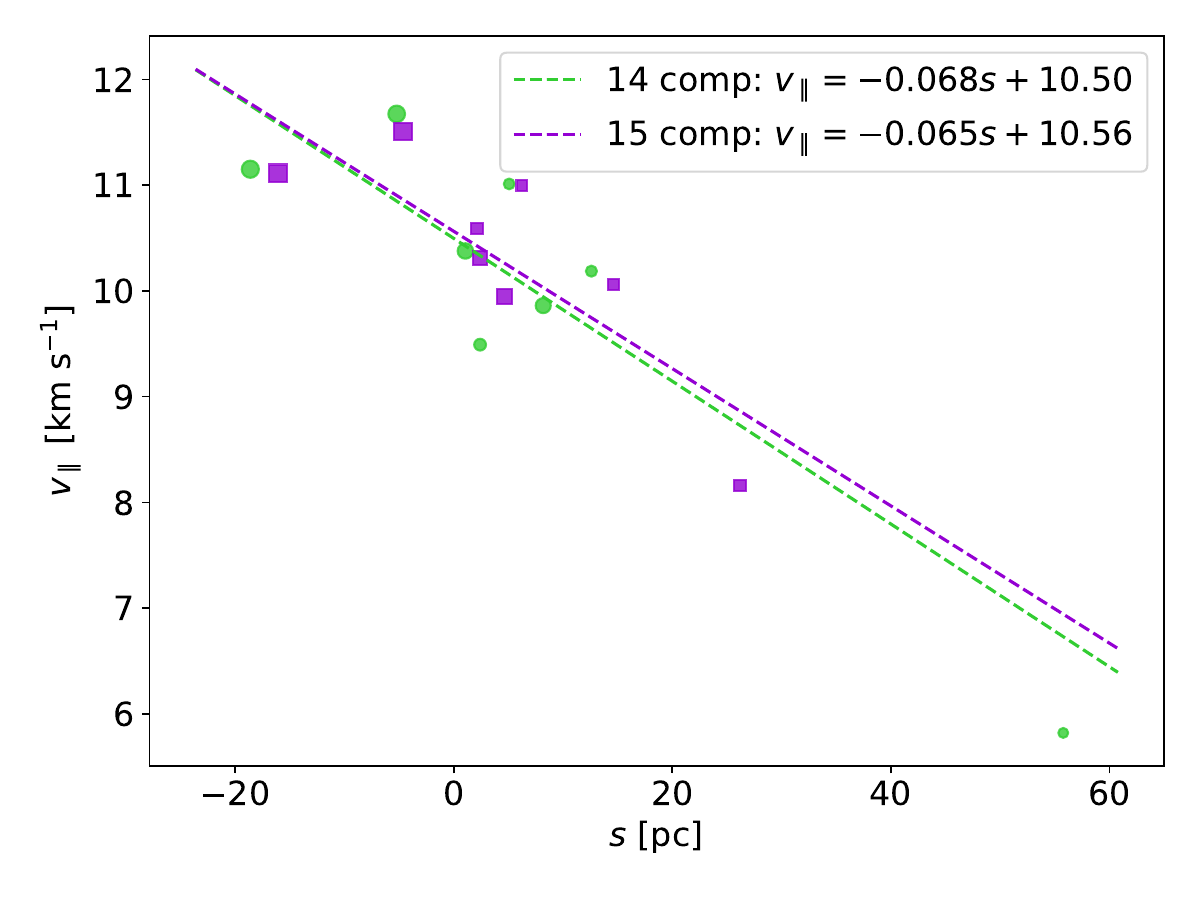}
    \caption{Velocity structure along the reconstructed major spatial axis at $\tau=11.6$~Myr. The component coordinate $s$ is measured along the first principal axis of the membership-weighted spatial covariance, and $v_\parallel$ is the component centroid velocity projected onto that same axis.
    Circles (green) and squares (purple) denote the 14-component full-sample and 15-component youth-selected fits, respectively. Marker area is proportional to posterior effective membership. Lines show the membership-weighted linear fits.}
    \label{fig:v_parallel}
\end{figure}

We fit a membership-weighted linear relation,
\begin{equation}
    v_{\parallel,j}
    =
    a_v+b_vs_j,
\end{equation}
to the non-background components in each model with $N_{\rm eff} > 10$. The fitted slopes
are
\begin{equation}
    b_{v,14}
    =
    -0.068 \pm 0.067~\mathrm{km\,s^{-1}\,pc^{-1}},
\end{equation}
for the 14-component full-sample fit and
\begin{equation}
    b_{v,15}
    =
    -0.065 \pm 0.086~\mathrm{km\,s^{-1}\,pc^{-1}},
\end{equation}
for the 15-component youth-selected fit. These results are shown in \autoref{fig:v_parallel}.

For our chosen coordinate system, negative slopes correspond to longitudinal convergence, i.e., to flows where the various components are converging along the filament. However, both slopes
are consistent with zero within their formal uncertainties. We therefore
find no statistically significant evidence for coherent longitudinal
compression or expansion along the reconstructed major axis. The residual weighted dispersions about the fitted relations are $0.37~\mathrm{km\,s^{-1}}$ and $0.26~\mathrm{km\,s^{-1}}$ for the 14-component and 15-component fits, respectively. These values describe the scatter of fitted component centroids about the linear relations and should not be interpreted as intrinsic stellar velocity dispersions.

% If the linear trend is interpreted as a coherent compression, its characteristic timescale is
% \begin{equation}
%     t_{\rm comp}
%     =
%     \frac{1}
%     {1.023|b_v|},
% \end{equation}
% giving $t_{\rm comp}=14.47$~Myr and $15.05$~Myr for the 14-component and 15-component fits, respectively.

The similar slopes obtained from the two independent decompositions suggest that the longitudinal velocity pattern is not strongly dependent on the removal of stars classified as old. Nevertheless, this result should be interpreted cautiously because the fit is performed on a small number of model components rather than on independent stellar velocity measurements and the signal-to-noise ratios are quite low.

% To search for a systematic transverse velocity pattern along the reconstructed major axis, we also consider the component of the velocity perpendicular to $\boldsymbol{e}_1$,
% \begin{equation}
%     \boldsymbol v_{\perp,j}
%     =
%     \boldsymbol\mu_{v,j}
%     -
%     (\boldsymbol\mu_{v,j}\cdot\boldsymbol e_1)\,\boldsymbol e_1,
% \end{equation}
% and plot its scalar projection $v_{\perp,j}$ as a function of $s_j$. We fit
% \begin{equation}
%     v_{\perp,j}
%     =
%     a_\perp + b_\perp s_j.
% \end{equation}
% A non-zero slope $b_\perp$ would indicate a systematic change of the transverse velocity with position along the major axis, corresponding to rotation or shear of the filament. The fitted slopes are
% \begin{equation}
%     b_{\perp,14}
%     =
%     -(0.005 \pm 0.067)~\mathrm{km\,s^{-1}\,pc^{-1}},
% \end{equation}
% for the 14-component fit and
% \begin{equation}
%     b_{\perp,15}
%     =
%     -(0.016 \pm 0.086)~\mathrm{km\,s^{-1}\,pc^{-1}},
% \end{equation}
% for the 15-component fit (bottom panel of \autoref{fig:v_parallel}). In both cases the slope is consistent with zero. We therefore find no significant evidence for a coherent transverse velocity gradient along the major axis.

We additionally test for solid-body rotation about each of the three principal axes by fitting the corresponding paired transverse velocity--position relations (e.g.\ $v_2=-\Omega_1 s_3$, $v_3=\Omega_1 s_2$ for rotation about $\boldsymbol{e}_1$, and analogously for $\boldsymbol{e}_2$ and $\boldsymbol{e}_3$). The best-fitting angular speeds are small ($\lesssim0.05~\mathrm{km\,s^{-1}\,pc^{-1}}$) and vary substantially between the 14-component and 15-component decompositions, with implied rotation timescales of order $20$--$120$~Myr. Given the lack of consistency between the two models and the small amplitudes relative to the internal velocity scatter, we do not interpret these as significant detections of coherent rotation. Full numerical results and diagnostic plots are provided in \aref{app:rotation}.

\subsubsection{Age}
\label{subsubsec:age}

We next examine whether the inferred component ages show a systematic ordering along the reconstructed major axis. Using the coordinate $s_j$ defined in \autoref{subsubsec:velocity_structure}, we fit a posterior-effective-membership-weighted linear relation
\begin{equation}
    t_j
    =
    a_t+b_ts_j
\end{equation}
to the kinematic ages of the robust non-background components.

The fitted slopes are
\begin{equation}
    b_{t,14}
    =
    0.083\pm0.067~
    \mathrm{Myr\,pc^{-1}}
\end{equation}
for the 14-component full-sample fit and
\begin{equation}
    b_{t,15}
    =
    0.086\pm0.086~
    \mathrm{Myr\,pc^{-1}}
\end{equation}
for the 15-component youth-selected fit.

The positive best-fitting slopes indicate that, under the adopted sign convention for the first principal axis, component ages increase towards positive $s$. The two fits yield similar central gradient values. However, both slopes are consistent with zero within their formal uncertainties. We therefore find no statistically significant evidence for a monotonic component-age gradient along the reconstructed major axis. The apparent trend is reported as a descriptive property of the fitted component decomposition, rather than as evidence for sequential star formation.

\begin{figure}
    \centering
    \includegraphics[width=\linewidth]{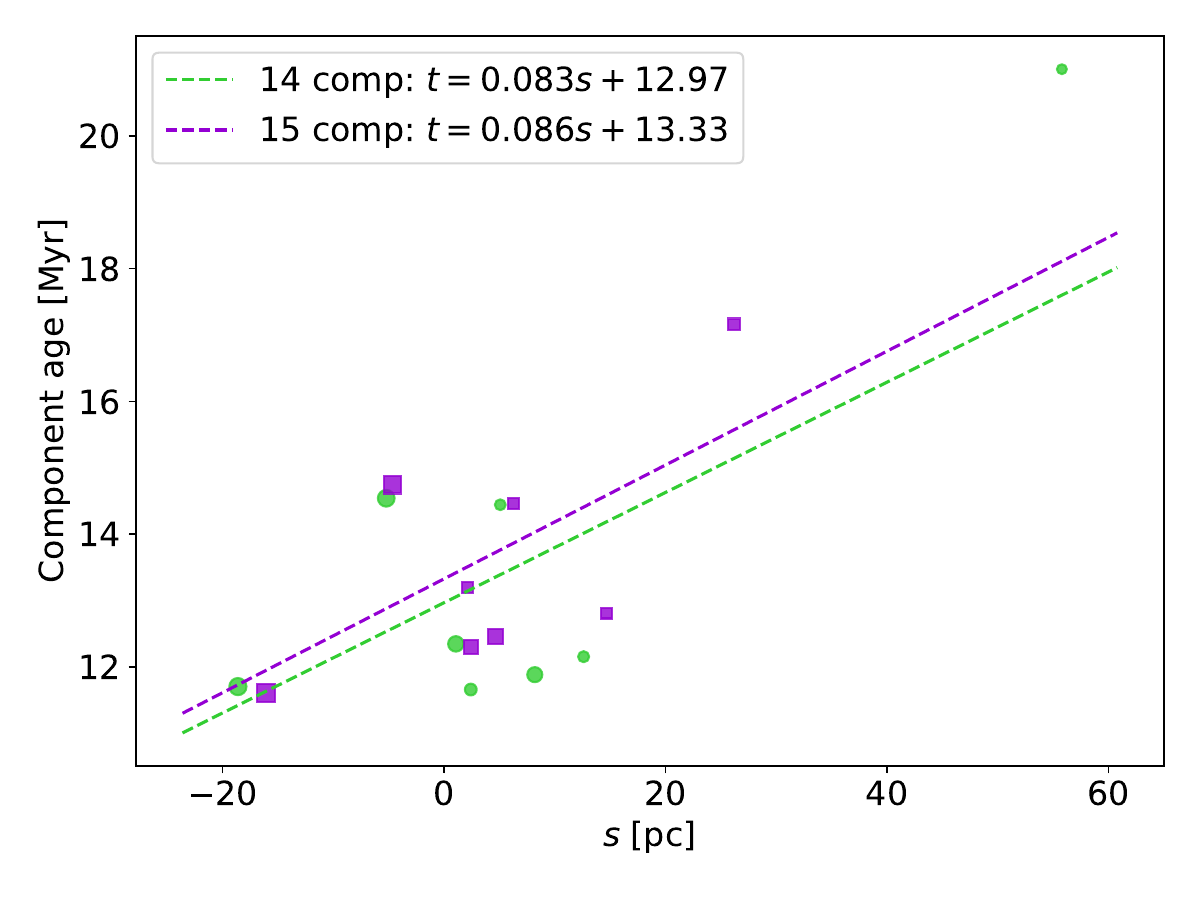}
    \caption{Fitted component kinematic age as a function of position along the first principal axis of the reconstructed distribution. Circles (green) and squares (purple) denote the 14-component full-sample and 15-component youth-selected fits, respectively. Marker area is proportional to posterior effective membership. Lines show the membership-weighted linear fits.}
    \label{fig:age_gradient}
\end{figure}

\subsubsection{Distribution of individual stellar ages and star formation history}

\begin{figure}
    \centering
    \includegraphics[width=\linewidth]{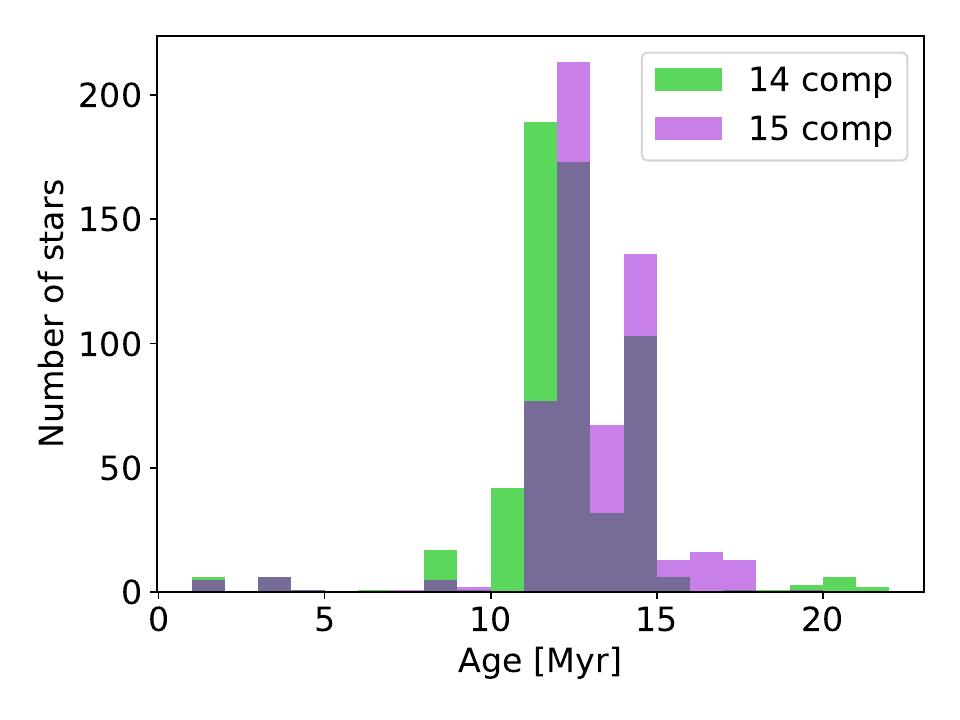}
    \caption{Posterior-weighted kinematic-age distributions for stars in the 14-component (full-sample; green) and 15-component (youth-selected sample; purple) models that have been assigned to the non-background components based on our criteria in \autoref{subsec:model_component}.}
    \label{fig:sfh}
\end{figure}

Since we have determined the kinematic ages for all stars in \autoref{ssec:stellar_ages}, we can also examine the distribution of posterior-weighted kinematic ages for individual stars, which we show in \autoref{fig:sfh}. Here we include all stars in the full sample for the 14-component fit and in the youth-selected sample for the 15-component fit that have been assigned to non-background components based on our criteria in \autoref{subsec:model_component}. The 14-component and 15-component models show a broadly similar dominant concentration at approximately 10--15 Myr. Differences in their fine-scale structure reflect the alternative component decompositions and should not be interpreted as independently resolved star-formation episodes. The median model-dependent scatter in posterior-weighted stellar age is $0.65$~Myr, with an interquartile range of $0.19$--$1.54$~Myr (\autoref{ssec:stellar_ages}). We therefore interpret only the broad age concentration shared by both decompositions, rather than fine-scale histogram features on approximately Myr scales.

The median posterior-weighted kinematic age is $Q_{50,14}=12.15$~Myr for the 14-component fit and $Q_{50,15}=12.81$~Myr for the 15-component fit, with corresponding 16th--84th percentile ranges of $11.31$--$14.35$~Myr and $11.91$--$14.68$~Myr. The fraction of stars in the $10$--$13$~Myr interval is $0.68$ for the 14-component fit and $0.5$ for the 15-component fit, while the fraction older than $16$~Myr is $0.02$ and $0.05$, respectively.

Non-kinematic age estimates for UCL span approximately $14$--$22$~Myr depending on the stellar models, membership definitions, and whether UCL is considered separately or together with Lower Centaurus-Crux (LCC). \citet{mamajek2002} find pre-main-sequence ages of $15$--$22$~Myr for UCL (with main-sequence turn-off ages of $17\pm1$ Myr, while \citet{pecaut_mamajek2016} determine an age of $16\pm2$ Myr from a spectroscopic survey of new K- and M-type members and note that the choice of evolutionary tracks shifts the inferred age by up to $\sim 2$ Myr. \textit{Gaia}-era photometric analyses broadly confirm this range: \citet{squicciarini2021}, using Gaia EDR3 astrometry and photometry, treat UCL/LCC as a $\sim20$~Myr reference population, with some subsets consistent with ages as young as $\sim14$~Myr if the bulk of UCL/LCC is assumed to be 20~Myr. \citet{kerr2021} identify UCL-related clusters with isochronal ages spanning $\sim 14-22$ Myr, with a prominent peak at $15-16$ Myr and an older peak at $20-22$ Myr. \citet{ratzenboeck2023} derive isochronal ages for 37 kinematically-identified clusters and reconstruct star-formation history dominated by a $\sim 15$ Myr burst that includes many UCL clusters.

Our dominant kinematic age concentration of approximately $11$--$15$~Myr, with a median of $12.1$--$12.8$~Myr depending on the component model, is broadly consistent with these non-kinematic estimates for UCL, lying toward the younger end of the classically quoted range. The small fraction of stars with inferred ages greater than $16$~Myr ($2$--$5$\%) suggests that our candidate sample is not substantially contaminated by an older background population (or by other components of UCL). Our reconstruction is smaller in scale than the $\sim\!100$-pc cluster chains reported across the wider Sco--Cen complex \citep[e.g.][]{kerr2021,ratzenboeck2023} and is derived from a spatially restricted candidate T-component sample rather than a complete Sco--Cen census, so direct structural comparisons should be made with caution.

%Taken together, the reconstructed morphology demonstrates that the candidate T-component population is elongated, approximately cylindrical, and preferentially aligned with the Galactic plane. The posterior-weighted kinematic ages identify a broad dominant age scale of approximately $11$--$15$~Myr, consistent with the UCL subgroup age inferred from non-kinematic methods. The fitted age and velocity gradients along the reconstructed major axis are not statistically significant within their formal uncertainties, so we refrain from interpreting the apparent trends as evidence for sequential star formation. These properties are qualitatively consistent with the established picture of Sco--Cen as a spatially and temporally structured association rather than a single coeval population. These results provide the phase-space context for interpreting the probabilistic association and kinematic age of PDS~70, which we consider next.

\subsection{PDS~70}
\label{subsec:pds70}
\subsubsection{Kinematic association and age}
We now consider PDS~70 in the context of the reconstructed past phase-space structure. PDS~70 has measured present-day six-dimensional phase-space coordinates, which were included in the membership analysis described in \autoref{sec:age_extraction}. The resulting posterior probabilities quantify its association with the fitted components in each component model. 

%As mentioned earlier, the \texttt{Chronostar} component distributions are known at their fitted birth states and evolved forward to the present day. 
%To compare PDS~70 with these reconstructed distributions at a common epoch, 
We begin our examination of PDS 70 in the context of the broader T component population by  propagating its present-day phase-space coordinates backwards to the same $\tau=11.6$ Myr lookback time we have used in examining the T component population near its formation time. We also propagate the PDS~70 phase-space covariance using the \textit{Gaia} astrometric covariance matrix and the adopted radial velocity uncertainty. We show the resulting uncertainty ellipse for PDS~70 in \autoref{fig:pds70_phase_space} and \autoref{fig:pds70_position}. 

Several features of this plot are noteworthy. First, we see that PDS~70's uncertainty ellipse is both large and extremely covariant -- and we remind readers that the uncertainty intervals shown are $1\sigma$ uncertainties, so the credible $2\sigma$ range extends well outside the frame of the plot. This structure and large extent is expected, particularly in the spatial coordinates, because present-day velocity uncertainties propagate into positional uncertainties, which become correlated and grow in time as the trajectory is traced back through the Galactic potential. This growth of uncertainty is the fundamental motivation for \texttt{Chronostar}'s forward-modelling approach, which evaluates membership probabilistically in the full six-dimensional phase space using birth components that are propagated forward to the present-day rather than stellar positions propagated backward into the past.

The components with the largest posterior probabilities of containing PDS~70 are highlighted as colour-filled ellipses and labelled by their component numbers and fitted ages in \autoref{fig:pds70_phase_space} and \autoref{fig:pds70_position}. In the projected views, these components are also those most consistent with the traceback position and velocity of PDS~70 within its propagated uncertainties. However, the figures are illustrative: the posterior membership probabilities are calculated from the complete six-dimensional \texttt{Chronostar} mixture model evaluated at the present day, not from the individual two-dimensional projections projected back into the past.

\begin{figure}
    \centering
    \includegraphics[width=\linewidth]{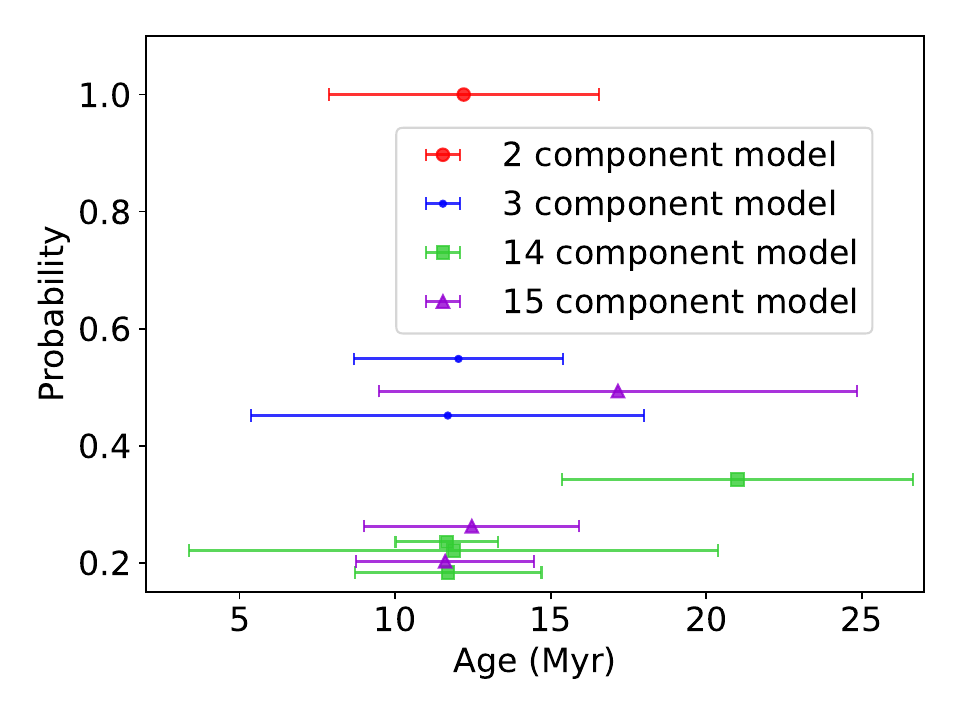}
    \caption{Component kinematic ages and uncertainties plotted against the posterior probability that PDS 70 belongs to each component. Only components with $P(\mathrm{PDS\,70}\mid j)\geq0.10$ are shown. Colours and symbols, as labelled, distinguish the models that the components belong to.}
    \label{fig:pds70_age}
\end{figure}

The inferred kinematic age of PDS~70 depends on the adopted model because its membership probability is distributed across multiple components in different models. \autoref{fig:pds70_age} shows the fitted ages and formal uncertainties of components with a posterior probability $P(\mathrm{PDS\,70}\mid j)\geq0.10$. Components below this threshold are omitted for clarity because they contribute little to the membership-weighted age and are predominantly background components with very small fitted ages.

Following the stellar kinematic age definition in \autoref{ssec:stellar_ages}, we calculate the posterior-membership-weighted age for PDS~70 for each model. The results are listed in the first two numerical columns of \autoref{tab:pds70_components} along with the uncertainty in the ages. 
The standard deviation of the fiducial membership-weighted ages across the four component models is $1.38$~Myr. This exceeds the 84th percentile ($1.33$~Myr) of the model-dependent stellar age scatter in our sample, indicating that the inferred age of PDS~70 is relatively sensitive to the adopted component decomposition. However, all fiducial models place PDS~70 in a population with a kinematic age of order $12$--$15$ Myr.

\begin{table*}
    \centering
    \caption{Posterior-membership-weighted kinematic ages of PDS~70 in each of the models for alternative radial velocity measurements. For each calculation, the fitted \texttt{Chronostar} components are held fixed and only the PDS~70 phase-space mean and covariance are recalculated. The \textit{Gaia} DR3 radial velocity is used in the fiducial analysis that defines the components in each model. The quoted uncertainties are propagated from the formal component-age uncertainties as defined in \autoref{ssec:stellar_ages}.}
    \label{tab:pds70_components}
    \begin{tabularx}{\textwidth}{l *{8}{>{\centering\arraybackslash}X}} 
        \hline
        &
        \multicolumn{2}{c}{Gaia DR3 (fiducial)} &
        \multicolumn{2}{c}{Gaia DR2} &
        \multicolumn{2}{c}{Žerjal et al. (2021)} &
        \multicolumn{2}{c}{Thanathibodee et al. (2020)} \\
        &
        \multicolumn{2}{c}{[$0.74\pm3.22$ km\,s$^{-1}$]} &
        \multicolumn{2}{c}{[$3.10\pm1.40$ km\,s$^{-1}$]} &
        \multicolumn{2}{c}{[$5.10\pm1.50$ km\,s$^{-1}$]} &
        \multicolumn{2}{c}{[$6.00\pm1.50$ km\,s$^{-1}$]} \\
        Model &
        $t_{\rm kin}$ &
        $\sigma_t$ &
        $t_{\rm kin}$ &
        $\sigma_t$ &
        $t_{\rm kin}$ &
        $\sigma_t$ &
        $t_{\rm kin}$ &
        $\sigma_t$ \\
        &
        [Myr]&[Myr]&[Myr]&[Myr]&[Myr]&[Myr]&[Myr]&[Myr]\\
        \hline
        2-component &
        12.20 & 4.34 &
        12.20 & 4.34 &
        12.20 & 4.34 &
        12.20 & 4.34 \\

        3-component &
        11.88 & 3.40 &
        11.82 & 4.17 &
        11.89 & 3.26 &
        11.92 & 3.09 \\

        14-component &
        14.92 & 2.79 &
        14.72 & 3.24 &
        11.91 & 2.90 &
        11.81 & 2.58 \\

        15-component &
        14.67 & 3.94 &
        15.87 & 5.65 &
        13.76 & 2.76 &
        12.97 & 2.09 \\
        \hline
    \end{tabularx}
\end{table*}

\subsubsection{Sensitivity to the adopted radial velocity}
\label{subsubsec:pds70_rv}
The radial velocity of PDS~70 remains uncertain and differs between published measurements. \textit{Gaia} DR2 reports a radial velocity of $3.1\pm1.4~\mathrm{km\,s^{-1}}$, while \citet{thanathibodee2020} and \citet{zerjal21} report $6.0\pm1.5~\mathrm{km\,s^{-1}}$ and $5.1\pm1.5~\mathrm{km\,s^{-1}}$, respectively . Our fiducial analysis adopts the \textit{Gaia} DR3 value of $0.74 \pm 3.22~\mathrm{km\,s^{-1}}$, following the radial velocity selection procedure described in \autoref{subsec:chronostar_data}. The earlier \texttt{Chronostar} analysis of Sco--Cen found that the membership of PDS~70 in the entire T component remained high when replacing the \textit{Gaia} DR2 radial velocity with the alternate values of \citet{zerjal21} and \citet{thanathibodee2020}.

To test the sensitivity of our kinematic age to the adopted radial velocity, we repeat the PDS~70 membership calculation using the \textit{Gaia} DR2, \textit{Gaia} DR3, \citet{zerjal21} and \citet{thanathibodee2020} radial velocities. For this test, the fitted \texttt{Chronostar} mixture decompositions are held fixed, and only the PDS~70 phase space mean and covariance are recalculated for each radial velocity measurement. This procedure therefore quantifies the conditional sensitivity of the PDS~70 memberships, and consequently its membership-weighted kinematic age, to the adopted radial velocity. It does not test the effect of refitting the complete T-component decomposition, though this effect is expected to be minimal given that this decomposition is based on hundreds of stars. 

The resulting membership-weighted kinematic ages are listed in \autoref{tab:pds70_components}. In the 2-component and 3-component models, the kinematic age is nearly insensitive to the adopted radial velocity, varying by less than $0.001$~Myr and $0.10$~Myr, respectively. In the more finely structured 14-component and 15-component models, the alternative radial velocities redistribute probability between distinct subcomponents, resulting in age ranges of $11.81$--$14.92$~Myr and $12.97$--$15.87$~Myr, respectively. The corresponding radial velocity-induced variations are $3.10$~Myr and $2.90$~Myr.

Thus, the inferred detailed age of PDS~70 is nearly independent of radial velocity in the simple models but has a non-negligible radial-velocity dependence in the more finely decomposed 14- and 15-component models. However, the resulting age changes are comparable to the corresponding formal membership-weighted age uncertainties ($2.1$--$5.6$~Myr), and all radial-velocity choices retain PDS~70 within the broad range of $12$--$16$ Myr.

\subsubsection{Implications for planet formation}
\label{subsubsec:pds70_planet_formation}
The age inferred above provides an independent kinematic constraint on the evolutionary state of the PDS~70 stellar system. This constraint is relevant to the interpretation of the directly imaged companions because the luminosity of a cooling planet depends on both its mass and its age, while the luminosity--age relation also depends on the initial entropy set by the formation pathway. 
We now examine how the adopted age and its uncertainty affect inferences drawn from the observed luminosities and dynamical mass estimates obtained from precise astrometry of PDS~70~\textit{b} and \textit{c}.

\begin{figure}
    \centering
    \includegraphics[width=\linewidth]{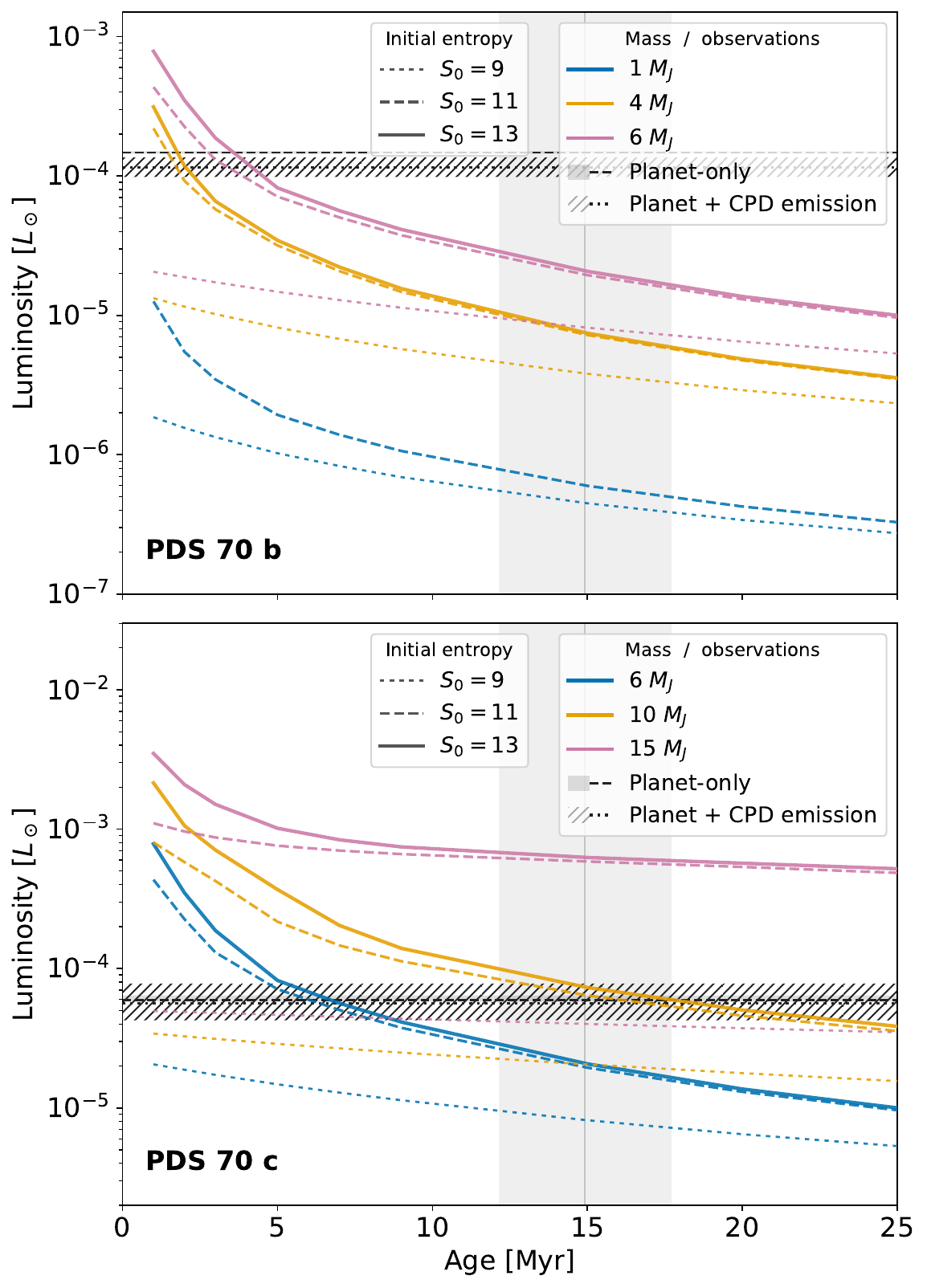}
    \caption{Integrated luminosities from the
    \citet{spiegel_burrows2012} evolutionary models as a function of
    planetary cooling age cooling age for PDS~70~\textit{b} (top) and PDS~70~\textit{c} (bottom). Solid, dashed, and dotted curves correspond to post-formation entropies $S_{\rm i}=13$, $11$, and $9~k_{\rm B}$ per baryon, respectively. Colours indicate the selected planetary masses, as labelled. The vertical shaded region marks the fiducial PDS~70 system age from the 14-component \textit{Gaia} DR3 model.  This range is shown as a representative kinematic age scale rather than a formal confidence interval. Horizontal lines and shaded regions show the luminosity inferred from the planet-only SED fit and the luminosity of the planetary blackbody component in the planet-plus-circumplanetary-disk SED fit of \citet{blakely2025}. The separately fitted circumplanetary blackbody component is not included in the latter luminosity. The mass tracks for each companion are selected from the median, $68\%$, and $95\%$ dynamical constraints of \citet{trevascus2025} under their two-planet orbital model.}
    \label{fig:planet}
\end{figure}

For this comparison, we use the age from the 14-component model with the \textit{Gaia} DR3 radial velocity value, $t_\text{star} = 14.92\pm2.79$ Myr. The central kinematic ages obtained from the other component models and alternative radial velocities span approximately $12$--$16$ Myr which is already (mostly) encompassed by the formal uncertainty of our fiducial model kinematic age (marked in the \autoref{fig:planet}). 

We compare this age range with the evolutionary models of \citet{spiegel_burrows2012}. These models provide flux densities at a distance of 10 pc for planets with masses $M_{\rm p}=1$--$15\,M_{\rm Jup}$, post-formation entropies $S_{\rm i}=8$--$13~k_{\rm B}$ per baryon, and ages from $1$ to $1000$~Myr. Following previous work on the PDS 70 planets \citep[e.g.,][]{trevascus2025}, we estimate the integrated model luminosity using
\begin{equation}
    L_{\rm mod}
    =
    4\pi(10~\mathrm{pc})^2
    \int_{0.8~\micron}^{15~\micron}
    F_\lambda^{10~\mathrm{pc}}\,\mathrm{d}\lambda .
\end{equation}
% The finite wavelength range means that these values should be described as integrated model luminosities over the model grid, rather than strictly bolometric luminosities, unless the omitted flux is negligible for all tracks considered.

The dynamical analysis of \citet{trevascus2025} provides the relevant mass scale against which these cooling models should be compared. We use their two-planet orbital solution, which assumes that PDS~70~\textit{b} and \textit{c} are the only planets included in the dynamical model. For each companion, we show the tracks near the posterior median mass and the upper masses of the 68\% and 95\% credible intervals. We consider post-formation entropies of $S_{\rm i}=9$, $11$, and $13~k_{\rm B}$ per baryon. 
These tracks permit a conservative comparison with the observed luminosity by considering high-mass solutions, but should not be interpreted as equally probable planet masses. 

The observed luminosity constraints are calculated from the effective temperatures  ($T_{\rm eff}$) and radii ($R$) reported by \citet{blakely2025}, who fit both a planet-only SED model and a model that includes an additional blackbody contribution that allows for circumplanetary emission. We calculate the luminosity as 
\begin{equation}
    L = 4\pi R^2\sigma_{\rm SB}T_{\rm eff}^4,
\end{equation}
where $\sigma_{\rm SB}$ is the Stefan--Boltzmann constant. We use the reported median $T_{\rm eff}$ and $R$ values to calculate the nominal luminosities, and use combinations of the reported 16th and 84th percentile limits to define approximate uncertainty envelopes that have been plotted in \autoref{fig:planet}.

The observed near- and mid-infrared emission need not arise entirely from the planetary photospheres. In particular, circumplanetary material and accretion shocks may contribute to the measured luminosity. We therefore distinguish between the luminosity inferred under a planet-only SED model and the planetary component recovered from a model with an additional circumplanetary-disk blackbody.
Schematically,
\begin{equation}
    L_{\rm obs}
    =
    L_{\rm p}
    +
    L_{\rm CPD}
    +
    L_{\rm acc},
\end{equation}
where $L_{\rm p}$ is the intrinsic planetary luminosity, $L_{\rm CPD}$ is the luminosity from circumplanetary material, and $L_{\rm acc}$ is the accretion contribution.
If $L_{\rm CPD}+L_{\rm acc}>0$, then the intrinsic planetary luminosity is lower than the luminosity inferred under a planet-only interpretation. This can shift the intersection with a cooling track towards an older cooling age. We therefore show the planet-only and the planet contribution in a planet-plus-disk luminosity estimates separately. In the latter case, the separately fitted circumplanetary blackbody contribution is excluded from the luminosity compared with the cooling tracks.
 % The temperatures and radii used for these constraints are $T_{\rm eff}=1331$~K, $R=2.22\,R_{\rm Jup}$ and $T_{\rm eff}=1402$~K, $R=1.77\,R_{\rm Jup}$, respectively.
 
The two companions provide different tests of the kinematic system age. For PDS~70~\textit{b} (\autoref{fig:planet}, top), the observed luminosity constraint (for both planet-only and planet-plus-circumplanetary-disk model consideration) is not reached by the selected cooling tracks within the system kinematic age range, even for the highest post-formation entropy considered. The tracks instead intersect the observed luminosity at substantially younger cooling ages. Under the assumptions of the adopted evolutionary models, this indicates that the cooling age of PDS~70~\textit{b} cannot be identified directly with the age of the host stellar system.

We can quantify this offset. Among the selected tracks, the observed luminosity of PDS~70~\textit{b} is reached only by the 4 and 6\,$M_\text{Jup}$ tracks with $S_\text{i} = 11$ and $13~k_\text{B}$ per baryon, at cooling ages between $\approx 1.5$ and $\approx 4.5$~Myr; the 1\,$M_\text{Jup}$ tracks and all tracks with $S_\text{i} = 9~k_\text{B}$ per baryon never reach it. The latest cooling age, $\approx 4.5$~Myr, occurs for the most massive track with the highest initial entropy, compared with the lower edge of the planet-plus-circumplanetary-disk luminosity. Combined with our fiducial system age of $14.92 \pm 2.79$~Myr, this implies that, PDS~70~\textit{b} reached its post-formation state $\approx 10.5$~Myr after the formation of its host star, and no less than $\approx 7.7$~Myr after it even at the lower $1\sigma$ bound of the system age. Because every other combination of mass and initial entropy crosses the observed luminosity earlier, this is a conservative lower limit on the delay within the adopted model grid.

The interpretation is less restrictive for PDS~70~\textit{c} (\autoref{fig:planet}, bottom) because its dynamical mass interval is substantially broader. In the selected model grid, the median-mass track intersects the observed luminosity at a cooling age of about $7$~Myr, while the upper part of the $68\%$ credible interval mass intersects it at approximately $18$~Myr (consistent with the kinematic system age-range). The upper end of the $95\%$ credible mass interval produces luminosities above the observed constraint over all of the plotted age range. Thus, PDS~70~\textit{c} does not provide the same strong evidence for a delayed cooling age as PDS~70~\textit{b}; its interpretation remains limited primarily by the current dynamical mass uncertainty.

% This comparison indicates that the planetary cooling ages inferred under a planet interpretation cannot be identified directly with the stellar-system age. 
While a non-zero delay between stellar birth and the onset of the modelled planetary cooling evolution (at least for PDS~70~\textit{b}) is one possible explanation for the data, this interpretation is not unique. Ongoing accretion can add luminosity that is not captured by isolated cooling tracks, while circumplanetary material can alter the SED and reduce the intrinsic luminosity attributed to the planetary photosphere. However, our planet-plus-circumplanetary-disk comparison already uses only the planetary blackbody component from the two-blackbody SED fit of \citet{blakely2025}, excluding the separately fitted extended circumplanetary-emission component. The persistence of the luminosity mismatch in this case indicates that subtracting this fitted circumplanetary contribution alone does not reconcile the planet-only cooling tracks with the kinematic system age. Ongoing accretion would reduce the intrinsic planetary luminosity and hence the inferred delay, but would itself imply that PDS~70~\textit{b} is still forming at the present system age; in either case, the formation of PDS~70~\textit{b} extends well beyond the formation of its host star.

The dynamical and SED analyses of PDS~70 therefore constrain a combination of stellar age, planetary cooling age, mass, entropy, accretion, and circumplanetary emission rather than a unique formation time.

\section{Conclusions}
\label{sec:conclusions}

We have combined new WiFeS spectroscopy with \textit{Gaia} astrometry and photometry to investigate the kinematic structure and age of the Scorpius--Centaurus T-component population containing the planet-hosting stellar system PDS~70. The spectroscopic measurements provide radial velocities for faint, low-mass candidate members and improve the median radial-velocity precision by a factor of approximately $5.5$ relative to the available \textit{Gaia} measurements. We also measure H$\alpha$ and lithium equivalent widths as independent youth diagnostics.

Our principal results are:
\begin{enumerate}
\item The observed candidate population is spectroscopically young. Lithium identifies $358/405$ observed stars ($88.4\%$) as young under our adopted temperature-dependent criterion, while $237/405$ stars ($58.5\%$) satisfy the H$\alpha$ emission criterion. The two diagnostics are positively correlated, supporting the use of the youth-selected sample as a test of contamination in the kinematic analysis.

\item We apply \texttt{Chronostar} to a 605-star full sample and a 580-star sample in which stars failing both youth diagnostics are removed. The BIC-preferred models contain 14 and 15 components, respectively. The fitted mixture components should be interpreted as mathematical phase-space building blocks rather than as a one-to-one catalogue of discrete star-formation events.

\item The reconstructed T-component population is spatially extended and elongated, with a Gaussian-equivalent major-axis scale of approximately $50$--$63$~pc, a major-to-transverse aspect ratio of $2.4$--$3.0$, and a major axis preferentially aligned with the Galactic plane. The comparable minor-axis dispersions indicate an approximately cylindrical, rather than sheet-like, configuration.

\item The posterior-weighted kinematic-age distribution of the candidate population is concentrated broadly at approximately $10$--$15$~Myr. This is consistent with the younger end of published non-kinematic age estimates for UCL and supports the view that Sco--Cen contains substantial spatial and temporal substructure.

\item We find no statistically significant monotonic age gradient, longitudinal velocity gradient, or coherent solid-body rotation within the reconstructed structure. Apparent component-level trends are consistent with zero within their formal uncertainties.

\item PDS~70 is probabilistically associated with the reconstructed young T-component population. Across the fiducial component models, its posterior-membership-weighted kinematic age spans $11.88$--$14.92$~Myr if we take radial velocity measurements from \textit{Gaia} DR3, or a slightly broader range of approximately $12$--$16$~Myr if we consider other published radial velocity estimates. 
%The detailed age is nearly insensitive to radial velocity in the low-component models but changes by up to approximately $3$~Myr in the more finely decomposed models.

\item When compared with warm-start cooling tracks under the adopted dynamical mass constraints, the planet-only and planet-plus-circumplanetary-disk planetary luminosity estimates for PDS~70~\textit{b} are not readily reproduced at the inferred stellar-system age. The cooling-track intersections occur at younger ages, indicating that the planetary cooling age cannot be directly identified with the stellar kinematic age under an isolated-cooling interpretation. If its luminosity is intrinsic, PDS~70~\textit{b} reached its post-formation state $\approx 10.5$~Myr after its host star formed, and no less than $\approx 7.7$~Myr after it.
The tension is weaker for PDS~70~\textit{c}. Its wider dynamical mass interval allows higher-mass cooling tracks that can reproduce the measured luminosity at ages comparable to the kinematic system-age scale.
\end{enumerate}

The central result is that the stellar-system age and planetary cooling age are not interchangeable, and that the formation of a giant planet can continue for of order 10~Myr after the birth of its host star. Ongoing accretion, circumplanetary emission, and limitations of current atmospheric and evolutionary models for actively forming planets could all modify this estimate. However, accretion contributions to the luminosity of PDS~70~\textit{b} would themselves imply that the planet is still forming at the present system age, so in either case the formation of PDS~70~\textit{b} extends well beyond the formation of its host star.

% This may indicate a delay between stellar birth and the post-formation cooling epoch of the planet. However, this interpretation is degenerate with ongoing accretion, circumplanetary emission, atmospheric modelling, and the limitations of isolated cooling tracks for actively forming planets.

PDS~70 emerges as a system where dynamical masses, direct luminosities, and an independent kinematic age can all be constrained simultaneously -- a combination that makes it uniquely suited to eventually close the gap between cooling models and formation theory. Future radial-velocity monitoring of PDS~70, improved dynamical mass constraints, and joint modelling of its planetary photospheres, accretion, and circumplanetary environments will be required to sharpen this quantitative constraint on how long after its host star a giant planet can finish forming.

% https://arxiv.org/pdf/2509.13607 : sco-cen velocity field vienna group

\section*{Acknowledgements}

AA would like to thank the GASP meeting participants at RSAA and Neelesh Amrutha for help with the processing of the observational data. AA and MRK acknowledge support from the Australian Research Council through Laureate Fellowship FL220100020. This research was undertaken with the assistance of resources from the National Computational Infrastructure (NCI Australia), an NCRIS enabled capability supported by the Australian Government, through award jh2.

%%%%%%%%%%%%%%%%%%%%%%%%%%%%%%%%%%%%%%%%%%%%%%%%%%
\section*{Data Availability}
Full electronic versions of \autoref{tab:full_sample} and \autoref{tab:membership_catalogue} are available at \url{https://github.com/aruuniima/T-subcomponent_PDS70}. The raw WiFeS observations from the ANU 2.3~m telescope and the corresponding extracted spectra are available from the authors upon reasonable request.



\bibliographystyle{mnras}
\bibliography{example} % if your bibtex file is called example.bib

%%%%%%%%%%%%%%%%%%%%%%%%%%%%%%%%%%%%%%%%%%%%%%%%%%

%%%%%%%%%%%%%%%%% APPENDICES %%%%%%%%%%%%%%%%%%%%%

\appendix

\section{Past phase-space structure in 14-component model}
\label{app:14comp}
\begin{figure}
    \centering
    \includegraphics[width=0.95\linewidth]{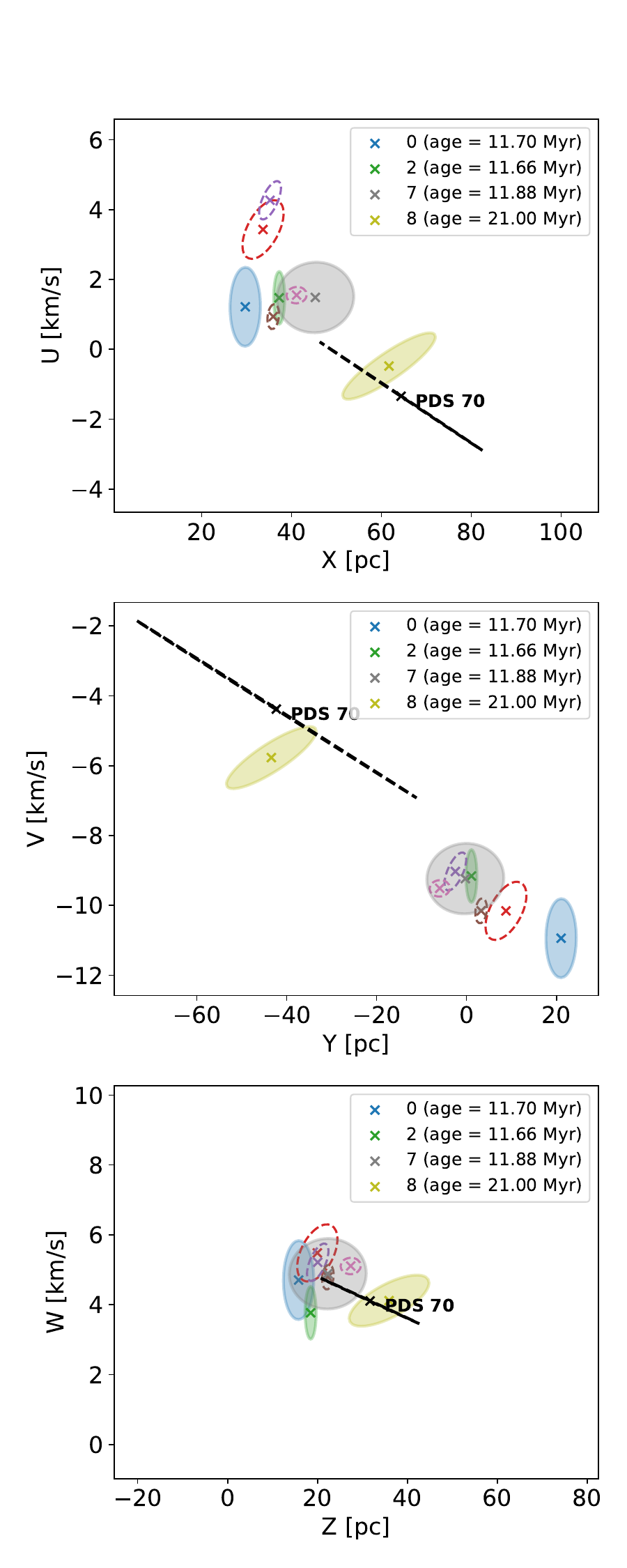}
    \caption{Same as \autoref{fig:pds70_phase_space}, but for the fourteen-component model computed on the full sample rather than the youth-selected one.}
    \label{fig:past_phasespace14}
\end{figure}

\begin{figure}
    \centering
    \includegraphics[width=0.95\linewidth]{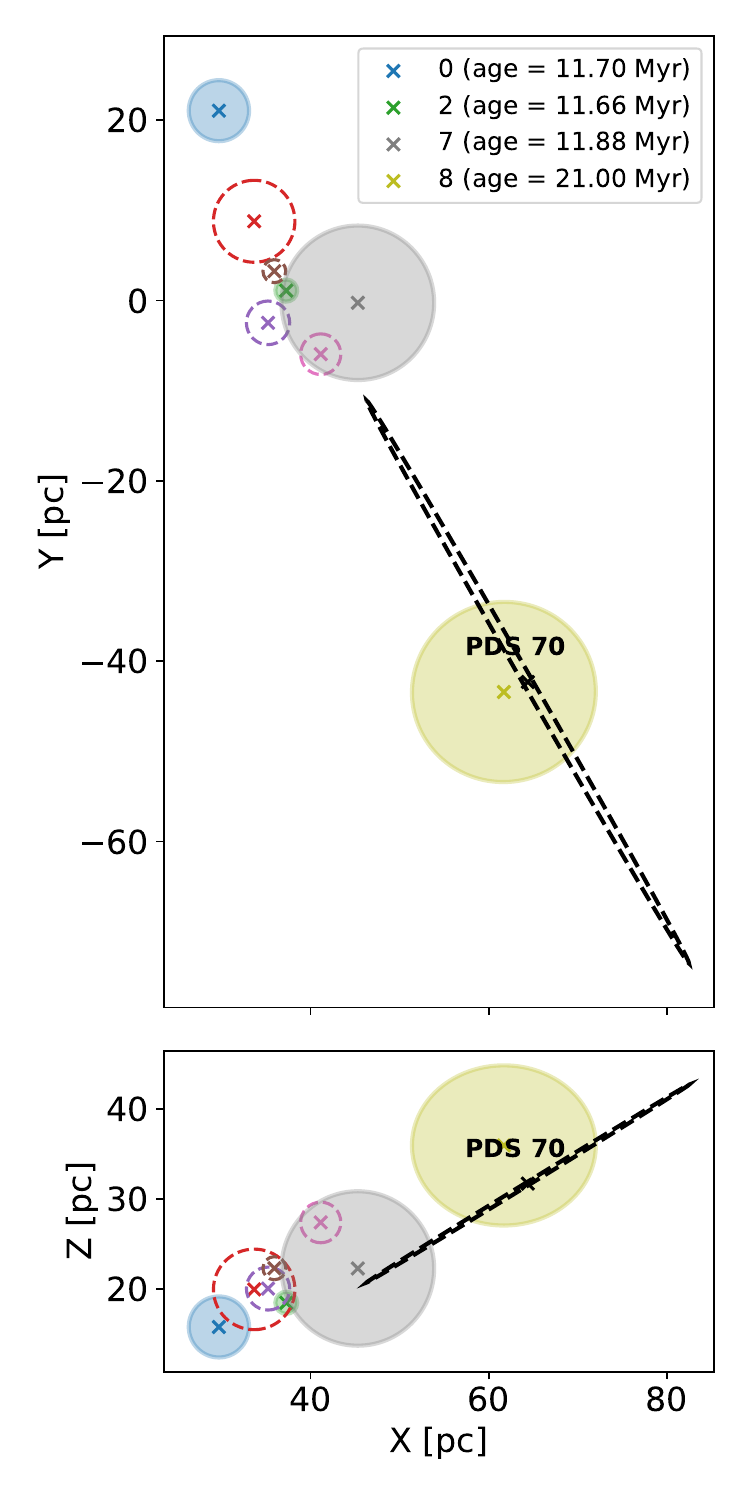}
    \caption{Same as \autoref{fig:pds70_position}, but for the fourteen-component model computed on the full sample rather than the youth-selected one.}
    \label{fig:past_position14}
\end{figure}

Here we provide the analogues of \autoref{fig:pds70_phase_space} and \autoref{fig:pds70_position} in \autoref{subsec:chronostar_past} for the 14-component model (\autoref{fig:past_phasespace14} and \autoref{fig:past_position14}).

\section{Test for solid-body rotation}
\label{app:rotation}

We test whether the reconstructed T-component population exhibits a solid-body rotation pattern about any of the three spatial principal axes. The orthonormal axes $\boldsymbol{e}_1$, $\boldsymbol{e}_2$, and $\boldsymbol{e}_3$ are defined by the principal-axis decomposition of the reconstructed spatial distribution, with $\boldsymbol{e}_1$ corresponding to the major axis. 

For component $j$, we define the position relative to the membership-weighted mixture mean as
\begin{equation}
    \Delta\boldsymbol{x}_j
    =
    \boldsymbol{\mu}_{x,j}
    -
    \bar{\boldsymbol{x}},
\end{equation}
and the velocity relative to the membership-weighted mean velocity as
\begin{equation}
    \Delta\boldsymbol{v}_j
    =
    \boldsymbol{\mu}_{v,j}
    -
    \bar{\boldsymbol{v}},
\end{equation}
where
\begin{equation}
    \bar{\boldsymbol{v}}
    =
    \sum_jw_j\boldsymbol{\mu}_{v,j}.
\end{equation}

The position and velocity coordinates in the principal-axis basis are then
\begin{equation}
    q_{1,j} = s_j
    =
    \Delta\boldsymbol{x}_j\cdot\boldsymbol{e}_1,
    \qquad
    q_{2,j}
    =
    \Delta\boldsymbol{x}_j\cdot\boldsymbol{e}_2,
    \qquad
    q_{3,j}
    =
    \Delta\boldsymbol{x}_j\cdot\boldsymbol{e}_3,
\end{equation}
and
\begin{equation}
    u_{1,j}
    =
    \Delta\boldsymbol{v}_j\cdot\boldsymbol{e}_1,
    \qquad
    u_{2,j}
    =
    \Delta\boldsymbol{v}_j\cdot\boldsymbol{e}_2,
    \qquad
    u_{3,j}
    =
    \Delta\boldsymbol{v}_j\cdot\boldsymbol{e}_3.
\end{equation}

For rotation about a single principal axis $\boldsymbol{e}_k$, the velocity field is
\begin{equation}
    \boldsymbol{u}_j
    =
    \boldsymbol{\Omega}_k \times \Delta\boldsymbol{x}_j,
    \qquad
    \boldsymbol{\Omega}_k = \Omega_k \boldsymbol{e}_k.
\end{equation}

Only the two velocity components perpendicular to that axis are non-zero, and each must vary linearly with the corresponding transverse position coordinate. For example, for rigid rotation about the major axis, $\boldsymbol{\Omega}_1=\Omega_1\boldsymbol{e}_1$, the transverse velocities satisfy
\begin{equation}
    u_{2,j}
    =
    -\Omega_1 q_{3,j},
    \qquad
    u_{3,j}
    =
    +\Omega_1 q_{2,j}.
    \label{eq:rigid_rotation}
\end{equation}
Thus, independent linear regressions in the $u_2$--$q_3$ and $u_3$--$q_2$ projections should recover slopes of equal magnitude and opposite sign if the reconstructed structure is undergoing coherent rigid rotation about $\boldsymbol{e}_1$. Physically, this corresponds to the filament rolling about its long axis.

For completeness, we also calculate the angular frequency that would be implied if the transverse motions were forced to follow a rigid-rotation model. We estimate
\begin{equation}
    \Omega_1
    =
    \frac{
        \sum_j w_j
        \left(
            q_{2,j}u_{3,j}
            -
            q_{3,j}u_{2,j}
        \right)
    }{
        \sum_jw_j
        \left(
            q_{2,j}^{2}
            +
            q_{3,j}^{2}
        \right)
    }.
    \label{eq:omega_major_axis}
\end{equation}
This expression is the membership-weighted angular momentum about
$\boldsymbol{e}_1$ divided by the corresponding moment-of-inertia term. Analogous expressions yield $\Omega_2$ and $\Omega_3$ for rotation about $\boldsymbol{e}_2$ and $\boldsymbol{e}_3$, respectively.

\begin{figure}
    \centering
    \includegraphics[width=\linewidth]{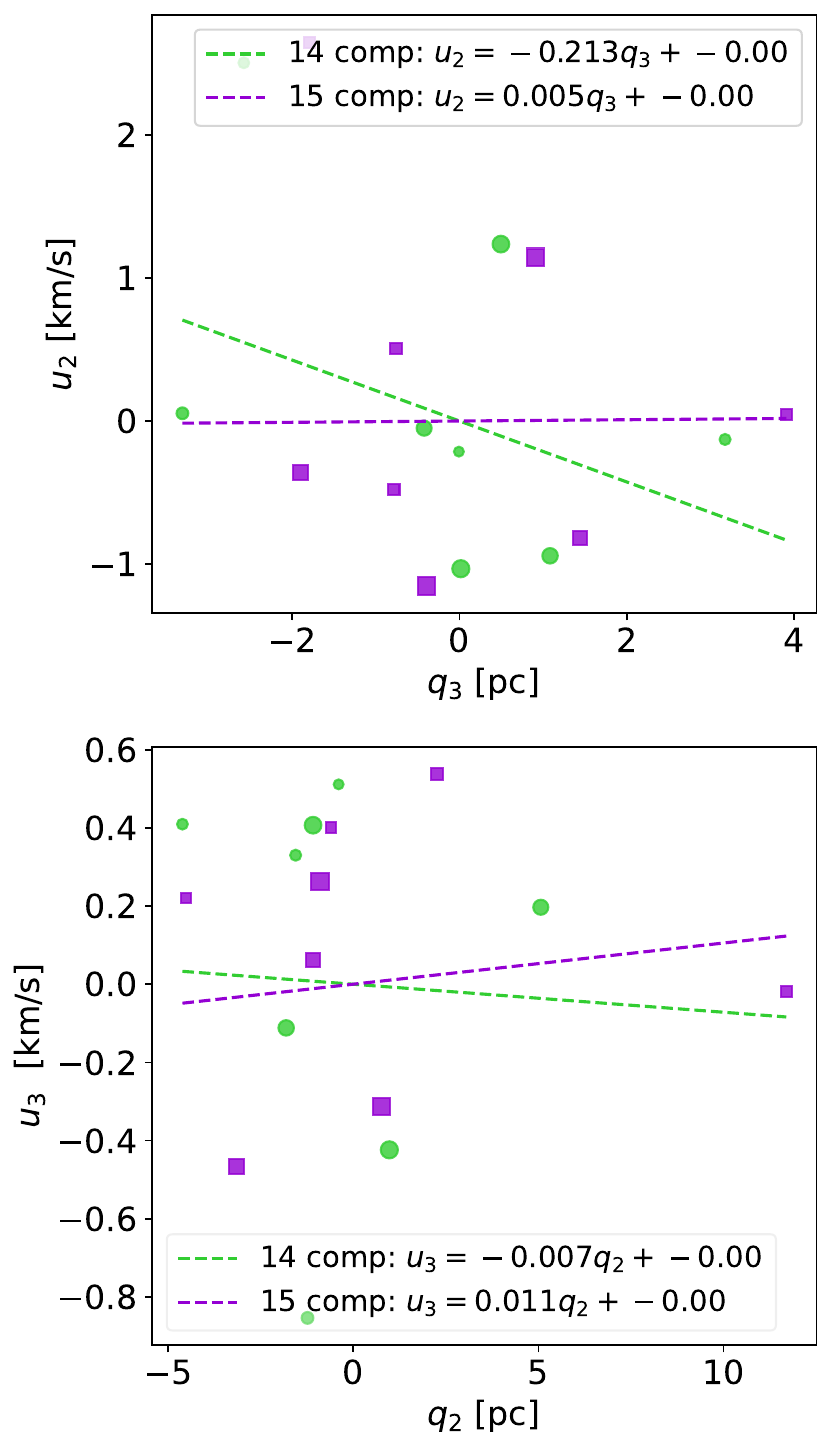}
    \caption{Test for solid-body rotation about the major axis
    $\boldsymbol{e}_1$ at $\tau=11.6$~Myr. The two panels show the transverse
    velocity--position relations required by $u_2=-\Omega_1q_3$ and
    $u_3=+\Omega_1q_2$. Green circles and purple squares correspond to the 14-component full-sample and 15-component youth-selected fits, respectively. Dashed lines show independent membership-weighted linear regressions for the same in the same colours. The fitted slopes do not provide the
    equal-amplitude, opposite-sign pattern expected for a common angular
    velocity.}
    \label{fig:rotation_e1}
\end{figure}

\begin{figure}
    \centering
    \includegraphics[width=\linewidth]{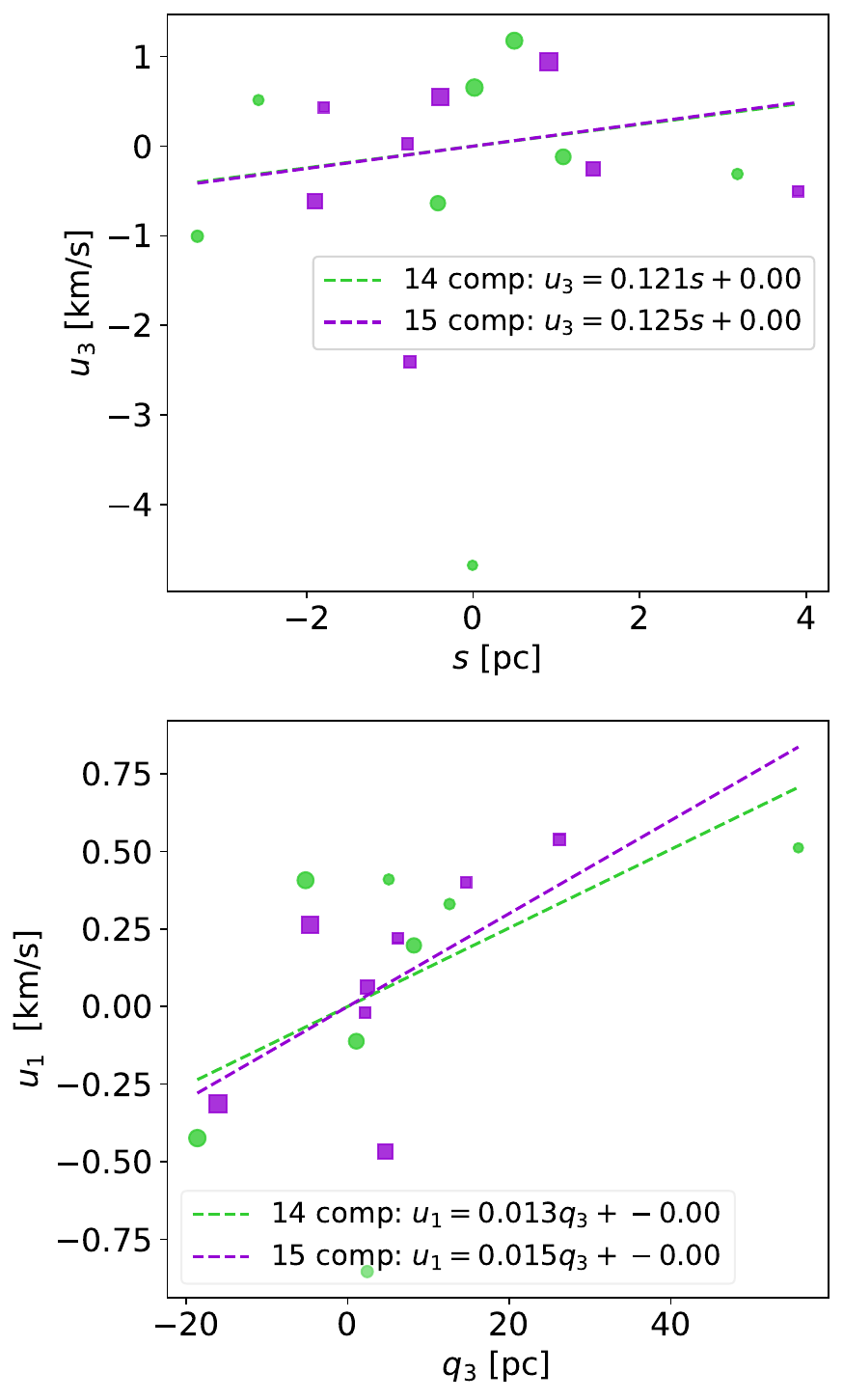}
    \caption{Test for solid-body rotation about the intermediate principal
    axis $\boldsymbol{e}_2$ at $\tau=11.6$~Myr. The two panels show the transverse
    velocity--position relations required by $u_3=-\Omega_2q_1=-\Omega_2s$ and $u_1=+\Omega_2q_3$. Green circles and purple squares correspond to the 14-component full-sample and 15-component youth-selected fits, respectively. Dashed lines show independent membership-weighted linear regressions for the same in the same colours. The fitted slopes are inconsistent with a single
    angular velocity.}
    \label{fig:rotation_e2}
\end{figure}

\begin{figure}
    \centering
    \includegraphics[width=\linewidth]{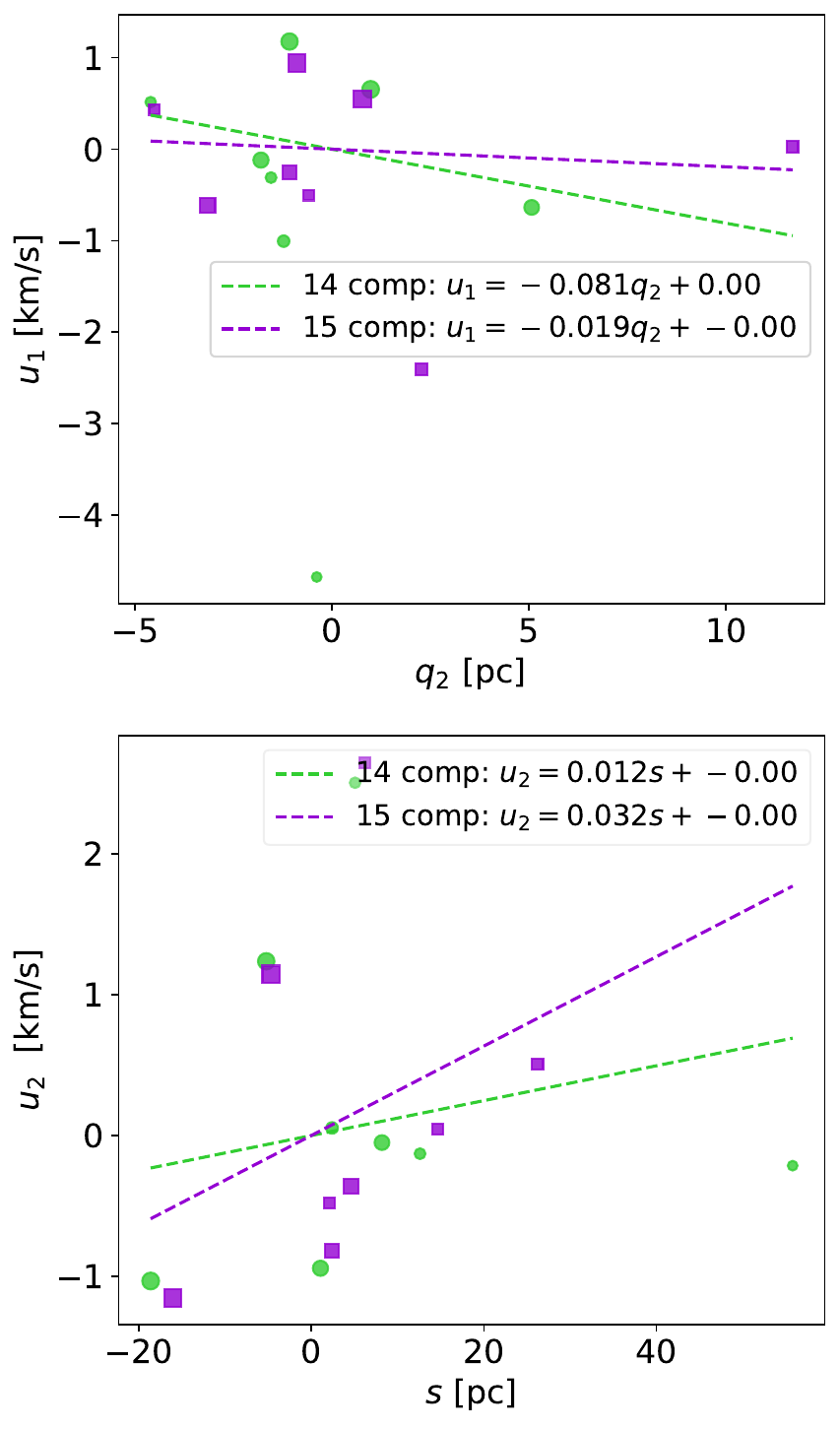}
    \caption{Test for solid-body rotation about the third principal axis
    $\boldsymbol{e}_3$ at $\tau=11.6$~Myr. The two panels show the velocity--position
    relations required by $u_1=-\Omega_3q_2$ and
    $u_2=+\Omega_3q_1=+\Omega_3s$. Green circles and purple squares correspond to the 14-component full-sample and 15-component youth-selected fits, respectively. Dashed lines show independent membership-weighted linear regressions for the same in the same colours. Although the central slopes have the expected
    opposite signs, they are individually consistent with zero and do not
    provide a significant common angular velocity.}
    \label{fig:rotation_e3}
\end{figure}

For each axis and each model (the 14- and 15-component models described in \autoref{subsec:chronostar_past}), we fit the two transverse velocity--position relations with linear equations, using posterior-effective-membership weights. We denote the fitted slopes by $b_{k,a}$ and $b_{k,b}$, corresponding to the first and second relations in \autoref{eq:omega_major_axis}. The fitted slopes, their formal uncertainties, and the combined angular-speed estimates $\Omega_k$ and nominal rotation times $T_{{\rm rot},k}=2\pi/\Omega_k$ for each axis and model are summarised in \autoref{tab:rotation_summary}. The corresponding diagnostic plots are shown in \autoref{fig:rotation_e1}, \autoref{fig:rotation_e2} and \autoref{fig:rotation_e3}.

\begin{table}
    \centering
    \caption{Summary of solid-body rotation tests about the three principal axes. Each slope is in $\mathrm{km\,s^{-1}\,pc^{-1}}$. For a given axis, solid-body rotation requires the two slopes to have equal magnitudes and the signs specified by the cross product (opposite signs for the two slopes). The combined angular-speed estimates $\Omega_k$ and nominal rotation times $T_{{\rm rot},k}=2\pi/\Omega_k$ are also listed.}
    \label{tab:rotation_summary}
    \begin{tabular}{lccccc}
        \hline
        Axis, model &
        $b_{k,a}$ &
        $b_{k,b}$ &
        $\Omega_k$ &
        $T_{{\rm rot},k}$ \\
        $\boldsymbol{e}_k$&{\small [km s$^{-1}$ pc$^{-1}]$}&{\small [km s$^{-1}$ pc$^{-1}$]}&{\small [km s$^{-1}$ pc$^{-1}$]}&{\small [Myr]}\\
        
        \hline
        $\boldsymbol{e}_1$, 14 comp.
        & $-0.213\pm0.662$
        & $-0.007\pm0.385$
        & $0.048$ & $21.1$ \\

        $\boldsymbol{e}_1$, 15 comp.
        & $+0.005\pm0.673$
        & $+0.011\pm0.276$
        & $0.008$ & $122.1$ \\

        $\boldsymbol{e}_2$, 14 comp.
        & $+0.121\pm0.662$
        & $+0.013\pm0.067$
        & $0.011$ & $90.5$ \\

        $\boldsymbol{e}_2$, 15 comp.
        & $+0.125\pm0.673$
        & $+0.015\pm0.086$
        & $0.013$ & $80.4$ \\

        $\boldsymbol{e}_3$, 14 comp.
        & $-0.081\pm0.385$
        & $+0.012\pm0.067$
        & $0.014$ & $71.0$ \\

        $\boldsymbol{e}_3$, 15 comp.
        & $-0.019\pm0.276$
        & $+0.032\pm0.086$
        & $0.031$ & $33.4$ \\
        \hline
    \end{tabular}
\end{table}

Across all three principal axes, the paired velocity--position relations do not provide a statistically significant, model-independent detection of solid-body rotation. The individual slopes are generally consistent with zero within their formal uncertainties, the two relations required for a common angular velocity are not robustly matched, and the inferred angular speeds $\Omega_k$ differ substantially between the 14-component and 15-component decompositions (with nominal rotation timescales of approximately $20$--$120$~Myr). We therefore find no compelling evidence that the reconstructed T-component population undergoes coherent solid-body rotation about any of its principal axes.

% https://www.aanda.org/articles/aa/pdf/2026/05/aa55519-25.pdf this paper has RV survey compilations for stars in Sco-Cen
% check actions and such and use our last papers stuff as well?
% this has everything from the project i think: https://github.com/mikeireland/chronostar/tree/master/projects/scocen 

%%%%%%%%%%%%%%%%%%%%%%%%%%%%%%%%%%%%%%%%%%%%%%%%%%

% Don't change these lines
\bsp	% typesetting comment
\label{lastpage}
\end{document}